\documentclass[aps,prd,twocolumn,times,graphicx,amssymb,tighten,showpacs,groupedaddress,superscriptaddress,floatfix]{revtex4-1}

\usepackage{graphicx,amssymb,amsmath}
\usepackage[dvips]{xcolor}
\usepackage{soul}
\usepackage{hyperref}
\usepackage{amsmath}
\usepackage{amssymb}
\newcommand{\be}{\begin{equation}}
\newcommand{\ee}{\end{equation}}

\definecolor{mycolor}{rgb}{0,0.71,0.05001}

\begin{document}
\title{Charged-current neutrino-nucleus reactions within the SuSAv2-MEC approach}

\author{G.D.~Megias}
\affiliation{Departamento de F\'{i}sica At\'omica, Molecular y Nuclear, Universidad de Sevilla, 41080 Sevilla, Spain}
\author{J.E.~Amaro}
\affiliation{Departamento de F\'isica At\'omica,  Molecular y Nuclear and Instituto Carlos I de F\'isica Te\'orica y Computacional, Universidad  de  Granada,  18071  Granada,  Spain}

\author{M.B.~Barbaro}
\affiliation{Dipartimento di Fisica, Universit\`{a} di Torino and INFN, Sezione di Torino, Via P. Giuria 1, 10125 Torino, Italy}

\author{J.A.~Caballero}
\affiliation{Departamento de F\'{i}sica At\'omica, Molecular y Nuclear, Universidad de Sevilla, 41080 Sevilla, Spain}

\author{T.W.~Donnelly}
\affiliation{Center for Theoretical Physics, Laboratory for Nuclear Science and Department of Physics, Massachusetts Institute of Technology, Cambridge, Massachusetts 02139, USA}

\author{I.~Ruiz~Simo}
\affiliation{Departamento de F\'isica At\'omica,  Molecular y Nuclear and Instituto Carlos I de F\'isica Te\'orica y Computacional, Universidad  de  Granada,  18071  Granada,  Spain}

\date{\today}
\begin{abstract}
We present a detailed study of charged-current (CC) neutrino-nucleus
reactions in a fully relativistic framework and comparisons with
recent experiments spanning an energy range from hundreds of MeV up to
100 GeV within the SuperScaling Approach, which is based on the
analysis of electron-nucleus scattering data and has been recently
improved with the inclusion of Relativistic Mean Field theory
effects. We also evaluate and discuss the impact of two-particle
two-hole meson-exchange currents (2p-2h MEC) on neutrino-nucleus
interactions through the analysis of two-particle two-hole axial and
vector contributions to weak response functions in a fully
relativistic Fermi gas. The results show a fairly good agreement with
experimental data over the whole range of neutrino energies.
\end{abstract}


\maketitle

\section{Introduction}

The enormous progress over recent years on neutrino oscillation
experiments have motivated many theoretical efforts to achieve a
consistent and accurate description of neutrino-nucleus scattering in
the GeV region. At these kinematics, several measurements of CC
neutrino-nucleus scattering cross sections have been performed by
different collaborations
(MiniBooNE~\cite{AguilarArevalo:2010zc,AguilarArevalo:2013hm},
NOMAD~\cite{Lyubushkin:2008pe},
T2K~\cite{T2Kincl,T2Kinclelectron,T2Kcc0pi},
SciBooNE~\cite{SciBooNEincl}, MINER$\nu$A~\cite{Minerva1,Minerva2}),
revealing the need of describing in a precise way the relevant
reaction mechanisms, mainly the quasielastic (QE) regime, one pion
production and 2p-2h MEC contributions. In particular, the CCQE
MiniBooNE results~\cite{AguilarArevalo:2010zc,AguilarArevalo:2013hm}
have stimulated many theoretical studies devoted to explaining the
apparent discrepancies between data and most theoretical predictions
based on the Impulse Approximation (IA). Based on results from
different groups, the inclusion of effects beyond IA, such as
multinucleon excitations, mainly 2p-2h MEC contributions, has allowed
one to explain these data without including any effective parameter
(such as the axial mass
$M_A$)~\cite{Martini:2009aa,Amaro:2010sd,Nieves:2011yp,Lalakulich:2012ac}.

In this context, a consistent evaluation of the ($e,e'$) cross section
in the same kinematical regime is crucial for a proper analysis of
neutrino-nucleus interactions as it provides a decisive benchmark for
assessing the validity of the theoretical description not only in the
QE regime, but also for the 2p-2h MEC contributions as well as at
higher energy transfers (nucleonic resonances, inelastic
spectrum). This has been recently studied in detail
in~\cite{Megias:2016ee}, where good agreement with ($e,e'$) data is
reached in the framework of the SuperScaling Approach (SuSA) for a
wide range of kinematics, covering from the QE regime to the deep
inelastic spectrum.

The SuSA approach~\cite{Day:1990,super,super1,Amaro:2004bs} assumes
the existence of universal scaling functions for both electromagnetic
and weak interactions. Analyses of inclusive ($e,e'$) data have shown
that at energy transfers below the QE peak superscaling is fulfilled
with very good accuracy  \cite{super,super1,Chiara1}:
this implies that the reduced cross section exhibits an independence
of the momentum transfer (first-kind scaling) and of the nuclear
target (second-kind scaling) when expressed as a function of the
appropriate scaling variable ($\psi$), itself a function of the energy
($\omega$) and momentum transfer ($q$). Nevertheless, at energies
above the QE peak both kinds of scaling are violated, which is
associated with effects beyond IA, such as 2p-2h MEC or with inelastic
contributions. An extension of this formalism, originally introduced
to describe the QE regime, to the $\Delta$-resonance domain and the
complete inelastic spectrum -- resonant, non-resonant and deep
inelastic scattering (DIS) -- has also been proposed in recent
works~\cite{Barbaro:2003ie,Maieron:2009an,Ivanov:2016Delta} .

Recently we have developed an improved version of the superscaling
prescription, called SuSAv2~\cite{Gonzalez-Jimenez:2014eqa}, by
incorporating relativistic mean field (RMF)
effects~\cite{Caballero:2005sj,Caballero:2006wi,Caballero:2007tz} in
the longitudinal and transverse nuclear responses, as well as in the
isovector and isoscalar channels. This is of great interest in order
to describe CC neutrino reactions that are purely-isovector.
Furthermore, a natural enhancement of the transverse nuclear response emerges from the RMF theory as a genuine relativistic effect.

As mentioned before, 2p-2h MEC play an important role in the analysis of 
neutrino oscillation experiments, being relevant especially in the ``dip'' region between the QE and the $\Delta$ peaks. These are added to our model in the so-called SuSAv2-MEC approach.

Although a comparison of neutrino scattering data with the SuSAv2-MEC predictions was already performed in~\cite{Megias:2014qva}, here two novelties are introduced in the model.
The first one concerns the implementation of RMF effects in the SuSA approach.
While the RMF works properly at low to intermediate $q$-values, where the final-state interactions (FSI) between the outgoing nucleon and the residual nucleus are significant, at higher momentum transfers these effects should become negligible and the Relativistic Plane Wave Impulse Approximation (RPWIA) - where the initial state is described by a mean field but FSI are neglected - is more appropriate to describe the nuclear dynamics in this regime.
The pure RMF model fails to reproduce the smooth transition to the RPWIA at high $q$ due to the strong energy-independent scalar and vector potentials included in the model.
Hence both approaches are incorporated in the present SuSAv2 model by using a $q$-dependent blending function, as described in~\cite{Megias:2016ee}, in such a way that the RMF dominates at low and intermediate $q$-values whereas the RPWIA contributions start to be relevant at higher momentum transfer.
The same approach has 
been applied 
not only to the QE but also to the inelastic regime and has been shown to provide a successful description of electron-nucleus inclusive data~\cite{Megias:2016ee} once 2p-2h excitations are also taken into account.

The second new aspect of the present calculation concerns the treatment of 2p-2h excitations. In~\cite{Megias:2014qva} we used the exact fully relativistic {\it vector} MEC evaluated in~\cite{DePace:2003xu,DePace:2004xu}.
In this work we include for the first time the fully relativistic weak (with {\it vector} and {\it axial} components) charged meson-exchange currents, 
in both longitudinal and transverse channels. These have been evaluated in~\cite{Simo:2014wka,Simo:2016aa,Simo:2016ab} from an exact microscopic calculation, where the two-body current is the sum of seagull, pion-in-flight, pion-pole and $\Delta$-pole operators and the basis wave functions are non-interacting Dirac spinors.

From this baseline, the SuSAv2-MEC predictions can be employed for the analysis of neutrino-nucleus reactions covering the entire energy spectrum once all the inelastic channels, already included for electron scattering, are also incorporated for neutrino reactions. This is presently in progress and results will be presented in a forthcoming publication. In this work we restrict ourselves to the contribution ascribed to the $\Delta$-resonance that in most of the cases plays a major role.


This paper is organized as follows. In Sect. II we briefly introduce the formalism for CCQE neutrino-nucleus scattering and describe our 2p-2h MEC calculations. In Sect. III we present a comparison of our QE and 2p-2h MEC predictions with all recent CCQE neutrino experimental data. An extension to the analysis of inclusive neutrino cross sections is shown in Sect. IV. Finally, in Sect. V we draw the conclusions of our study, including some remarks related to further work.

\section{Theoretical formalism}


The general formalism describing CC neutrino-nucleus scattering 
has been detailed in previous works~\cite{Day:1990,Amaro:2005sr}, where the double differential ($\nu_l,l'$) cross section is given as the sum of longitudinal (L) and transverse (T) channels, each of them composed of pure vector (VV) and axial components (AA), and the interference transverse (T') vector-axial channel (VA), which is constructive (+) for neutrino scattering and destructive (-) for antineutrino one:

\begin{equation}
\frac{d\sigma}{dk' d\Omega}=\sigma_0\left(\hat V_LR_L+\hat V_TR_T\pm2\hat V_{T'}R_{T'}\right) \,,
\end{equation}
where 
\begin{equation}
\hat V_LR_L=\hat V_{CC}R_{CC}+
2 \,\hat V_{CL}R_{CL}+\hat V_{LL}R_{LL} \,,
\label{eq:vlrl}
\end{equation}
$\hat V_K$ are kinematical factors, $R_K$ are the nuclear response
functions and
\begin{equation}
\sigma_0=
\frac{G_F^2\cos^2\theta_c}{2\pi^2}
\left(k^\prime \cos\frac{\tilde\theta}{2}\right)^2
\end{equation}
depends on the Fermi
constant $G_F$, the Cabibbo angle $\theta_c$, the outgoing lepton
momentum $k^\prime$, and the generalized scattering angle
$\tilde\theta$.

As anticipated in the Introduction, in this work we evaluate the nuclear responses by employing a set of purely isovector scaling functions based on the RMF and the RPWIA models (SuSAv2 model) in order to account properly the FSI between the ougoing nucleon and the residual nucleus, as described in~\cite{Gonzalez-Jimenez:2014eqa,Megias:2016ee}. 

Concerning the description of the 2p-2h MEC, we employ a calculation performed within the relativistic Fermi gas model in which a fully Lorentz covariant analysis can be achieved~\cite{Simo:2014wka,Simo:2016aa}. In the present study we include for the first time the axial contribution in both longitudinal and transverse channels. 

As it has been analyzed in previous works, a fully relativistic
calculation of the 2p-2h MEC response functions involves a non-trivial
calculation of all the many-body MEC diagrams, which implies more than
100,000 terms and subsequent seven-dimensional integrations. To reduce
the computational time as well as to ease the implementation of the
model in MonteCarlo generators used in the analysis of current
neutrino oscillation experiments, where a broad range of kinematics
are involved, we make use of a parametrization of the MEC
responses. The functional form employed for the parametrization of the
transverse electromagnetic vector response was detailed
in~\cite{Megias:2014qva}. In the present work, we follow this
prescription and extend it to the different axial and vector
components involved in the analysis of CC neutrino reactions as well
as considering both transverse and longitudinal contributions.

Finally, we consider an extension of the SuSAv2 model to the region
where the $\Delta$-excitation dominates, as presented
in~\cite{Ivanov:2016Delta}. This approach has been carried out by
subtracting the QE+MEC contribution from the experimental ($e,e'$)
cross section in a similar way as done in the  SuperScaling
model for the QE regime. Therefore, we obtain a new scaling function
$f^\Delta$ which is suited to the $\Delta$-resonance region and can be
applied to analyze inclusive neutrino-nucleus cross sections, such as
recent results from the T2K and SciBooNE collaborations.

\subsection{2p-2h MEC responses
}
\label{sec:2p2h}



In this section we illustrate and discuss in detail some properties of the 2p-2h MEC response functions.

\begin{figure}[H]
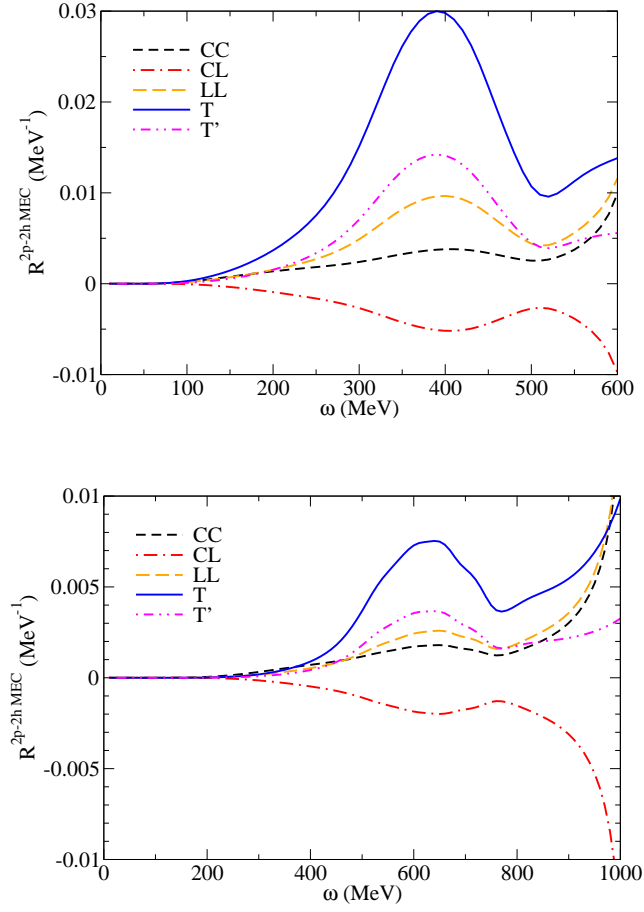

\begin{center}\vspace{-0.2cm}
\includegraphics[scale=0.32, angle=0]{RMEC_q600.eps}\vspace{0.85cm}\\
\includegraphics[scale=0.32, angle=0]{RMEC_q1000.eps}
\begin{center}
\vspace{-1cm}
\end{center}
\end{center}
\caption{(Color online) Comparison between 2p-2h MEC transverse ($T=T_{VV}+T_{AA}$ and $T'=T'_{VA}$) response functions and the longitudinal ones ($CC$, $CL$ and $LL$) at $q=600$ MeV/c (top panel) and $q=1000$ MeV/c (bottom panel).}\label{mec1}
\end{figure}

\begin{figure}[H]
\begin{center}\vspace{-0.2cm}
\includegraphics[scale=0.32, angle=270]{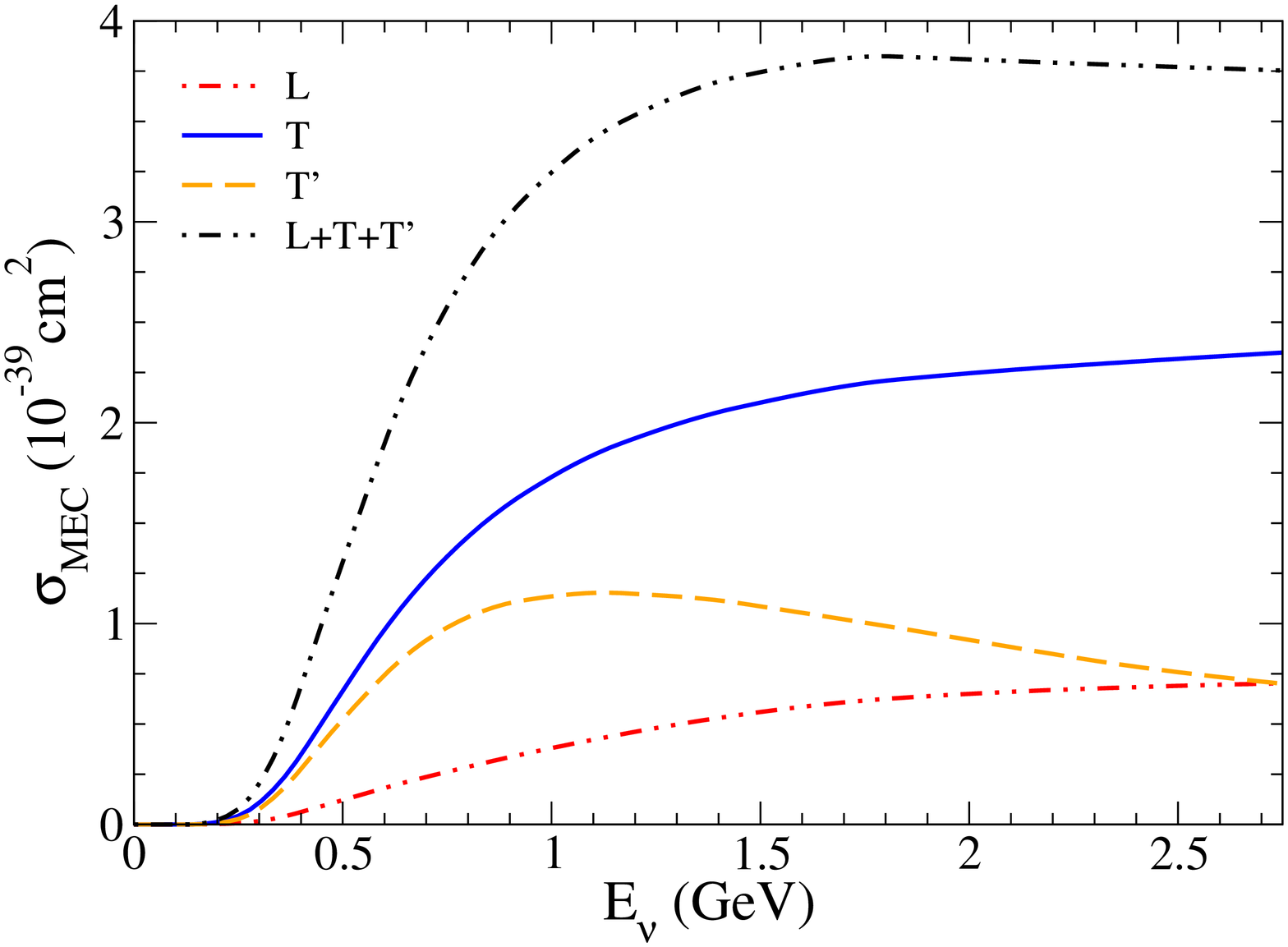}\\
\includegraphics[scale=0.32, angle=270]{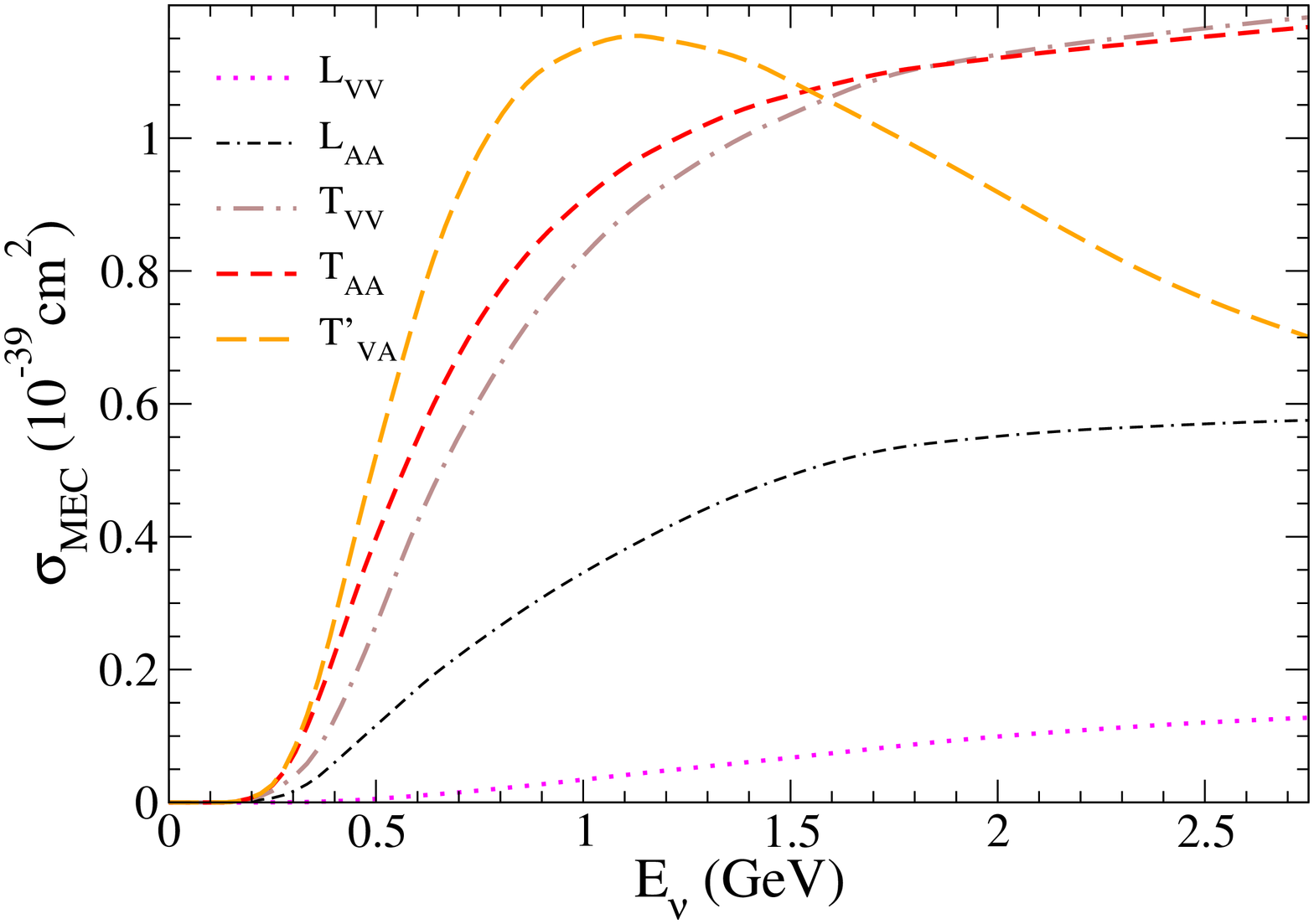}
\begin{center}
\vspace{-1cm}
\end{center}
\end{center}
\caption{(Color online) Separation  into  components  of  the
total 2p-2h MEC $\nu_\mu$ cross section  displayed  versus  neutrino energy $E_\nu$. The total longitudinal ($L$), transverse ($T$) and transverse interference ($T'$) contributions are shown (top panel) as well as the total 2p-2h MEC cross section ($L+T+T'$). Longitudinal and transverse channels are decomposed into vector and axial contributions (bottom panel). }\label{sigmamec}
\end{figure}

In Fig. \ref{mec1} we compare the contributions of the different 2p-2h
MEC responses as functions of the energy transferred to the nucleus
for two values of the momentum transfer, $q$=600 and 1000 MeV/c.  Note
that in general the 5 responses are comparable in size, depending on
the specific kinematics. However, in the cross section the
contribution of the CC and LL is roughly compensated by that of the
negative CL response, so that for neutrino energies below $\sim$ 1 GeV
the net longitudinal contribution plays a minor role in the total MEC
response.  This is illustrated in Fig. \ref{sigmamec}, where the L, T
and T' contributions to the 2p-2h MEC cross section are displayed
versus the neutrino energy.  At higher energies the L and T'
contributions become comparable, both being much smaller that the
dominant T one. 

The balance between the longitudinal and transverse 2p-2h channel discussed above is somehow different from the one emerging in the electromagnetic case.
As described in a recent work~\cite{Megias:2016ee}, the longitudinal electromagnetic MEC response is indeed negligible with regard to the transverse one. 
However, as illustrated in Fig. 2b, we notice that when computing the total 2p-2h MEC weak cross section the longitudinal contribution is dominated by the axial channel and thus it plays a more relevant role compared with the EM case.


Concerning the transverse responses, it is noticeable that the magnitude of the pure axial and vector channels to the cross section are very similar.
Moreover, the vector-axial interference contribution reaches its maximum around $E_\nu \sim $1 GeV and decreases at higher energies as a consequence of the behavior of the leptonic factor $V_{T'}$.

\begin{figure}[H]
\begin{center}\vspace{-0.2cm}
\includegraphics[scale=0.32, angle=270]{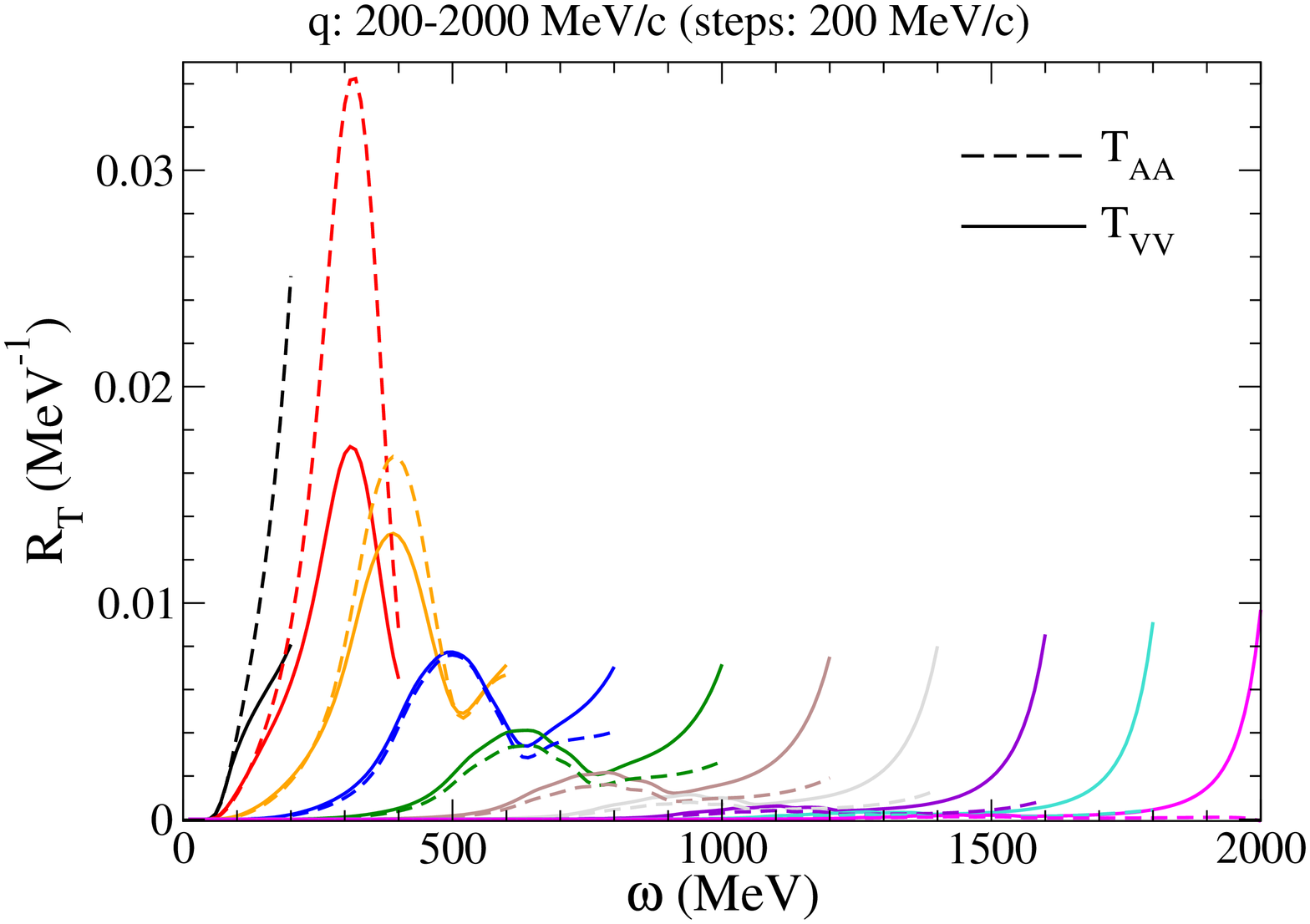}\hspace*{-0.15cm}\\
\includegraphics[scale=0.32, angle=270]{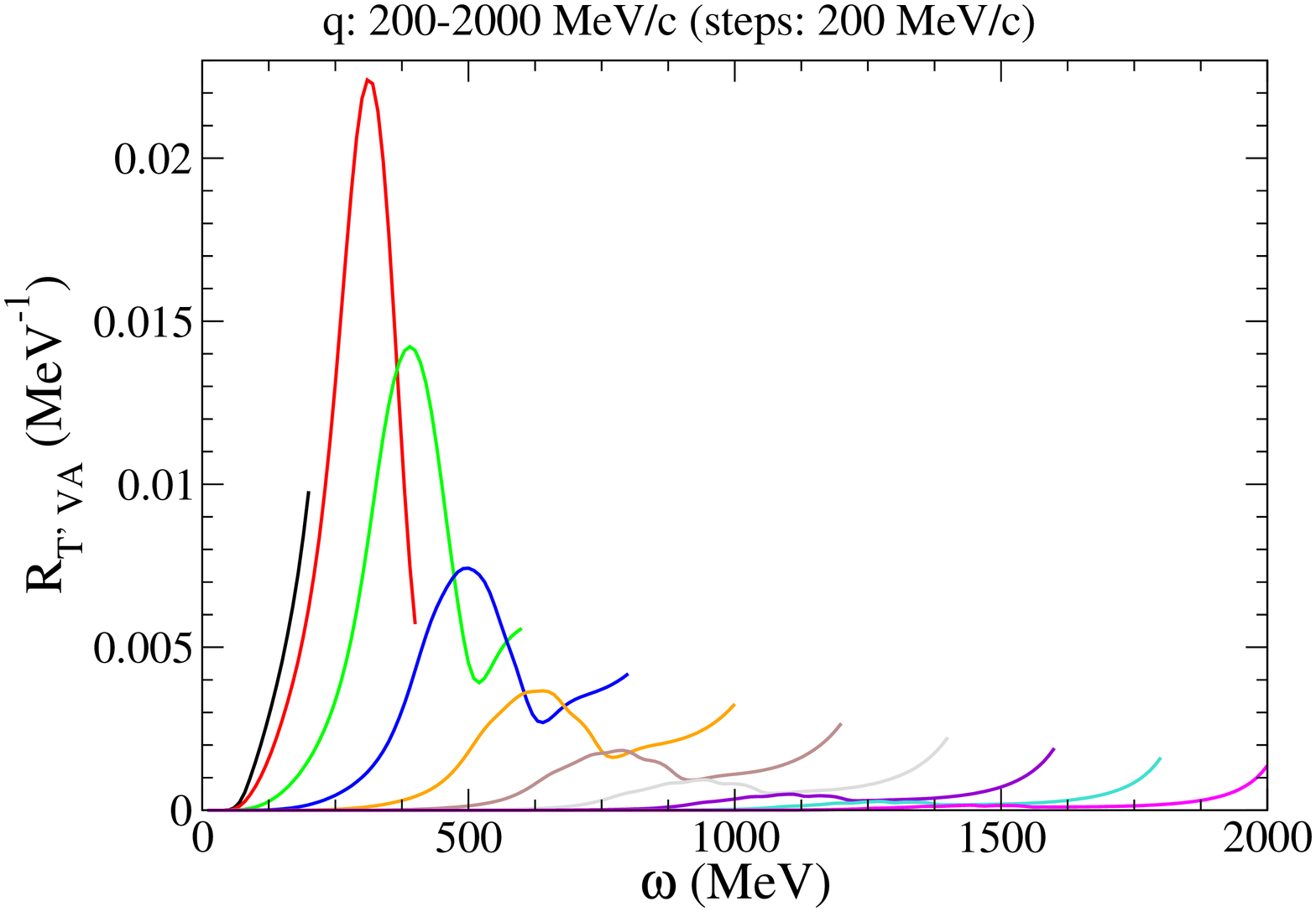}\hspace*{-0.15cm}
\begin{center}
\vspace{-1cm}
\end{center}
\end{center}
\caption{(Color online) Comparison between 2p-2h MEC axial-axial transverse ($T_{AA}$) response functions and the vector-vector ones ($T_{VV}$) versus $\omega$ (top panel). The transverse vector-axial interference term ($T'_{VA}$) is also shown (bottom panel). The curves are displayed
from left to right in steps of $q=200$ MeV/c.}\label{mec2}
\end{figure}

The analysis of the evolution with $q$ of the individual transverse components (see Fig. \ref{mec2}) shows that the axial term is larger than the vector one at low-intermediate kinematics ($q<800$ MeV/c) whereas the opposite occurs at higher kinematics. 



\begin{figure}[H]
\begin{center}\vspace{-0.2cm}
\includegraphics[scale=0.225, angle=0]{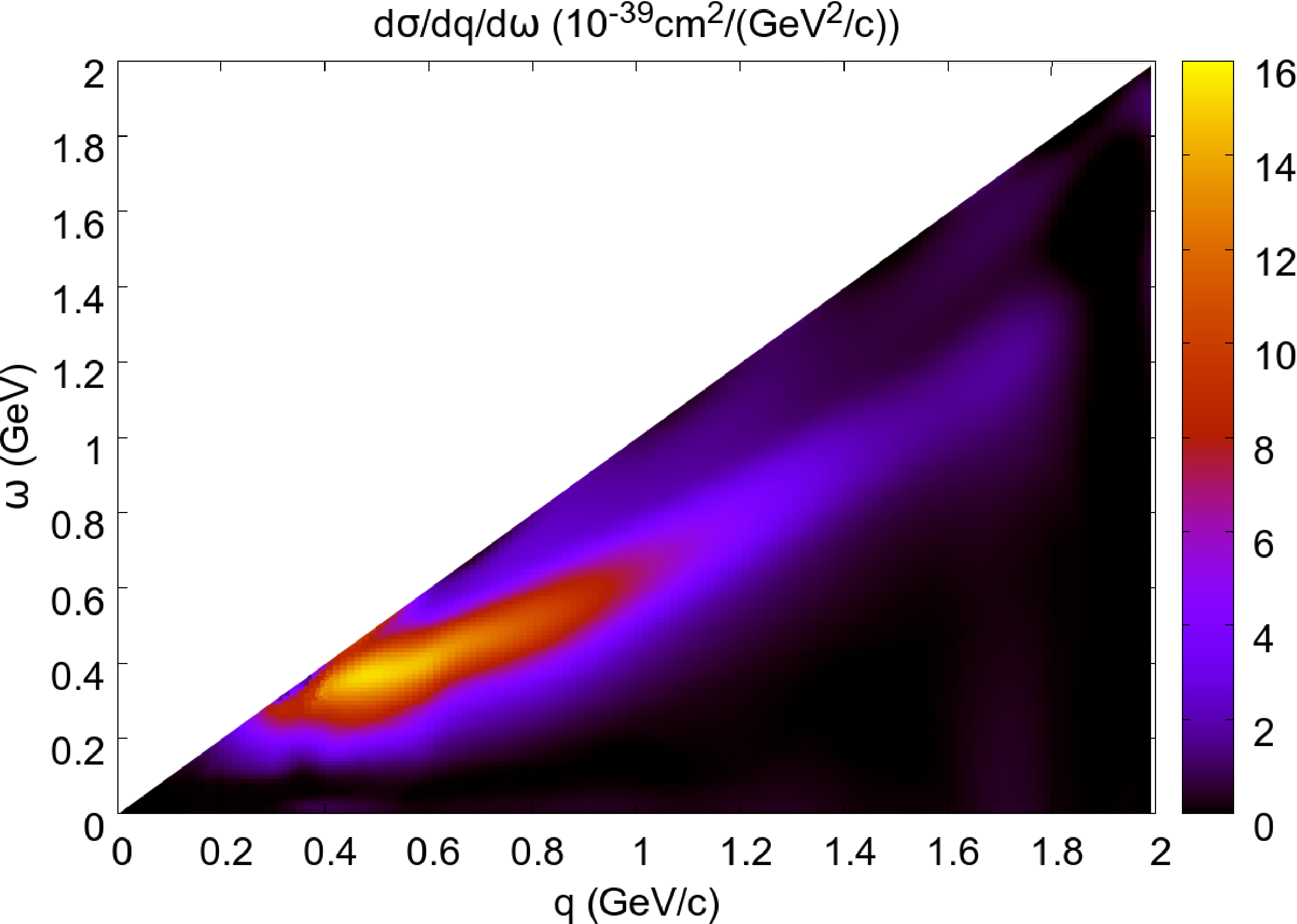}\\ \vspace{0.4cm}
\includegraphics[scale=0.292, angle=0]{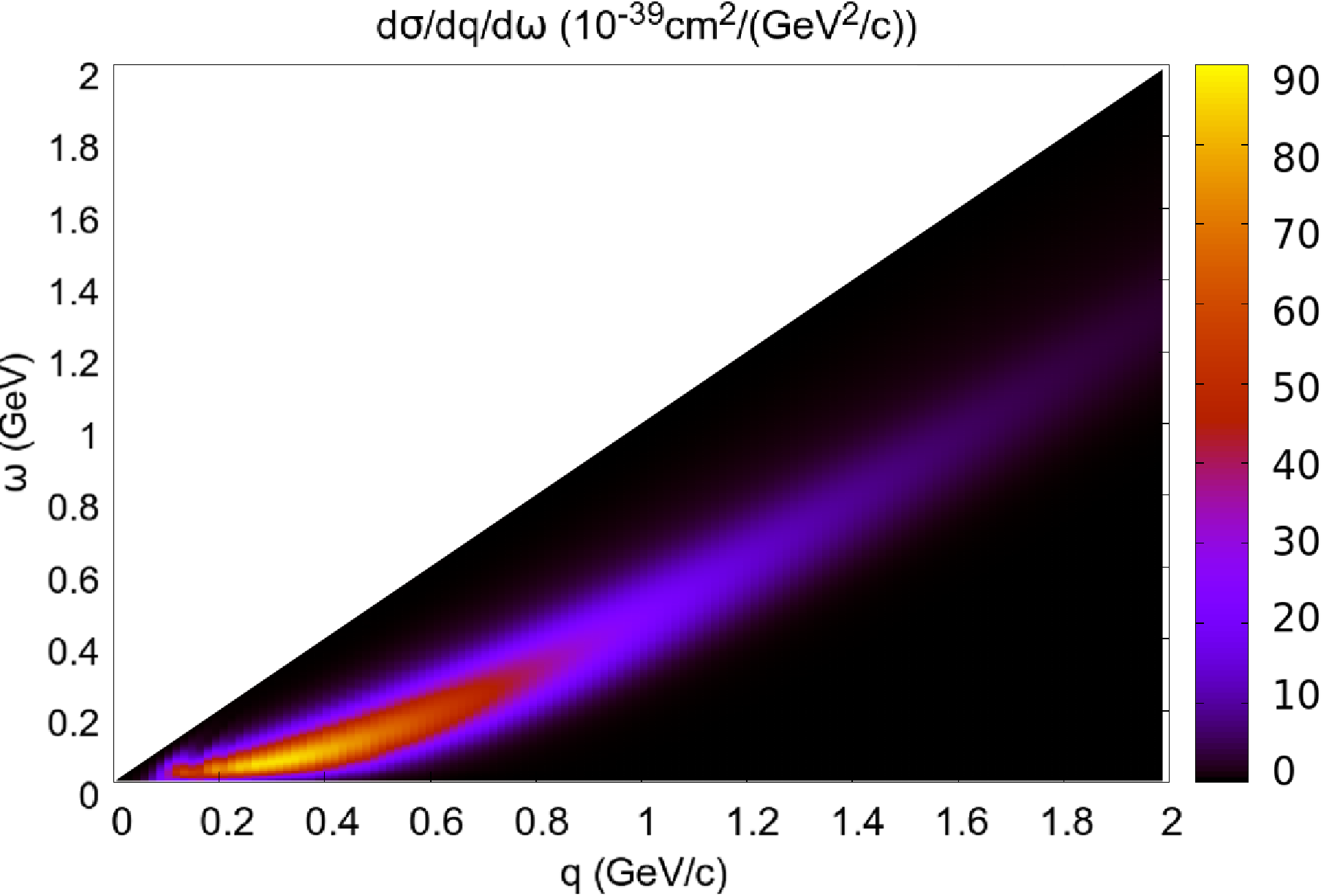}
\begin{center}
\vspace{-1cm}
\end{center}
\end{center}
\caption{(Color online) 2p-2h MEC (top panel) and QE (bottom panel) density plots of the double-differential cross section per neutron of $^{12}$C versus $\omega$ and $q$ at $E_\nu=3$ GeV.}\label{2p2h-region}
\end{figure}

\begin{figure}[H]
\begin{center}\vspace{-0.2cm}
\includegraphics[scale=0.27, angle=0]{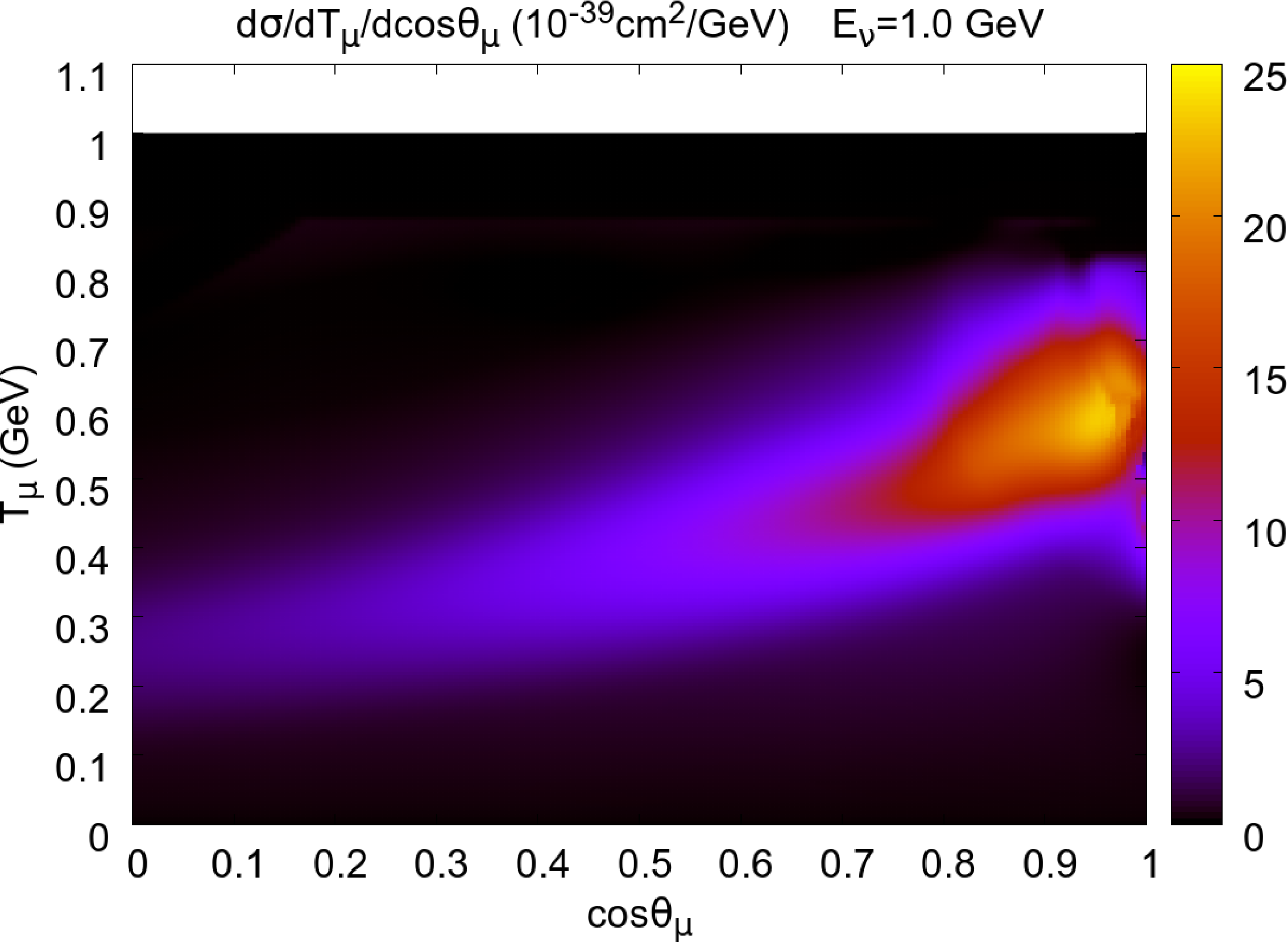}\\
\includegraphics[scale=0.27, angle=0]{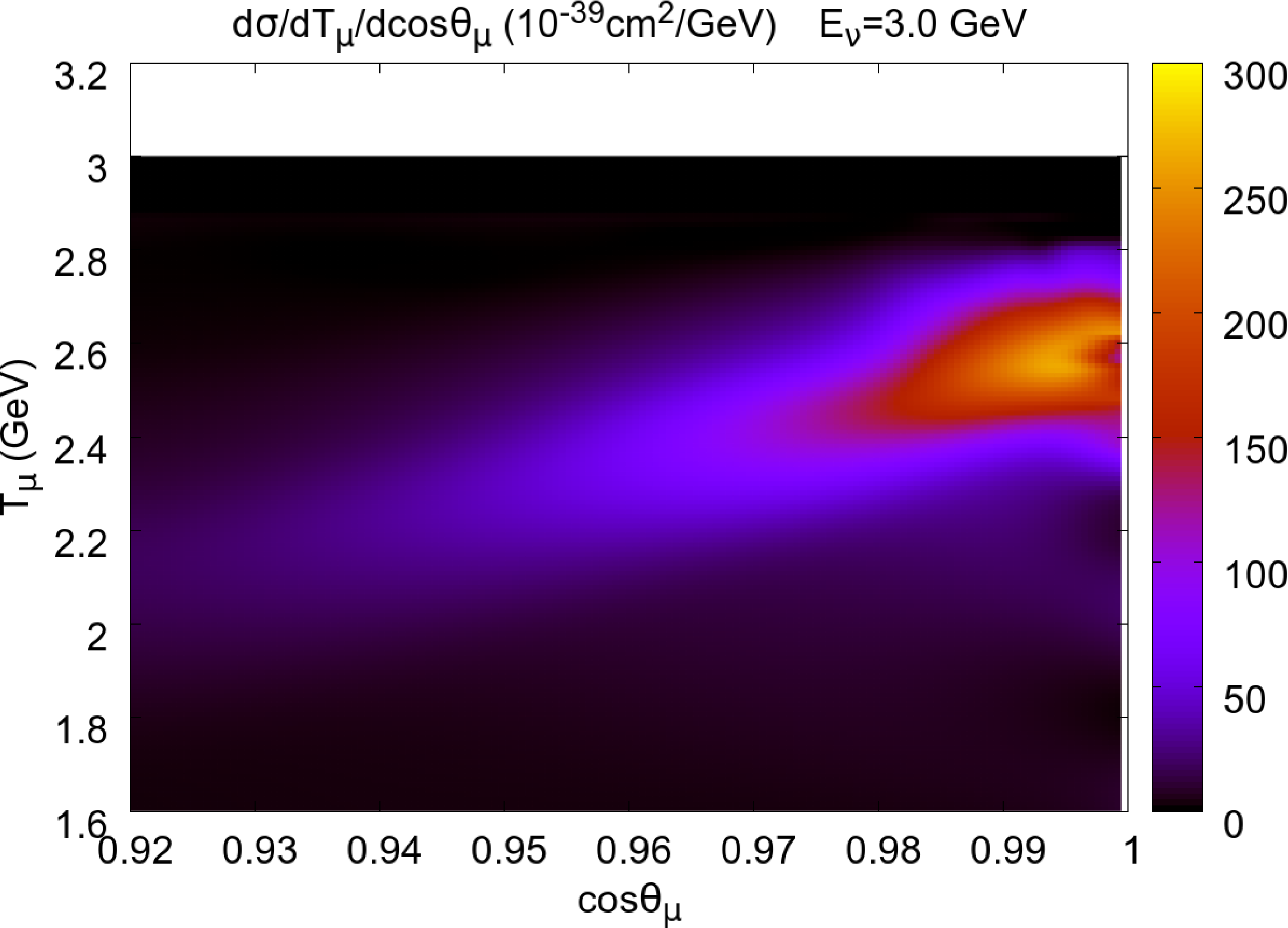}\\
\includegraphics[scale=0.27, angle=0]{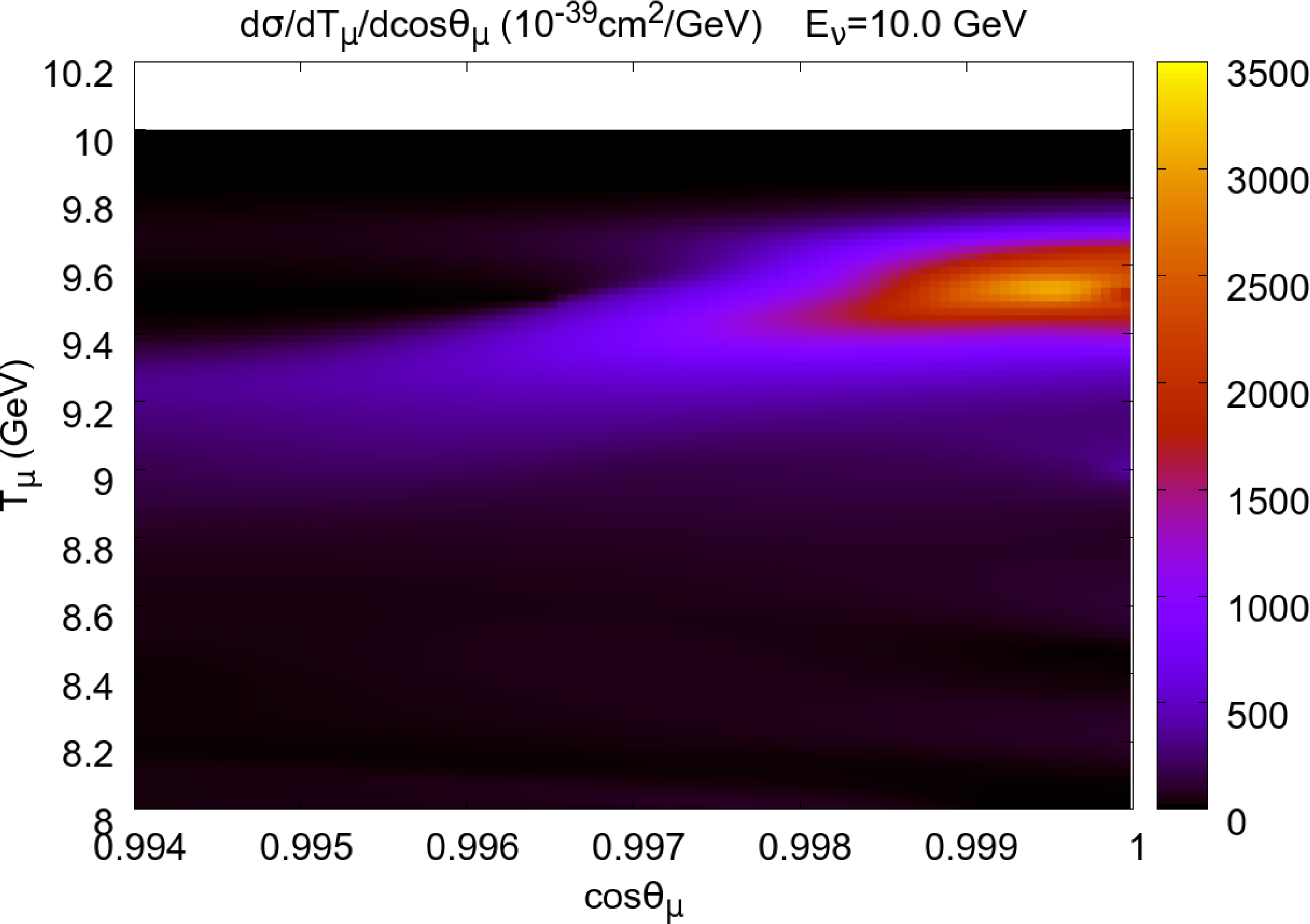}
\begin{center}
\vspace{-1cm}
\end{center}
\end{center}
\caption{(Color online) 2p-2h MEC density plots of the double-differential cross section per neutron of $^{12}$C at three different neutrino energies $E_\nu$ versus $T_\mu$ and $\cos\theta_\mu$.
}\label{dens2}
\end{figure}

To conclude this section, it is also important to identify the
kinematical region where the 2p-2h MEC responses attain their maximum
values. This is clearly illustrated in the density plot of
Fig.\ref{2p2h-region}, which represents the double differential cross
section in terms of $\omega$ and $q$ at $E_\nu=3.0$ GeV; here the top
(bottom) panel corresponds to the 2p-2h MEC (pure QE)
contributions. As shown in the figure, the main contribution to the
MEC cross section comes from $q\in(0.3,1.0)$ GeV/c and
$\omega\in(0.3,0.8)$ GeV. On the contrary, the QE peak is moved to
lower values of $\omega$. Both the one-body and two-body responses die
with the momentum transfer $q$, but their ratio is rather constant
(see \cite{Simo:2016aa}). Although results in
Fig.\ref{2p2h-region} correspond to a fixed incident neutrino energy,
3 GeV, similar results are obtained for larger $E_\nu$ values.  It is
important to point out the differences between our predictions and
those ones based on the model of Nieves~\cite{Nieves13PRD} that show
the 2p-2h MEC contribution to be shifted to slightly bigger values of
the energy and momentum transfer. For completeness we also show in Fig.\ref{dens2} the
density plots for the 2p-2h MEC contributions in terms of the values
of the muon kinetic energy and the scattering angle for three values
of neutrino energy: 1 GeV (top panel), 3 GeV (middle) and 10 GeV
(bottom). As observed, the main contribution resides in the region of
very small angles, close to zero.

\section{Results}

In this section we show the predictions of the SuSAv2-MEC model compared with data from different collaborations: MiniBooNE, MINER$\nu$A, T2K and SciBooNE. Our study is mainly restricted to the quasielastic (QE) regime where the impulse approximation in addition to the effects linked to the 2p-2h meson-exchange currents play a major role. However, some results that incorporate the contribution of the $\Delta$ excitation are also compared with data. As shown in \cite{Megias:2016ee} the SuSAv2-MEC model has been applied to the inelastic region for electron scattering. Its extension to neutrino
reactions is in progress and their predictions will be shown in a forthcoming publication. However, the resonant pion production, that in most cases is the largest contribution, is computed following our previous investigations in \cite{Ivanov:2016Delta}.

In the case of the QE regime, our study includes the analysis of neutrino and antineutrino scattering reactions corresponding to MiniBooNE as well as to MINER$\nu$A experiments. In the latter we consider muon and electron neutrinos. Results for T2K are also analyzed in detail. In this case, in order to make the discussion that follows simpler, we first restrict ourselves to the QE domain, and we extend the discussion later to inclusive charged-current (CC) neutrino reactions where high inelasticities are of significance. Our main interest is to show the capability of the present model, SuSAv2-MEC, to describe successfully a large variety of neutrino scattering data corresponding to different experiments with a wide range of kinematics explored. The model, that was already proven to be capable of reproducing $(e,e')$ data, is now extended to neutrinos with emphasis on the crucial role played by 2p-2h MEC effects. These have been computed for the first time within a fully relativistic formalism and without resorting to any particular assumption on the different responses: vector-vector, axial-axial and vector-axial interference.

\subsection{CCQE experimental cross sections}


\begin{figure}[H]
\begin{center}\vspace{-1.8cm}
\includegraphics[scale=0.22, angle=270]{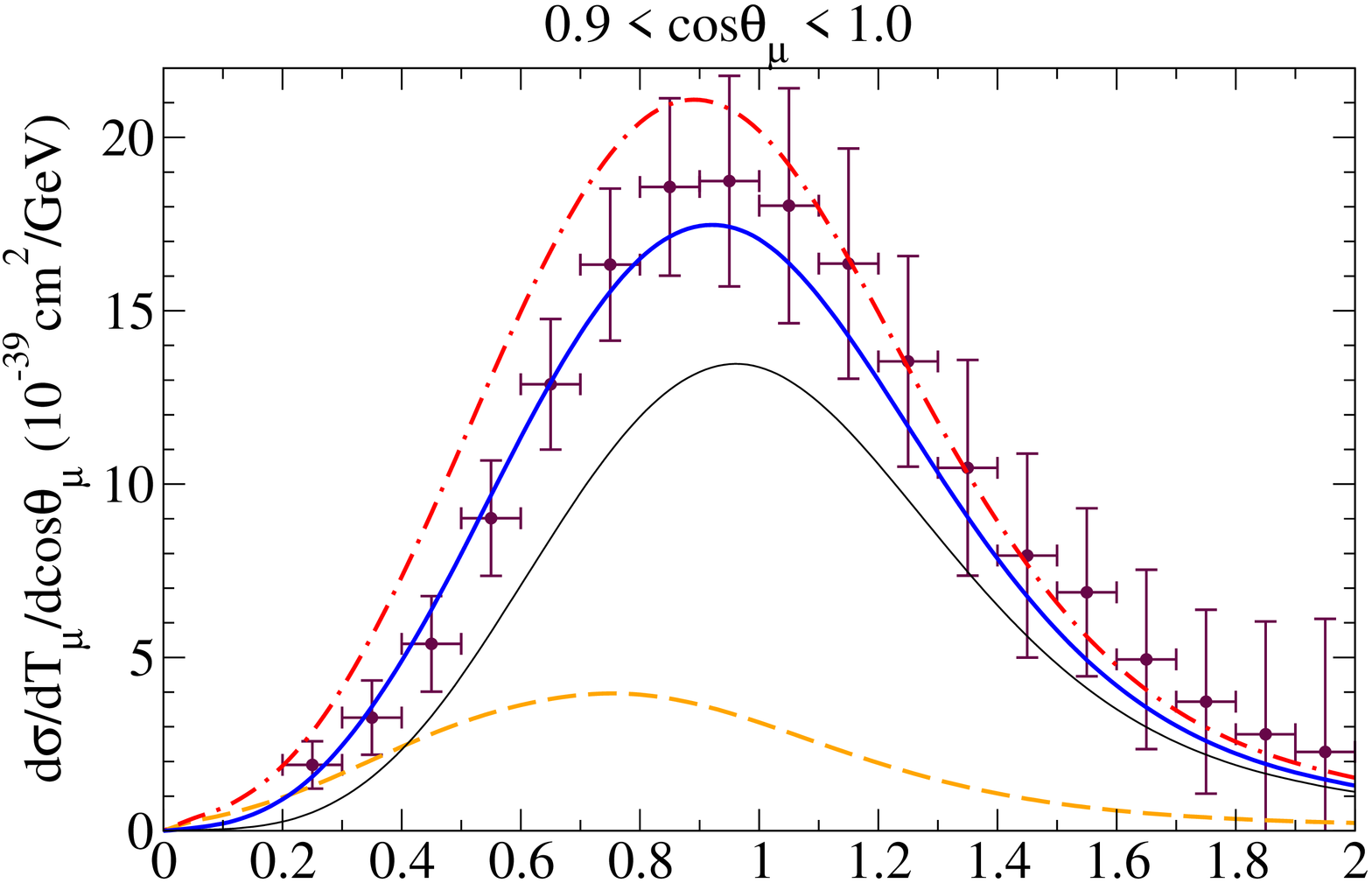}\hspace*{-0.15cm}\includegraphics[scale=0.22, angle=270]{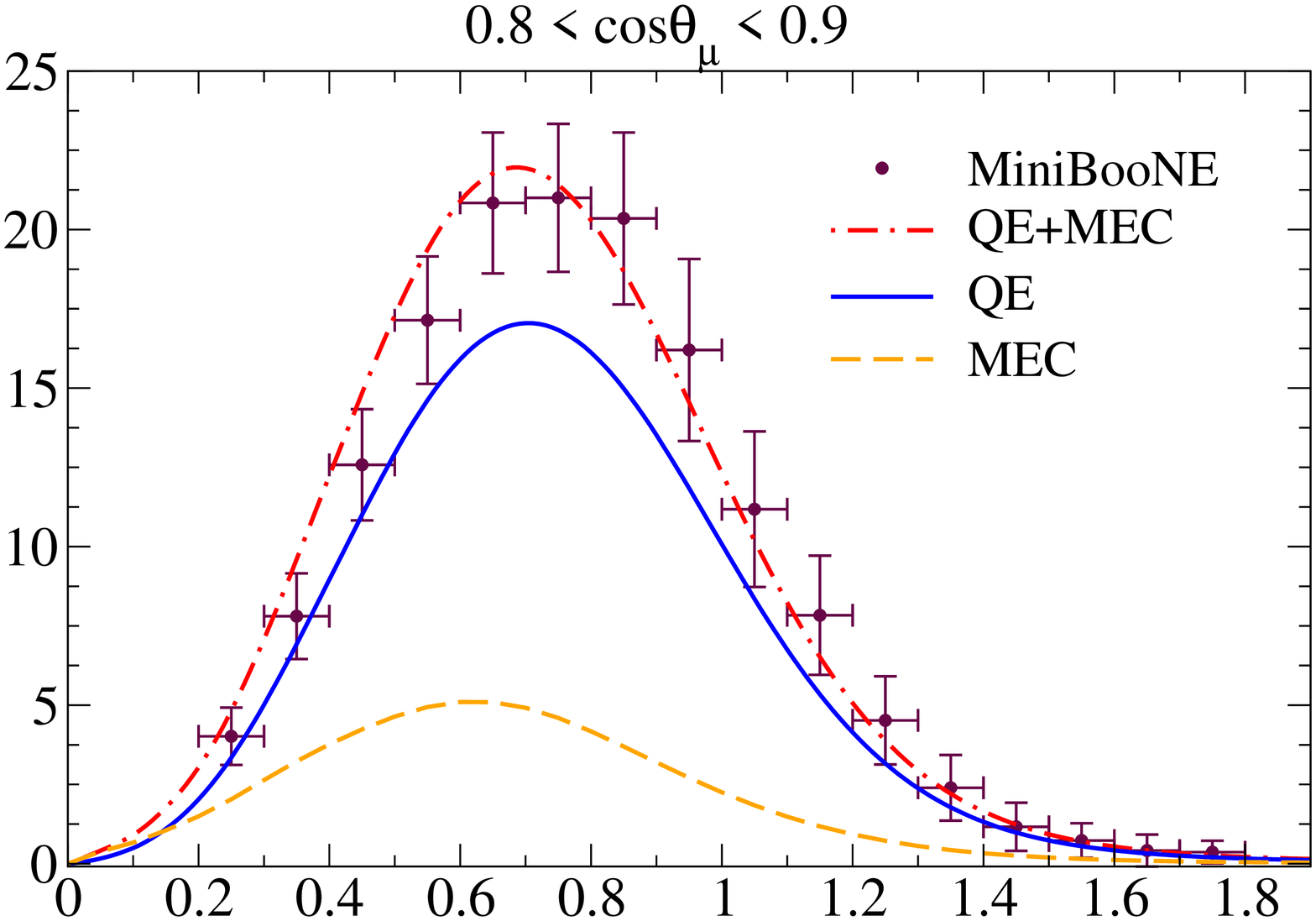}\hspace*{-0.15cm}\includegraphics[scale=0.22, angle=270]{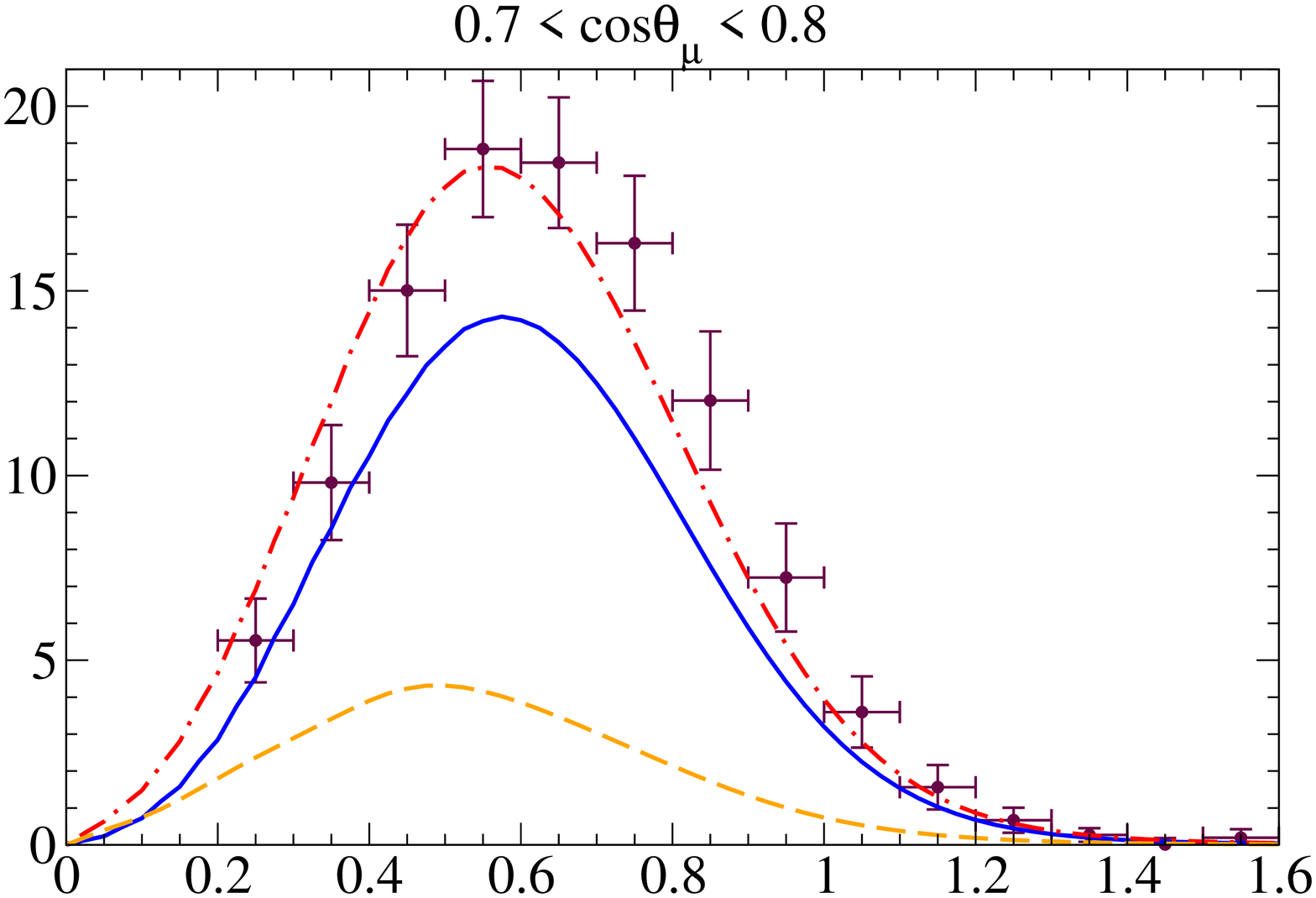}\\
\includegraphics[scale=0.22, angle=270]{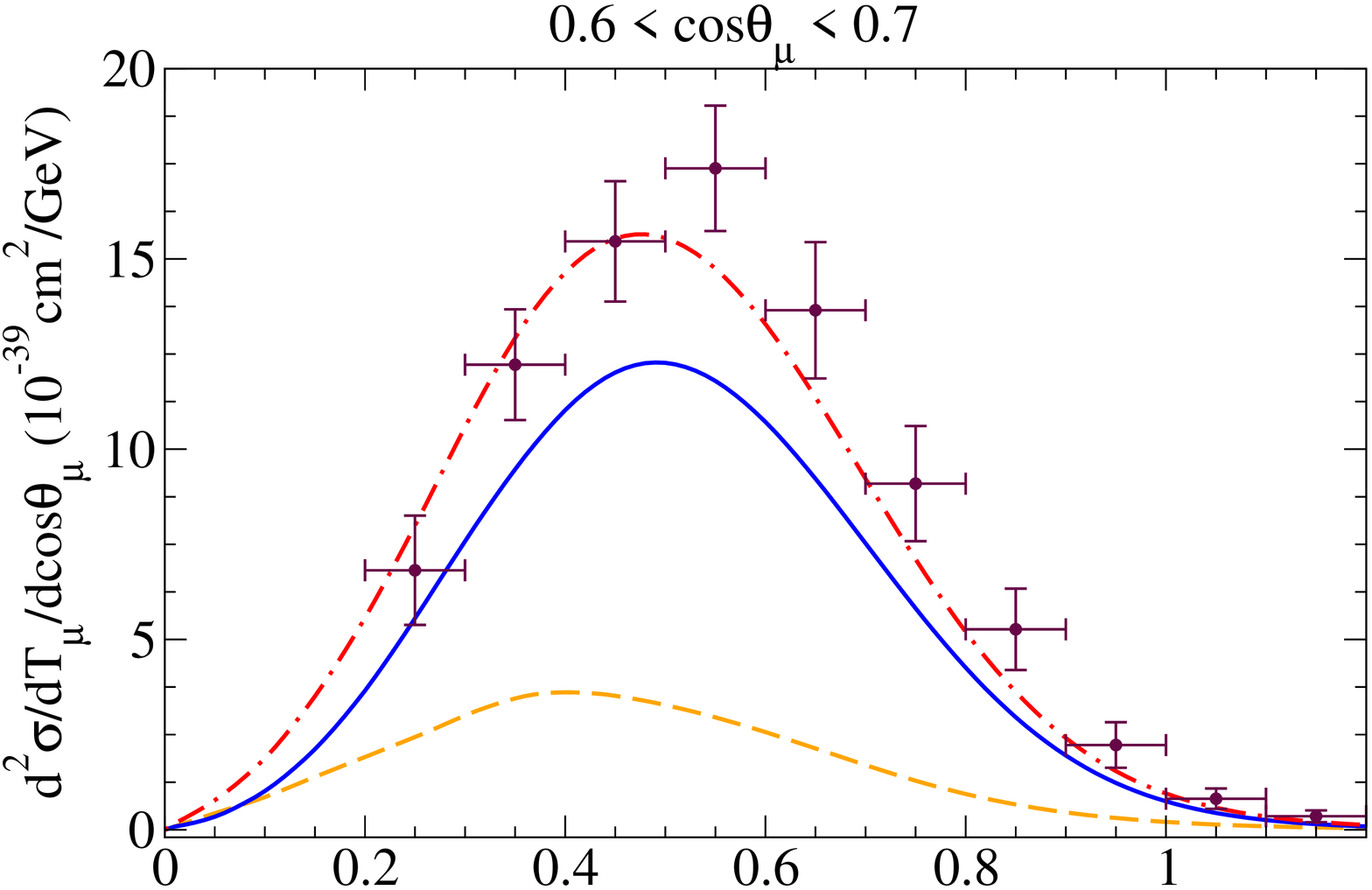}\hspace*{-0.15cm}\includegraphics[scale=0.22, angle=270]{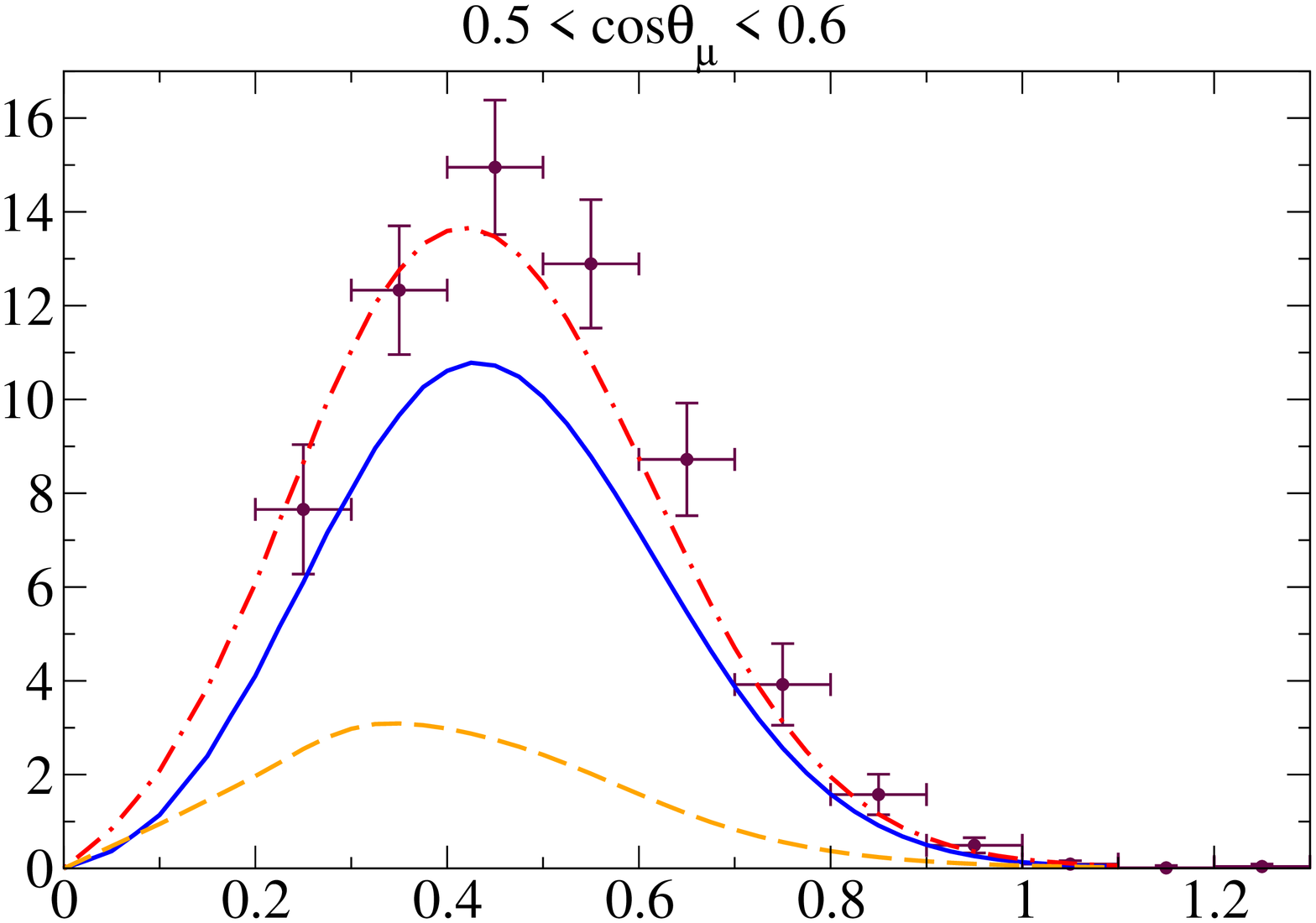}\hspace*{-0.15cm}\includegraphics[scale=0.22, angle=270]{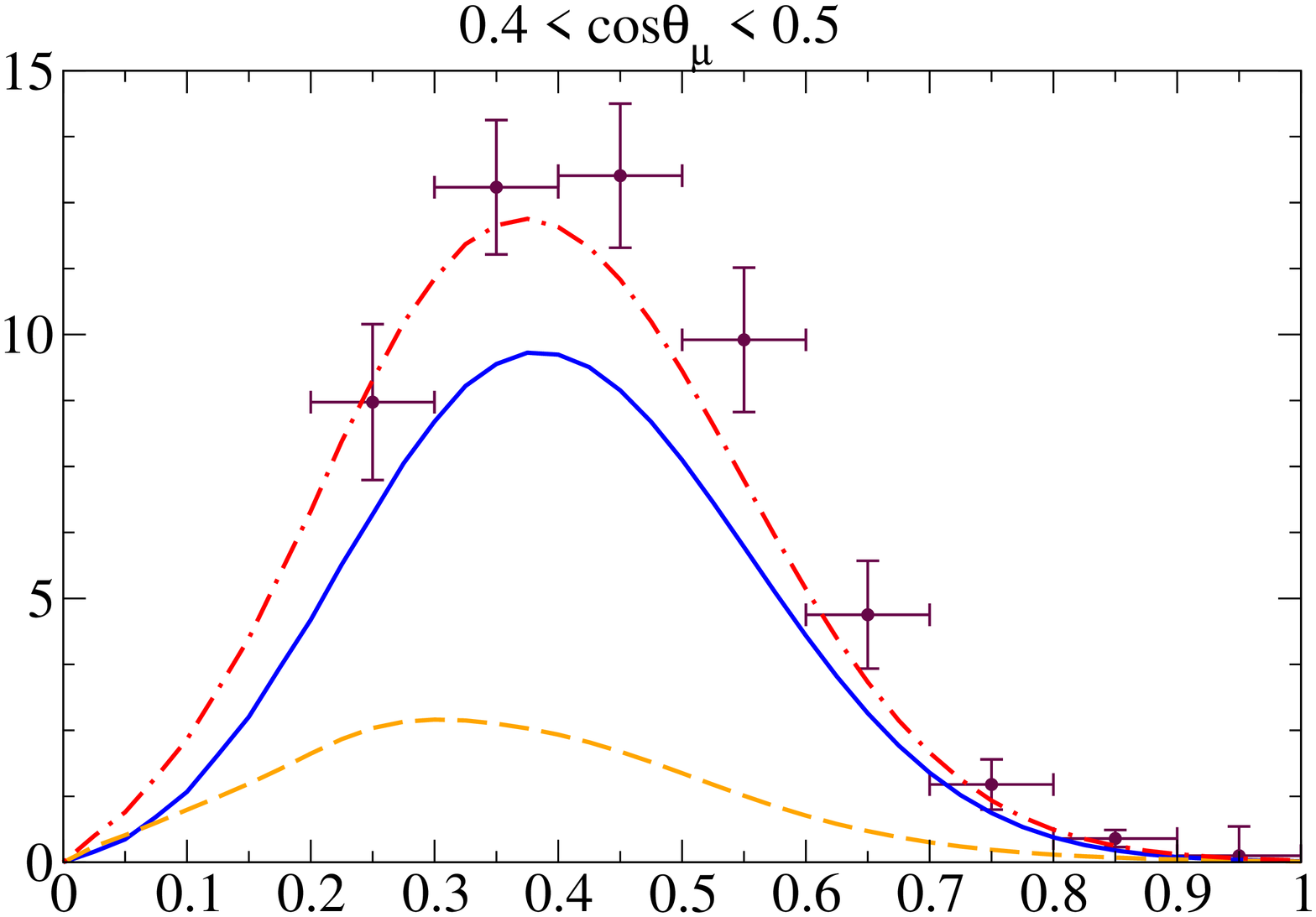}\\
\includegraphics[scale=0.22, angle=270]{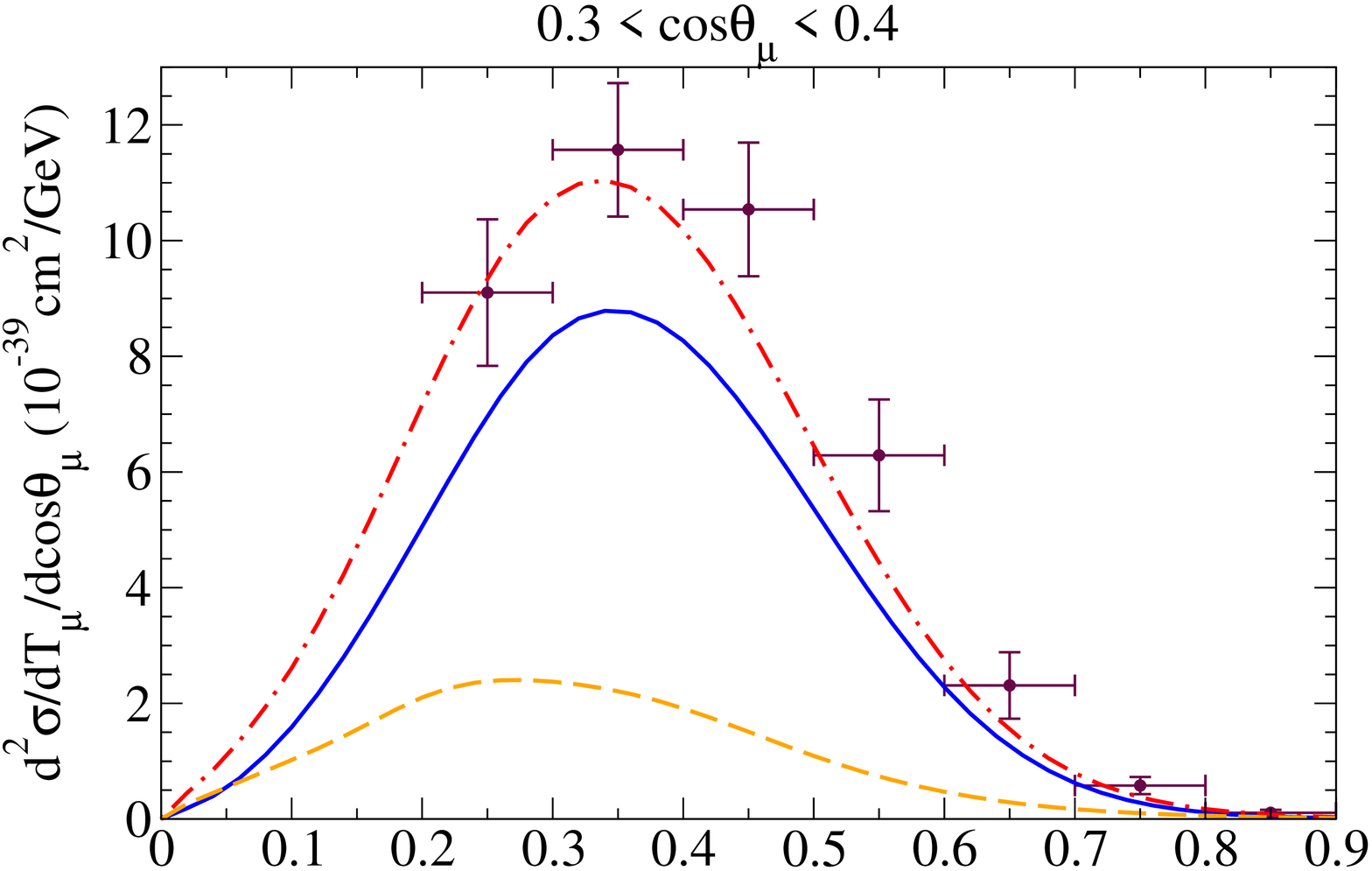}\hspace*{-0.15cm}\includegraphics[scale=0.22, angle=270]{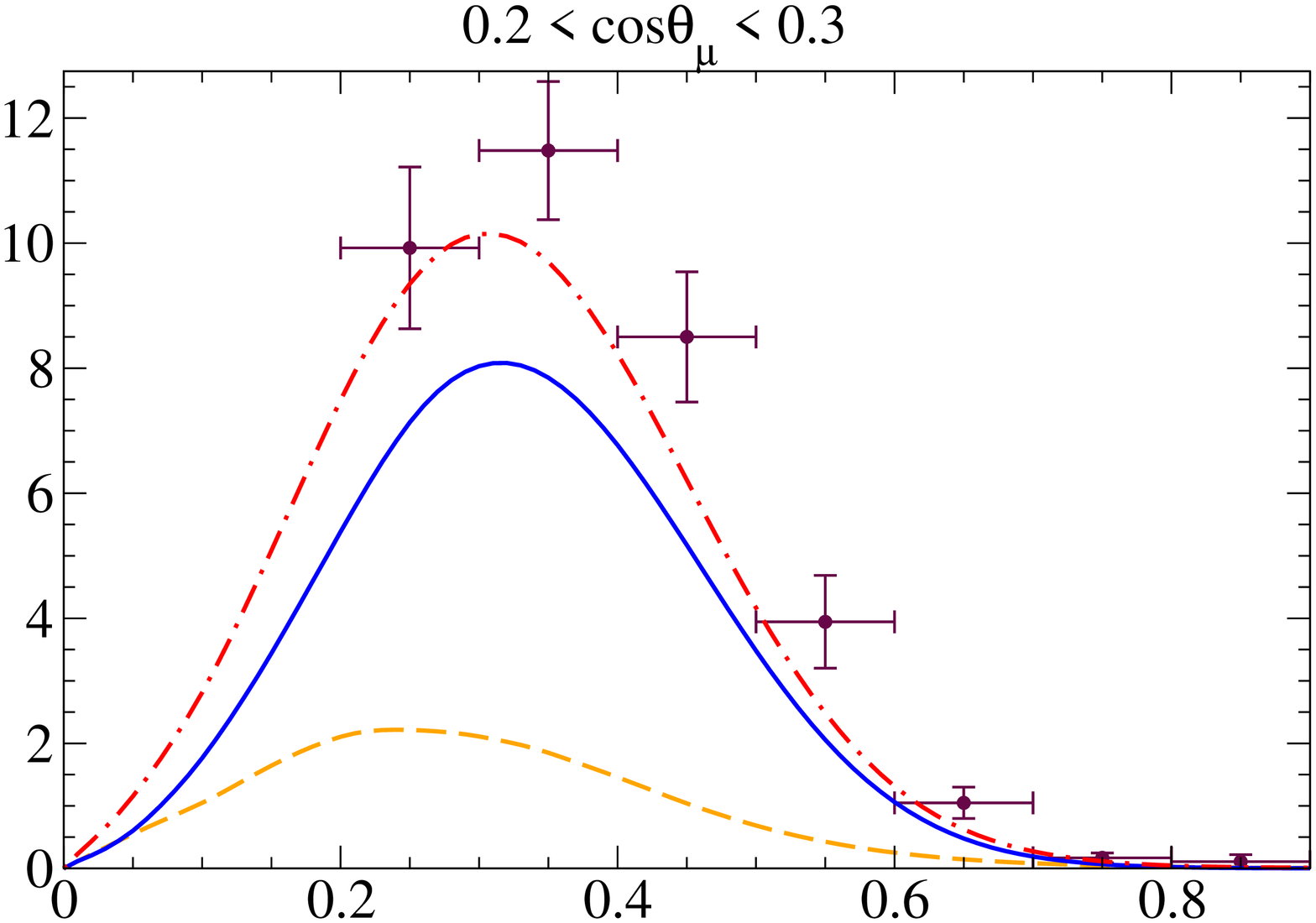}\hspace*{-0.15cm}\includegraphics[scale=0.22, angle=270]{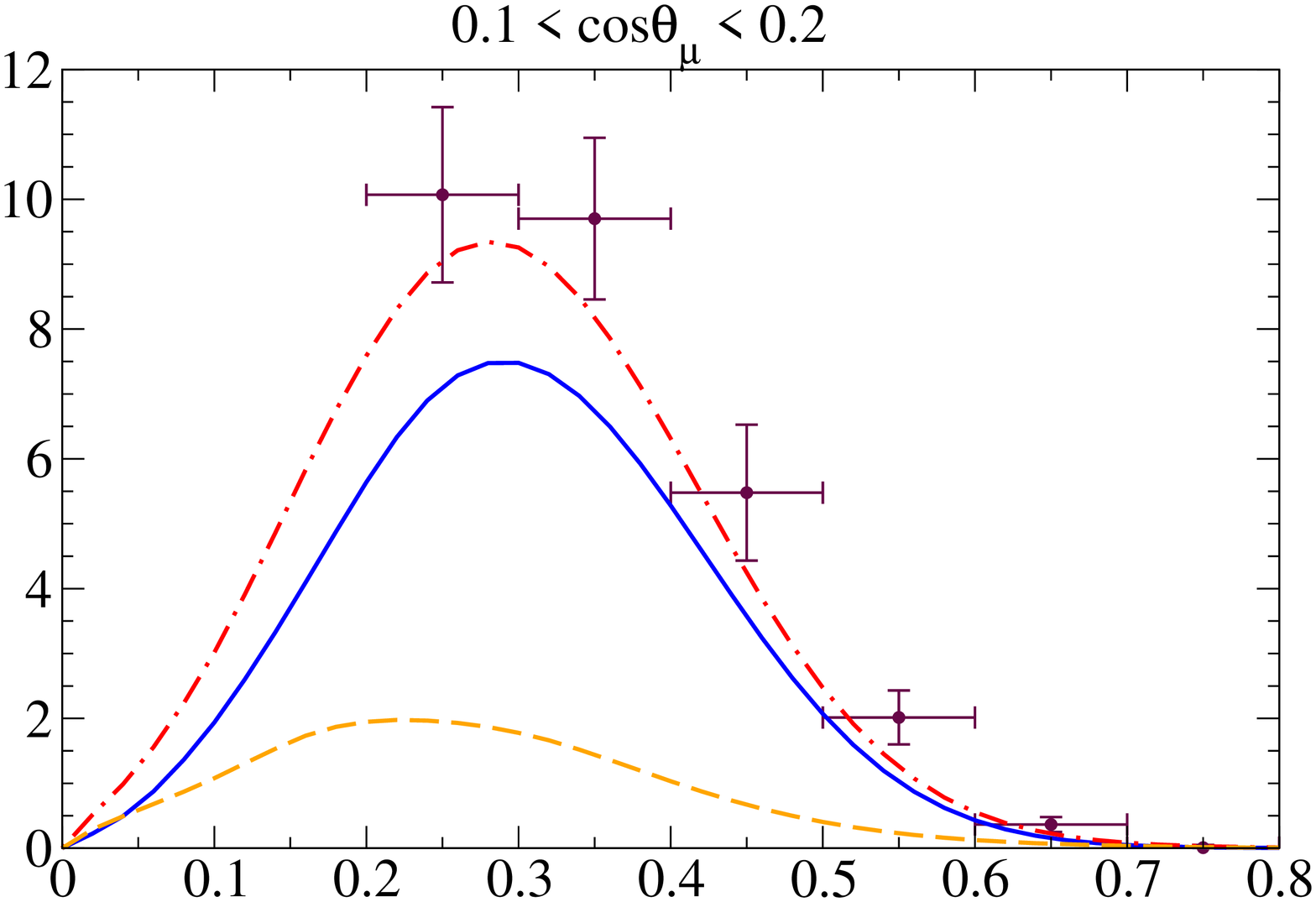}\\
\includegraphics[scale=0.22, angle=270]{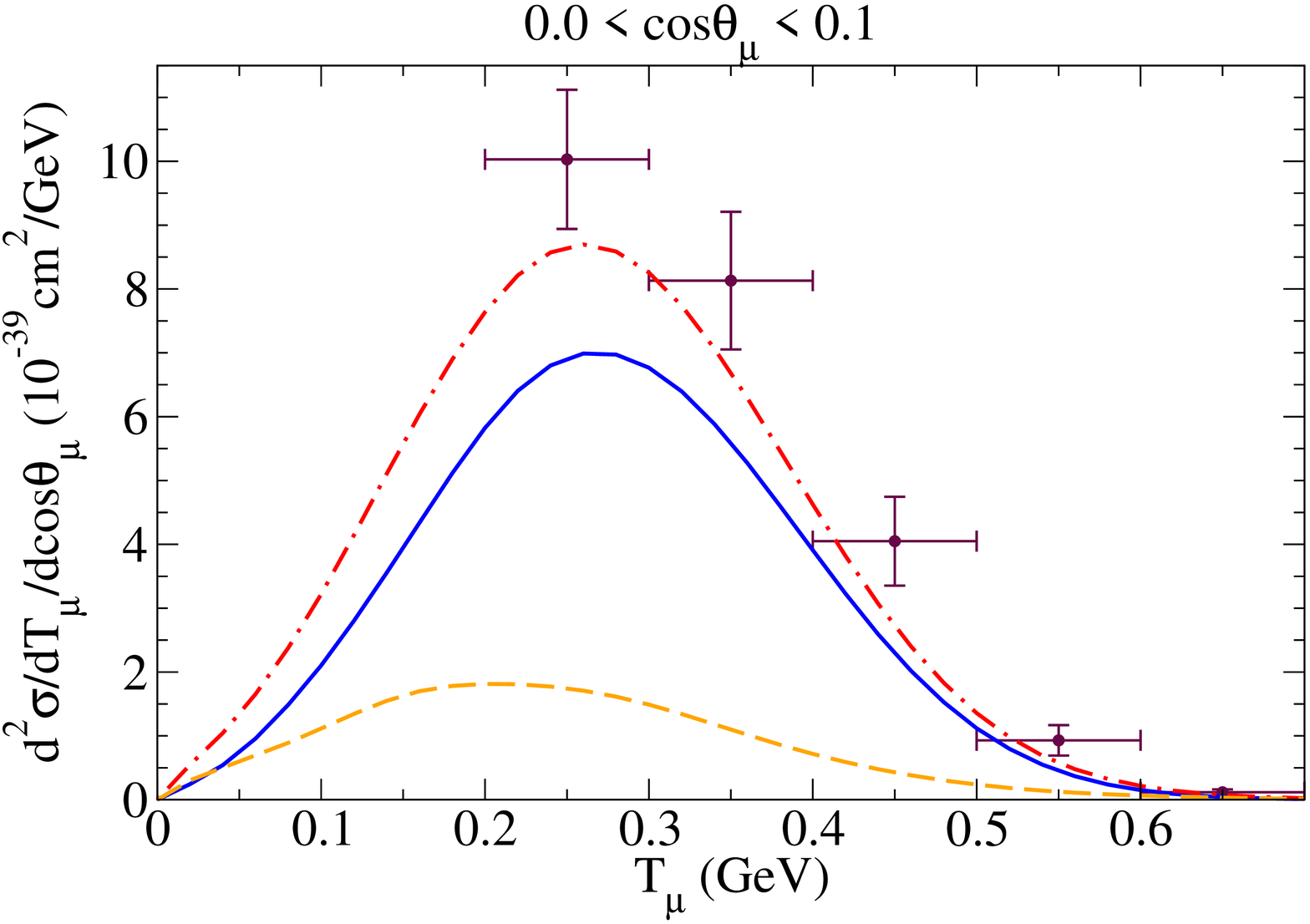}\hspace*{-0.15cm}\includegraphics[scale=0.22, angle=270]{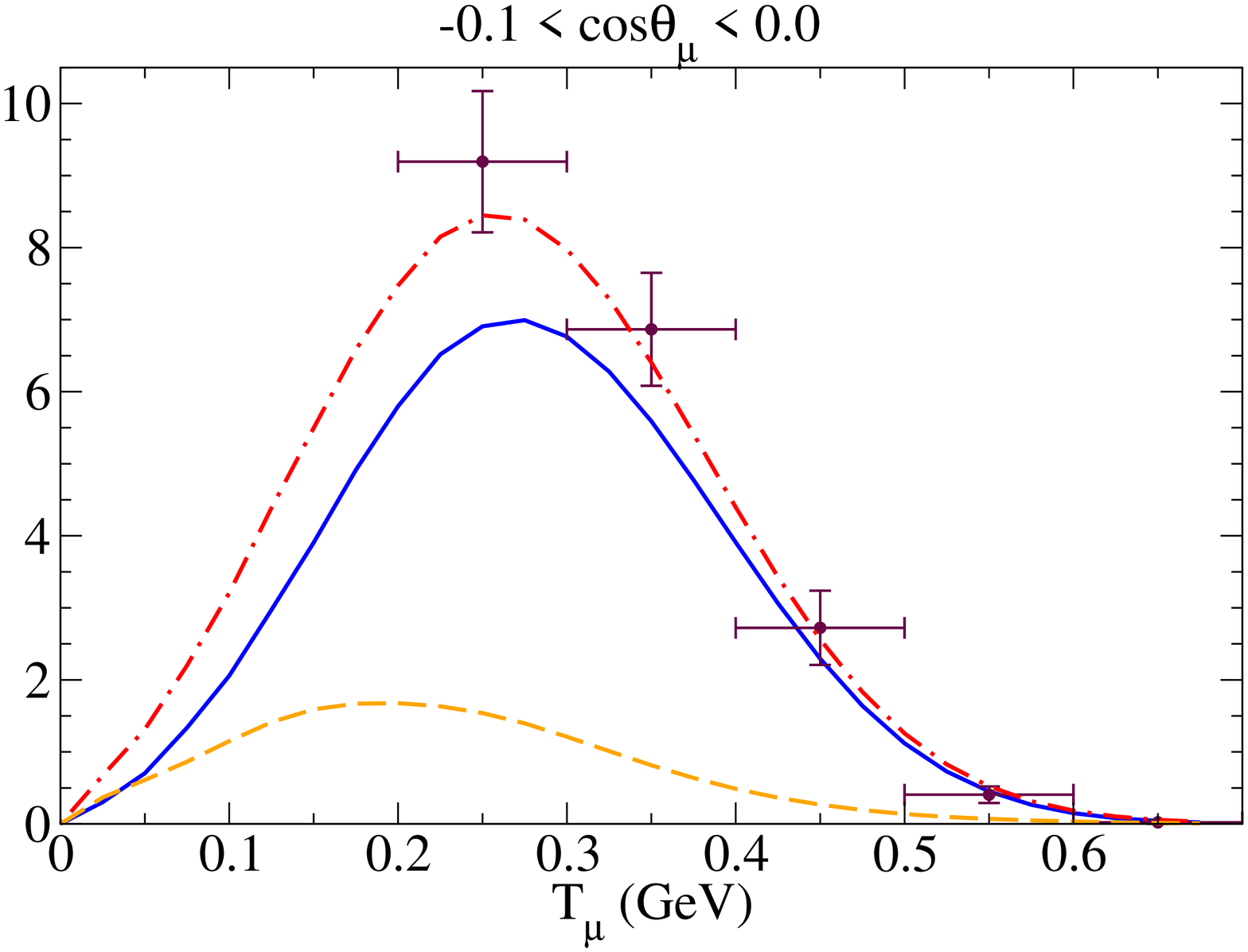}\hspace*{-0.15cm}\includegraphics[scale=0.22, angle=270]{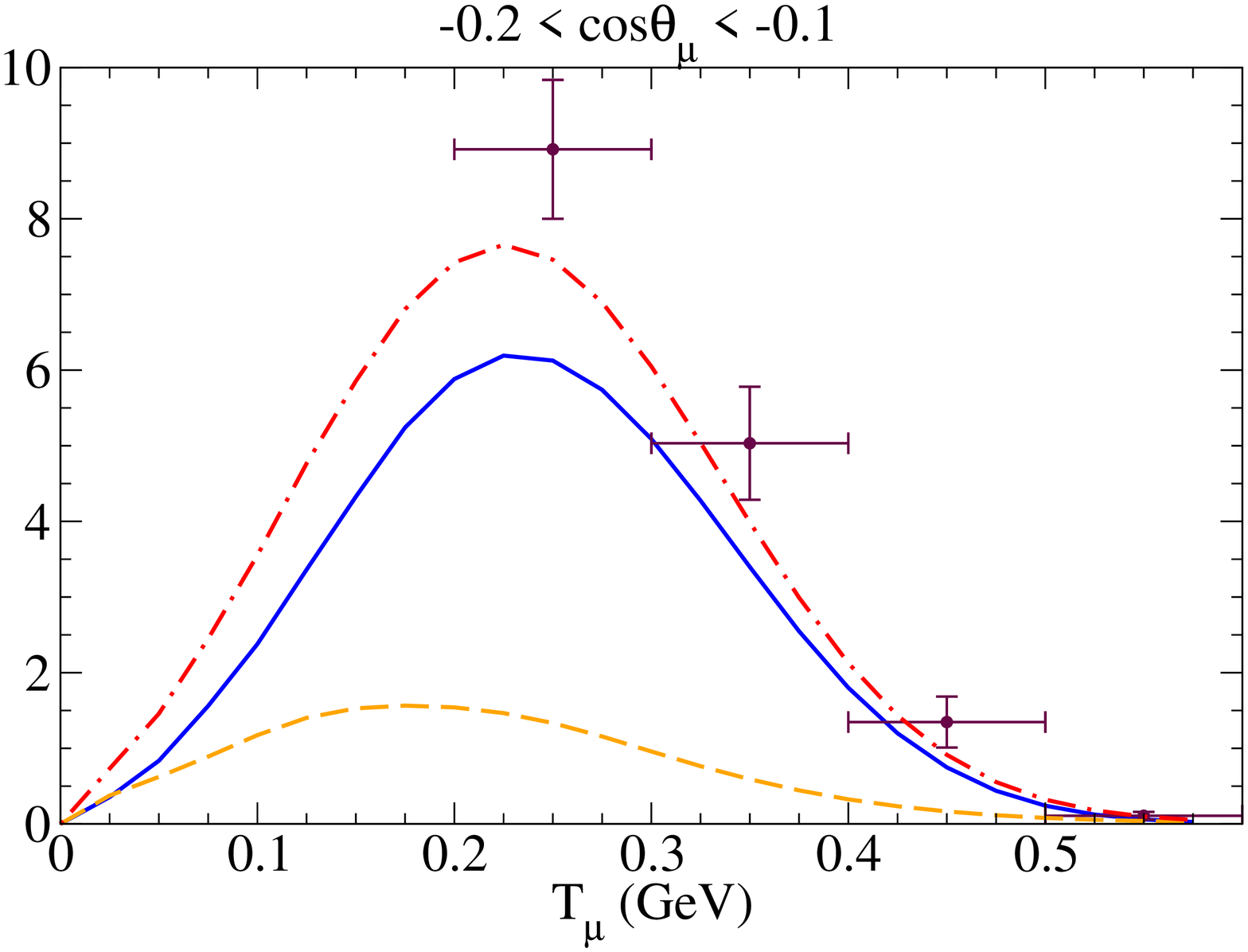}
\begin{center}
\vspace{-1cm}
\end{center}
\end{center}
\caption{(Color online) MiniBoone flux-folded double differential cross section per target nucleon for the $\nu_\mu$ CCQE process on $^{12}$C displayed versus the $\mu^-$ kinetic energy $T_\mu$ for various bins of $\cos\theta_\mu$ obtained within the SuSAv2+MEC approach. QE and 2p-2h MEC results are also shown separately. Data are from~\cite{AguilarArevalo:2010zc}.}\label{Miniboone_nu}
\end{figure}

\begin{figure}[H]
\begin{center}\vspace{-1.8cm}
\includegraphics[scale=0.22, angle=270]{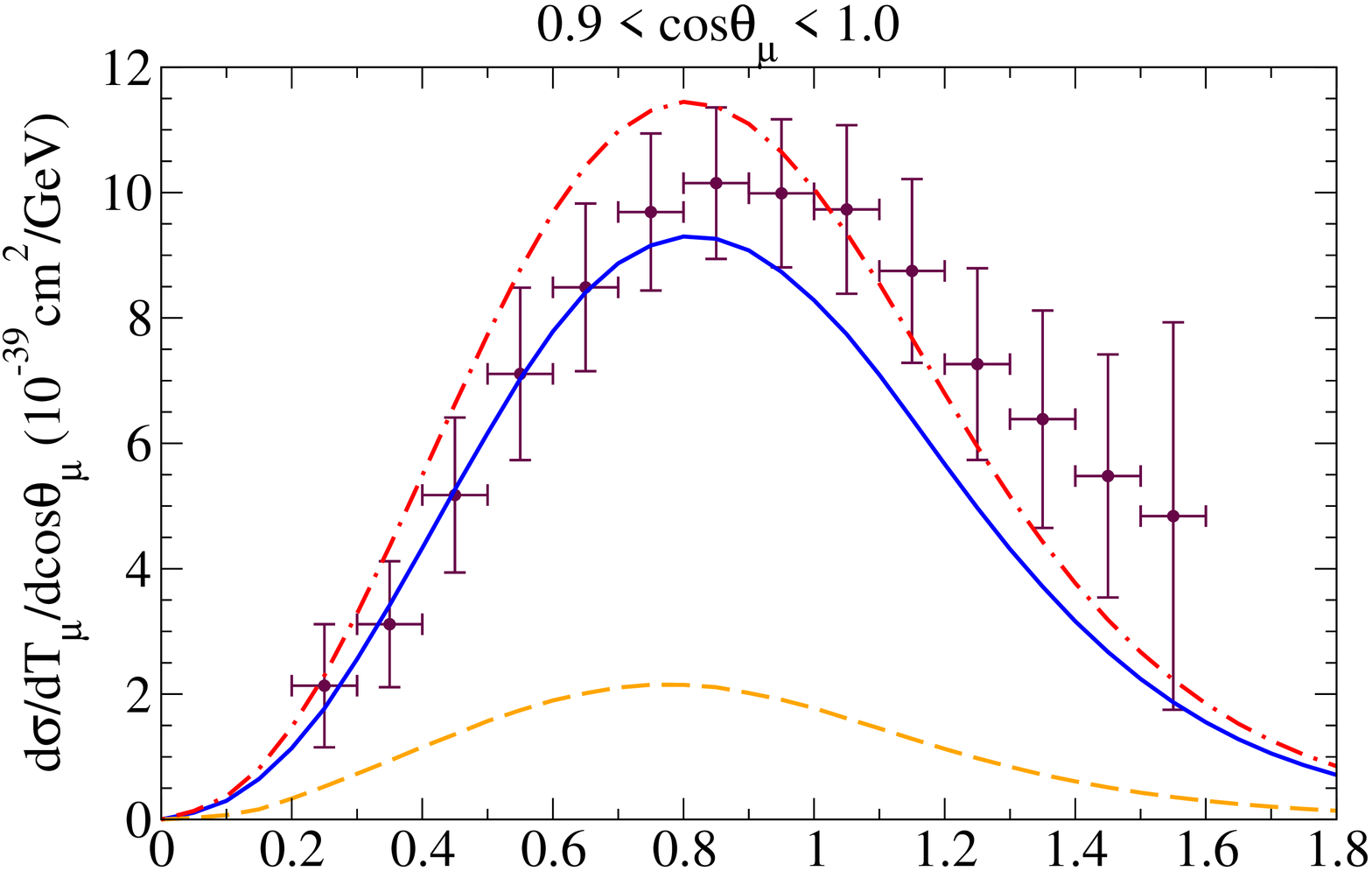}\hspace*{-0.15cm}\includegraphics[scale=0.22, angle=270]{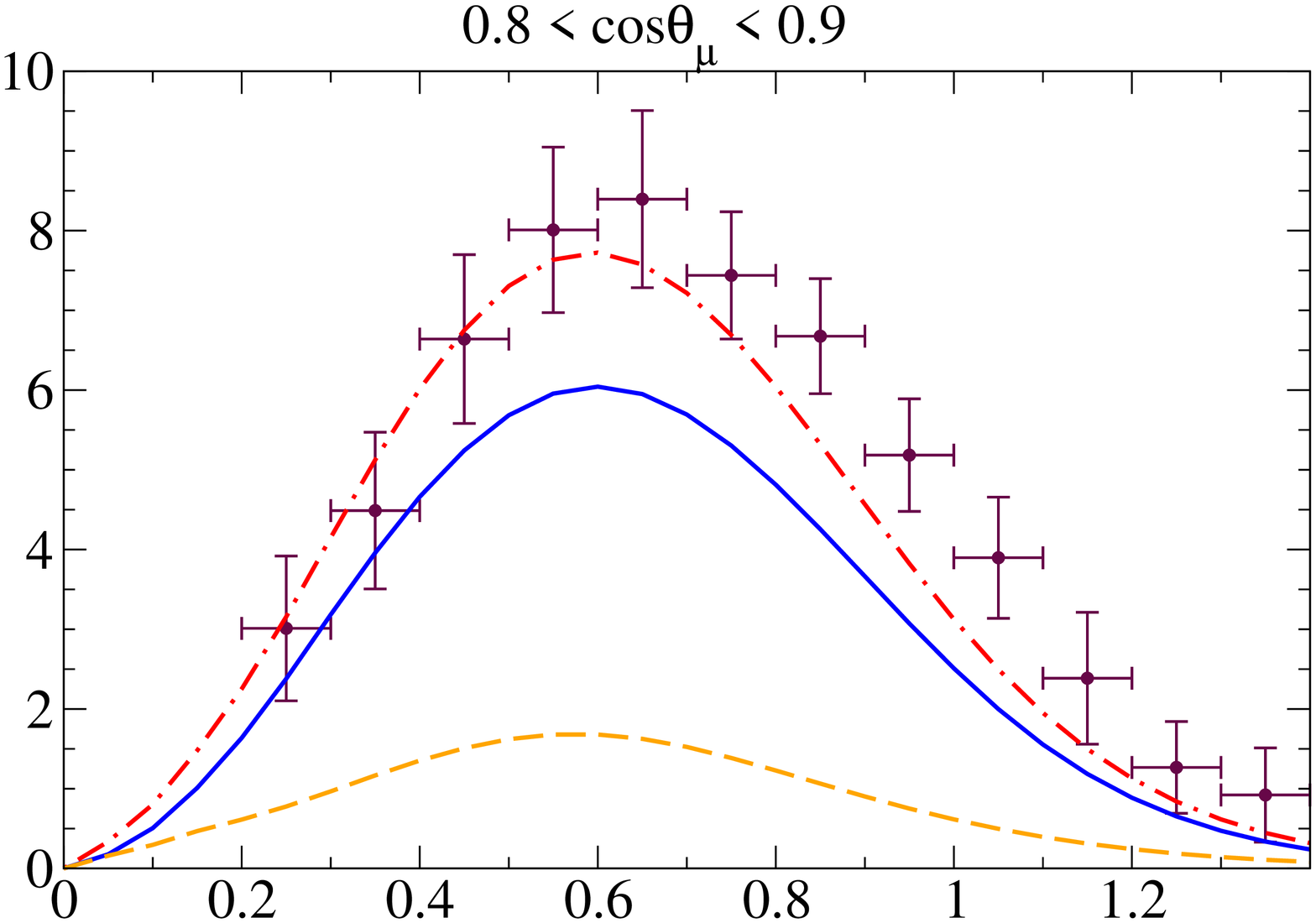}\hspace*{-0.15cm}\includegraphics[scale=0.22, angle=270]{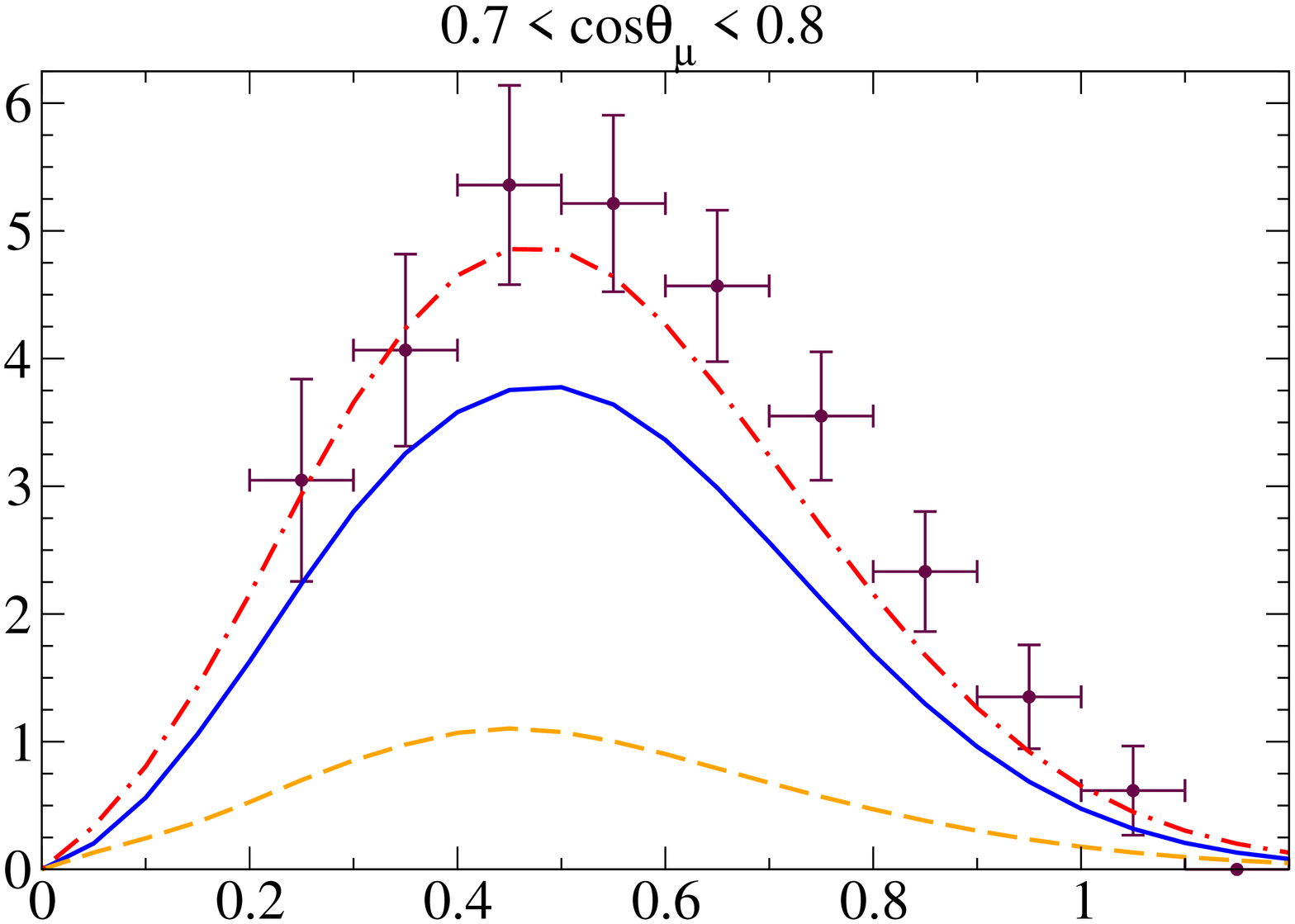}\\
\includegraphics[scale=0.22, angle=270]{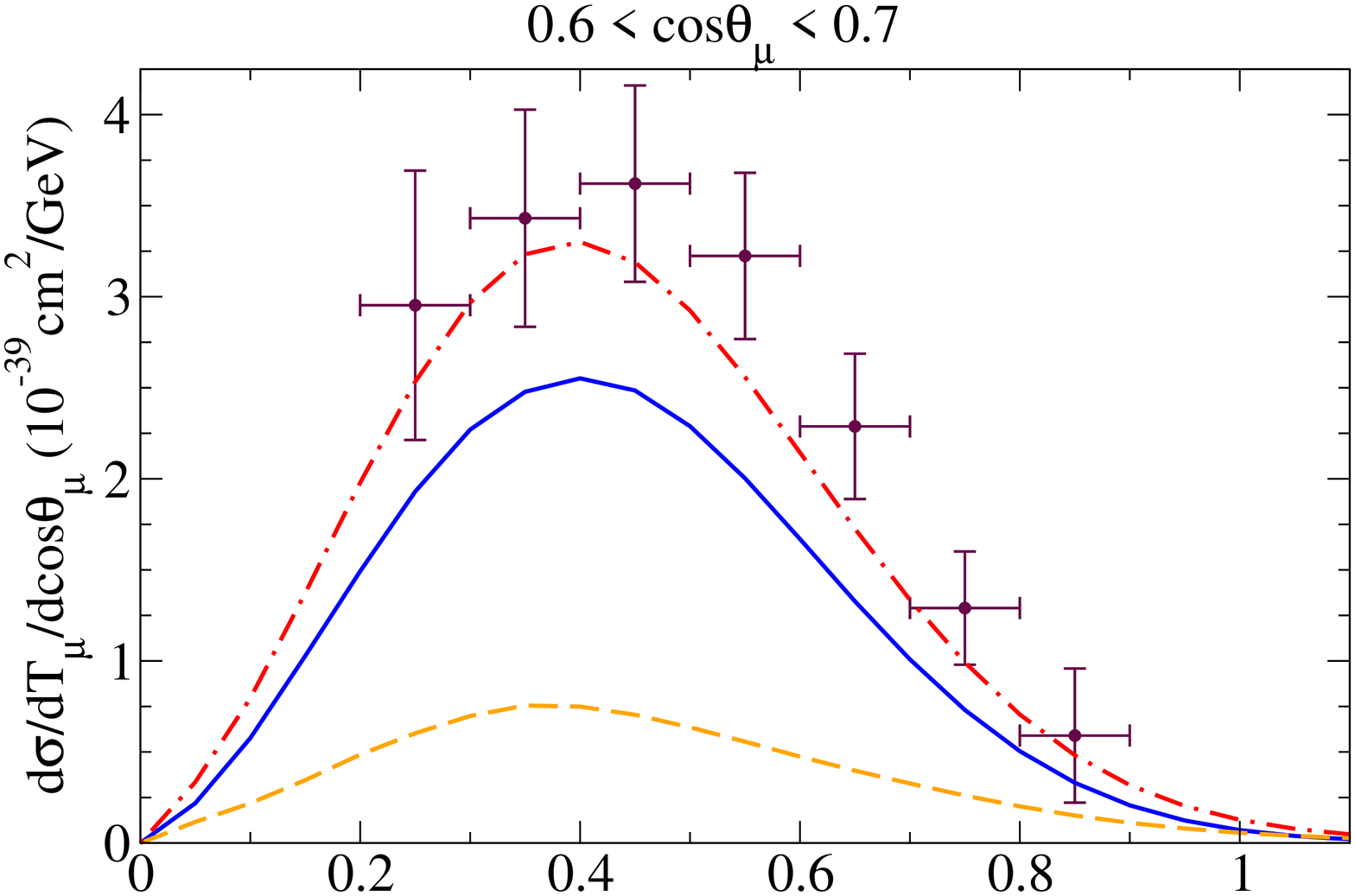}\hspace*{-0.15cm}\includegraphics[scale=0.22, angle=270]{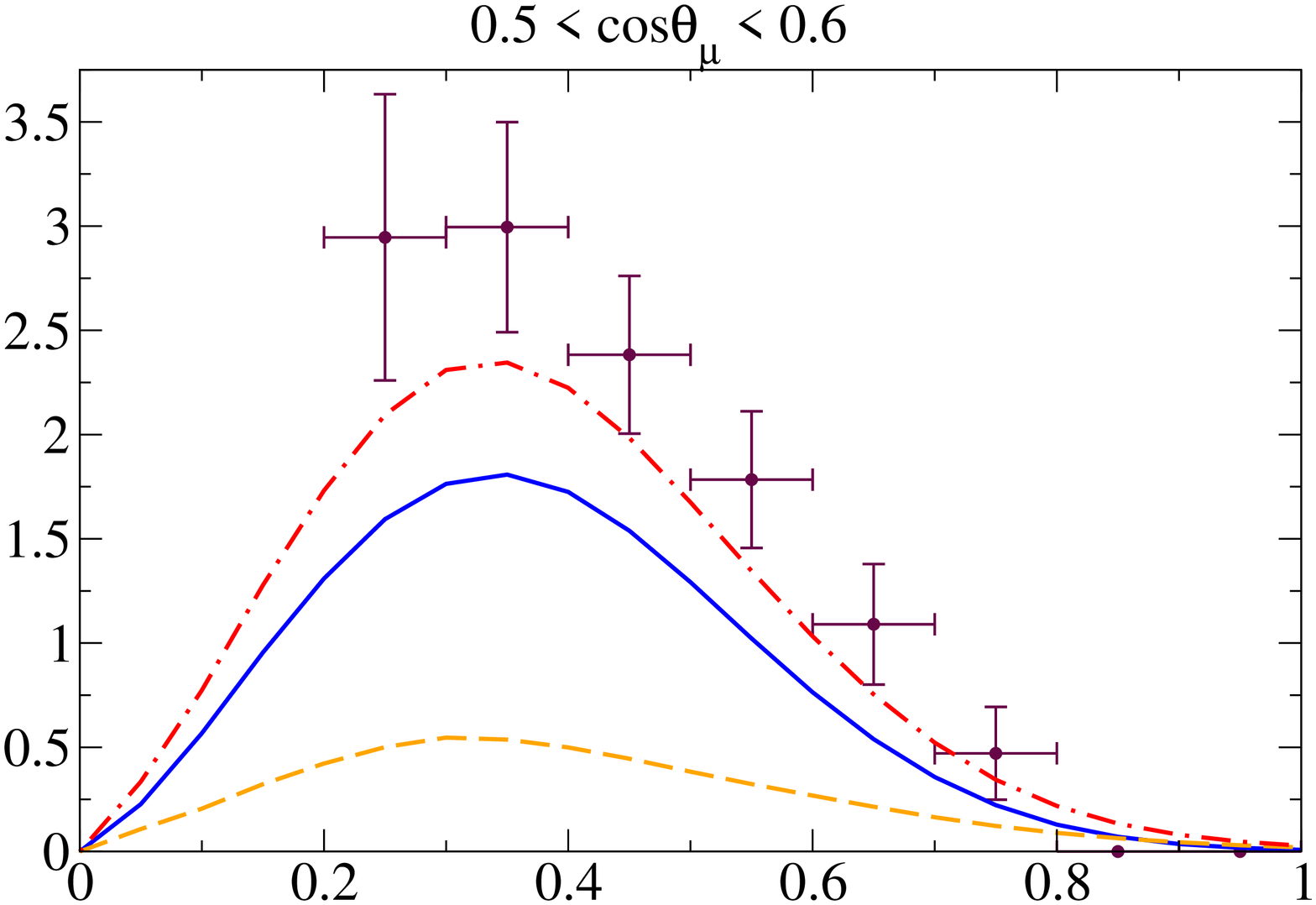}\hspace*{-0.15cm}\includegraphics[scale=0.22, angle=270]{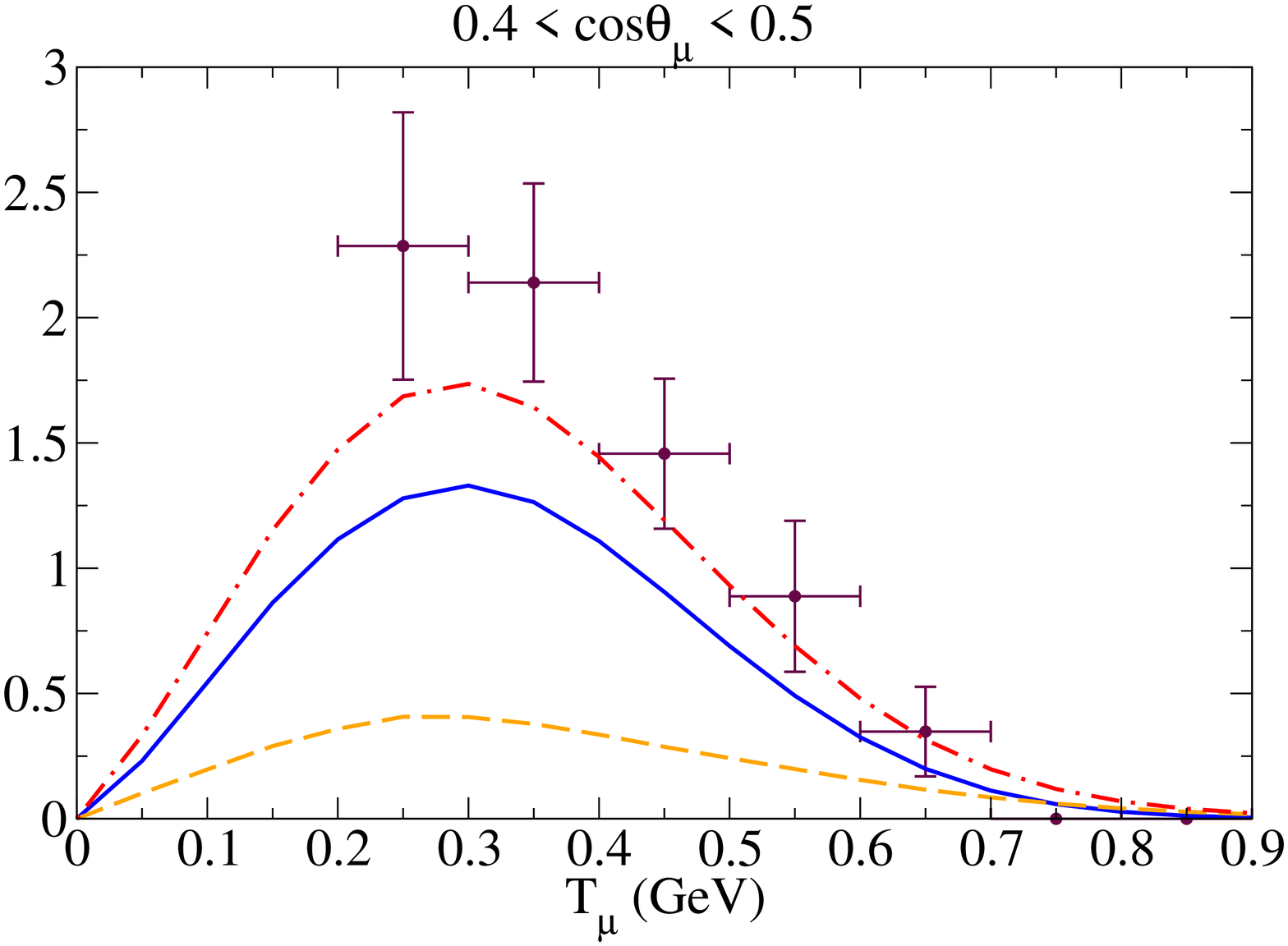}\\
\includegraphics[scale=0.22, angle=270]{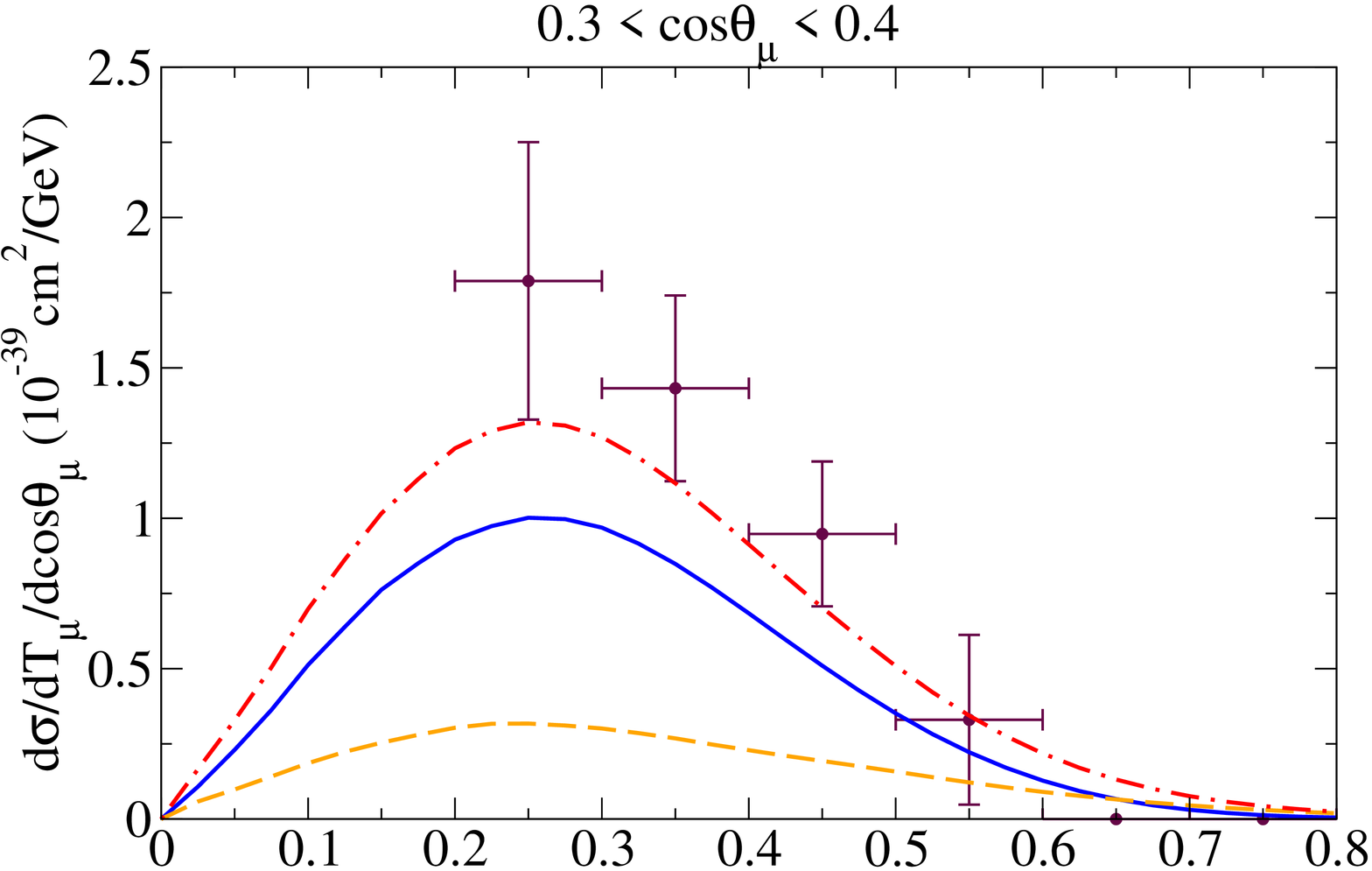}\hspace*{-0.15cm}\includegraphics[scale=0.22, angle=270]{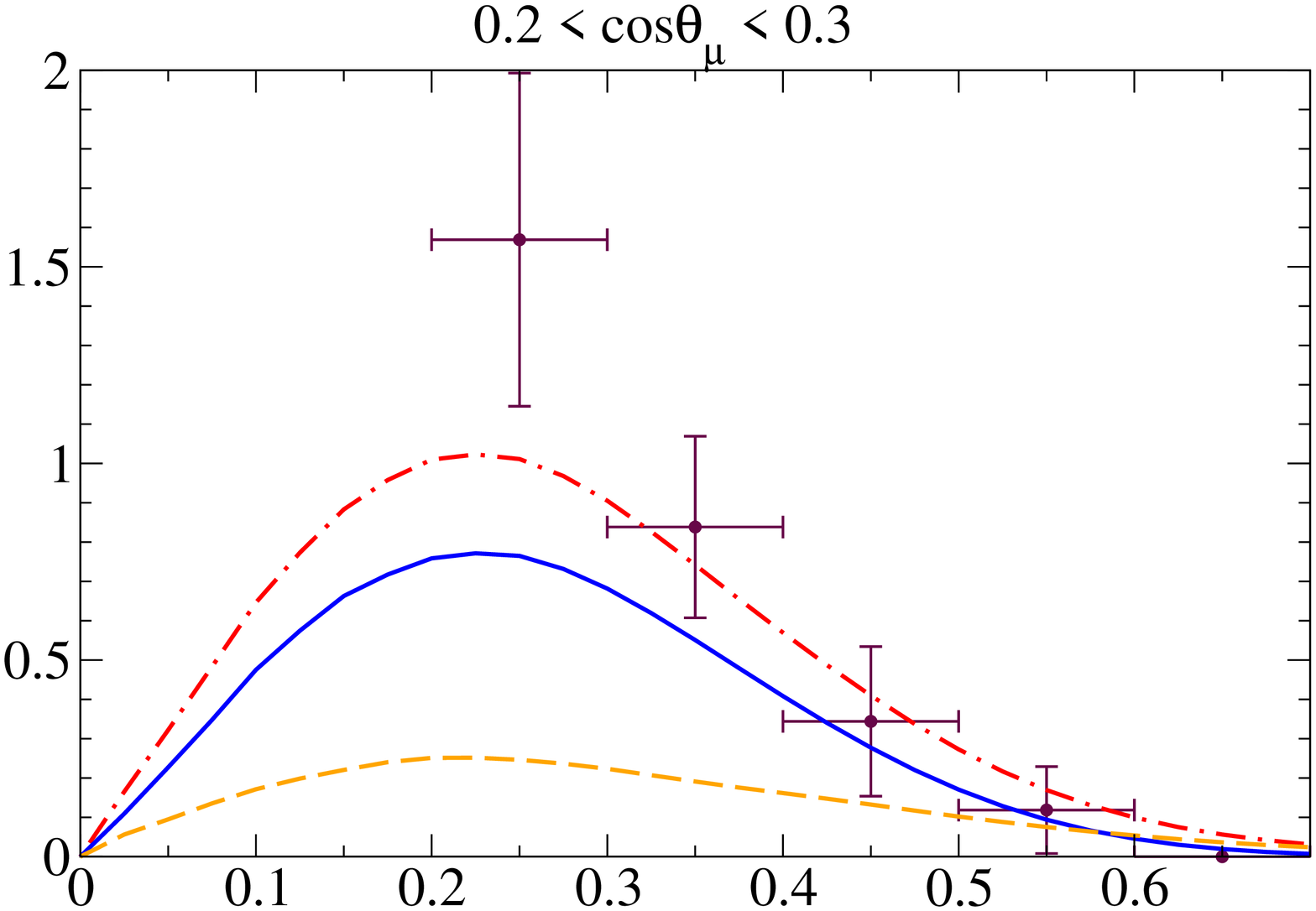}\hspace*{-0.15cm}\includegraphics[scale=0.22, angle=270]{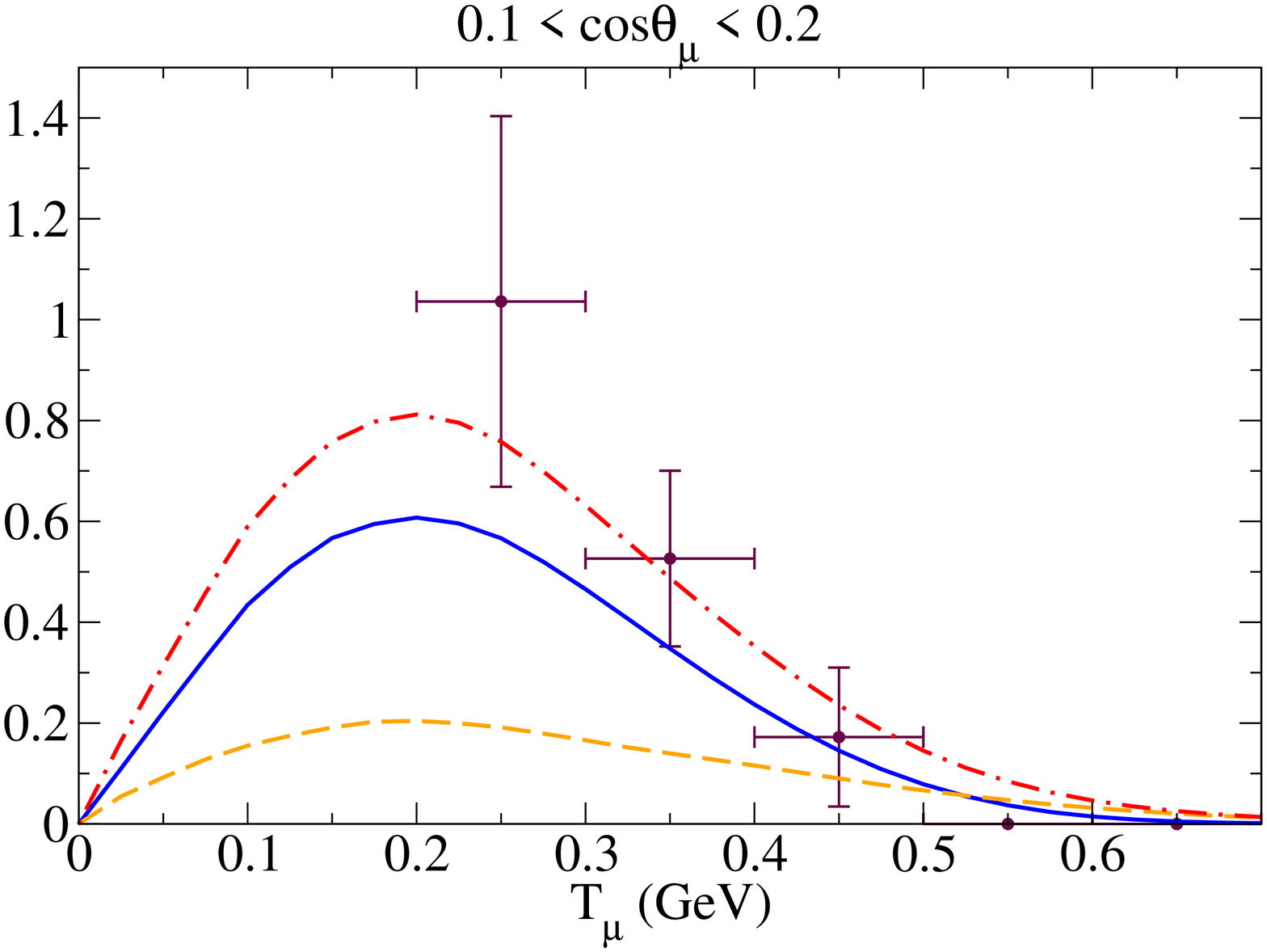}\\
\includegraphics[scale=0.22, angle=270]{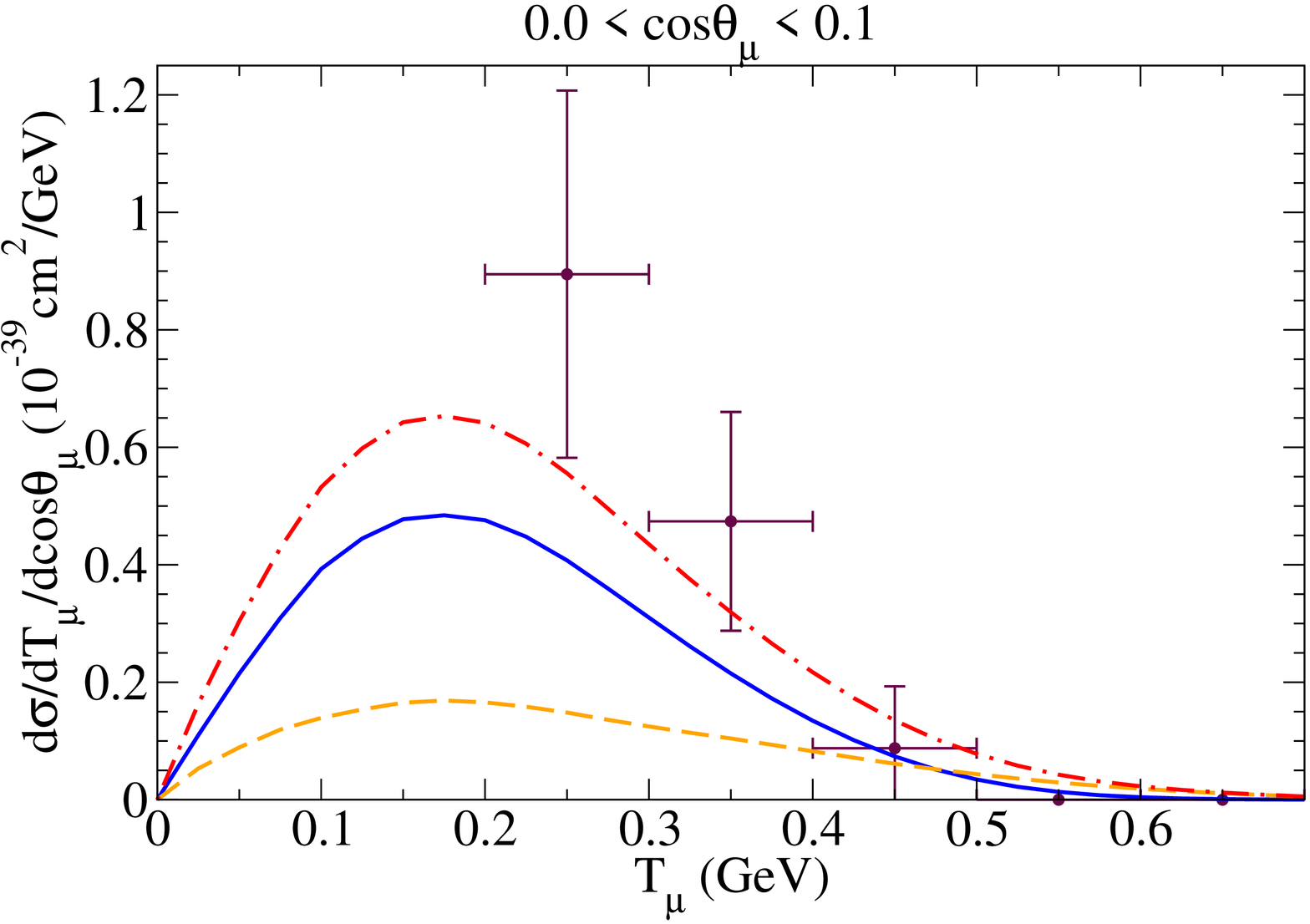}\hspace*{-0.15cm}\includegraphics[scale=0.22, angle=270]{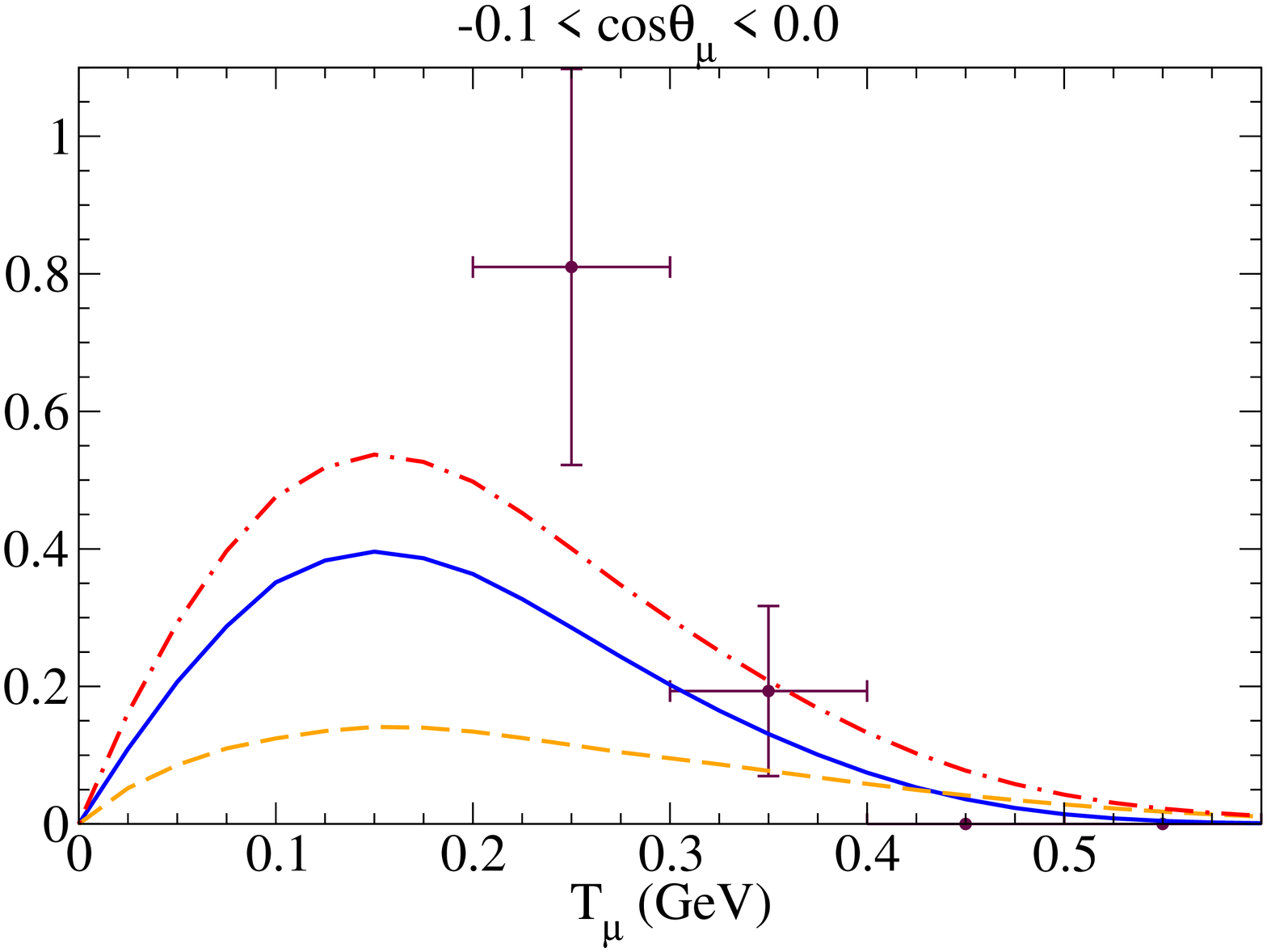}\hspace*{-0.15cm}\includegraphics[scale=0.22, angle=270]{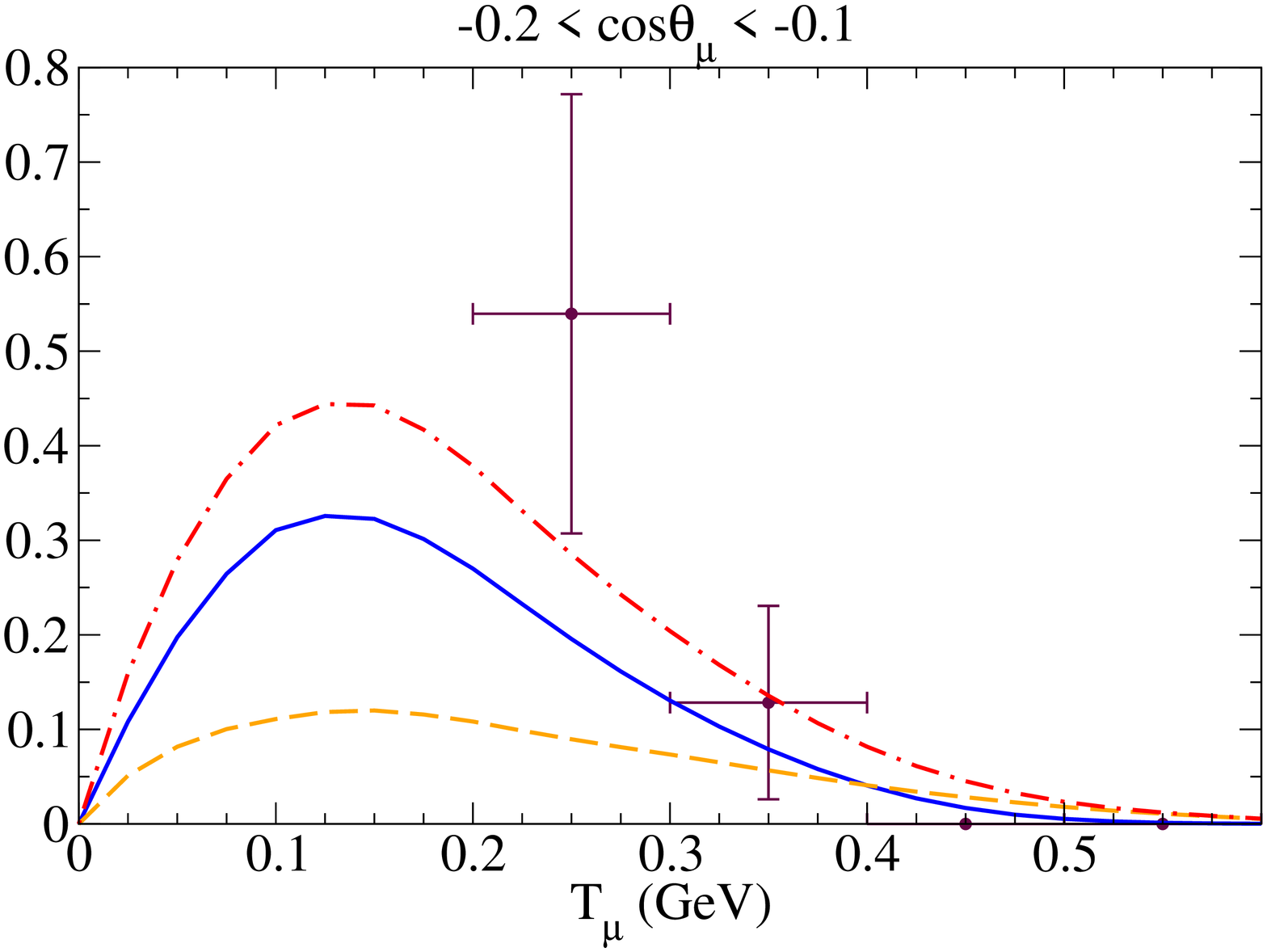}
\begin{center}
\vspace{-1cm}
\end{center}
\end{center}
\caption{(Color online) As for Fig.~\ref{Miniboone_nu}, but now for the $\bar\nu_\mu$ CCQE process on $^{12}$C. Data are from~\cite{AguilarArevalo:2013hm}.}\label{Miniboone_nubar}
\end{figure}


In Figs.~\ref{Miniboone_nu} and \ref{Miniboone_nubar} we show the double differential cross section averaged over the neutrino (antineutrino) energy flux against the kinetic energy of the final muon. Fig.\ref{Miniboone_nu} (Fig.\ref{Miniboone_nubar}) corresponds to neutrino (antineutrino) scattering on $^{12}$C.
Data are taken from the MiniBooNE collaboration~\cite{AguilarArevalo:2010zc,AguilarArevalo:2013hm}. We represent a large variety of kinematical situations where each panel refers to results averaged over a particular muon angular bin. Notice that the mean energy of the MiniBooNE $\nu_\mu$ ($\overline{\nu}_\mu$) flux is 788 (665) MeV. These high energies require a fully relativistic treatment of the process. In Figs.~\ref{Miniboone_nu} and \ref{Miniboone_nubar} we show results for the pure QE response (blue line), the total contribution of the 2p-2h MEC (orange dashed line), {\it i.e.,} including vector and axial terms in the three responses, L, T and T'. Finally, the total response (QE+2p-2h MEC) is represented by the red dot-dashed line.

As observed, the model tends to overpredict the data for the most
forward angles, {\it i.e.,} $0.9\le \cos\theta_\mu \le 1$. This
corresponds to very small energy and momentum transfers, a kinematic
situation where ``quasi-free" scattering is highly
questionable. However, note how well the pure QE response (blue line)
fits the data, in particular, for neutrinos. As the scattering angle
increases, the theoretical prediction including both the QE and the
2p-2h MEC effects agrees well with the data. This is the case for
neutrinos (Fig.\ref{Miniboone_nu}) at all angles. On the contrary, in
the case of antineutrinos (Fig.\ref{Miniboone_nubar}) the discrepancy
between theory and data tends to increase as $\theta_\mu$ gets
larger. Notice, however, that in these situations only a small number
of data points with large uncertainties exist and the cross section is
much smaller.  Results in Figs.~\ref{Miniboone_nu} and
\ref{Miniboone_nubar} clearly show the relevant role played by effects
beyond the impulse approximation. In particular, 2p-2h MEC
contributions are essential in order to describe data. This has been
discussed at length in previous
works~\cite{Martini:2009aa,Martini:2011wp,Martini12PRD} but using different kinds of non-relativistic
approximations and some assumptions on the behavior of the
responses entering in neutrino reactions, {\it i.e.,} assuming the
axial-axial contributions being equal to the vector-vector ones and
the interference T' response to be proportional to the pure
transverse vector-vector one. Here we calculate explicitly all the
contributions within a fully relativistic framework (see results in
Figs.~\ref{Miniboone_nu} and \ref{Miniboone_nubar}). As shown, the
contribution of the 2p-2h MEC effects is very relevant for both
neutrinos and antineutrinos, their relative percentage at the maximum,
compared with the pure QE response, being of order $25-35\%$. The
relative strength associated with 2p-2h MEC gets larger for increasing
values of the angle, particularly, in the case of antineutrinos.  Note
that, in spite of the quite different neutrino and antineutrino energy
fluxes, the quality of the agreement with data is rather similar in
the two cases.


\begin{figure}[H]
\begin{center}\vspace{-1.8cm}
\includegraphics[scale=0.22, angle=270]{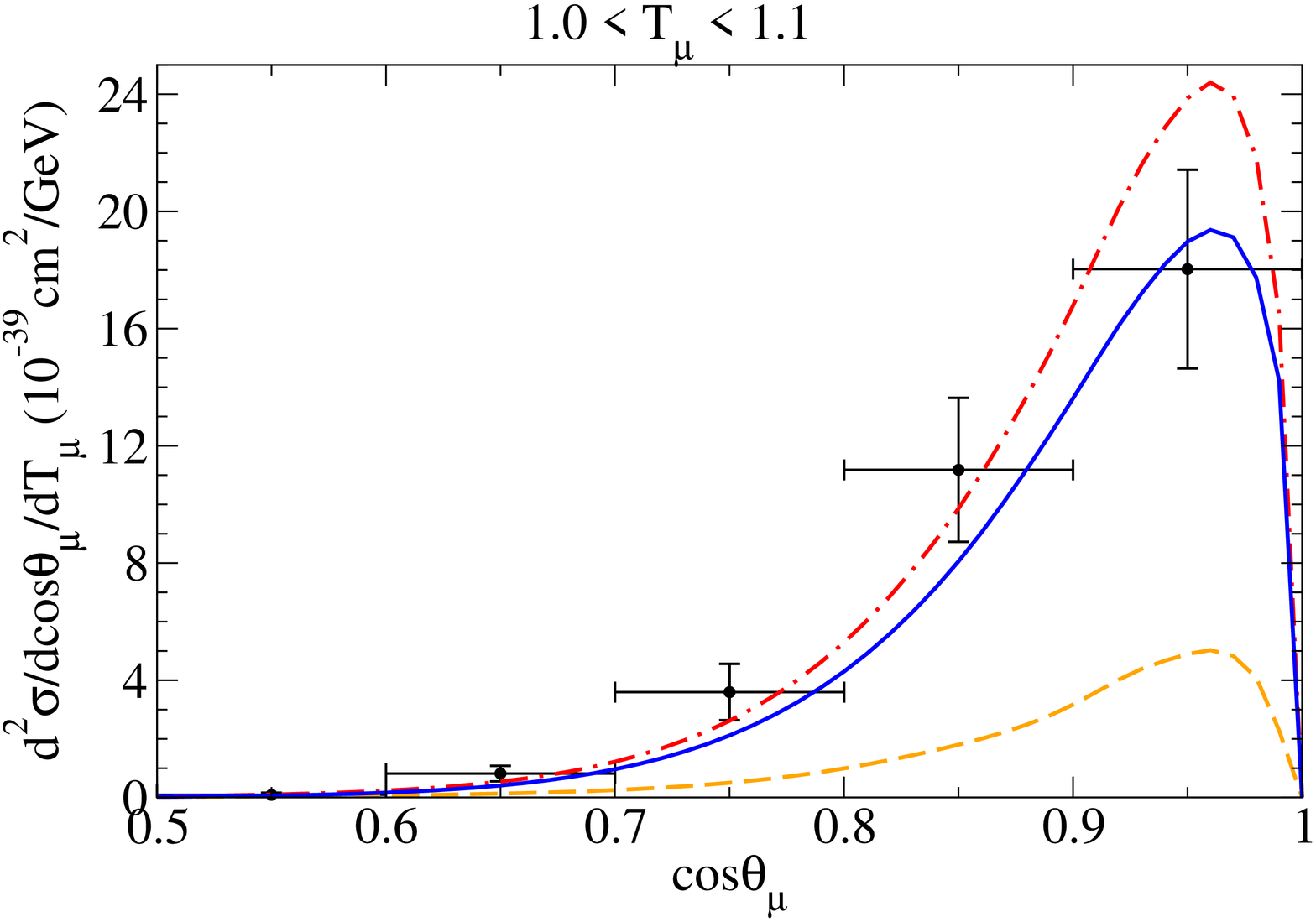}\hspace*{-0.05cm}\includegraphics[scale=0.22, angle=270]{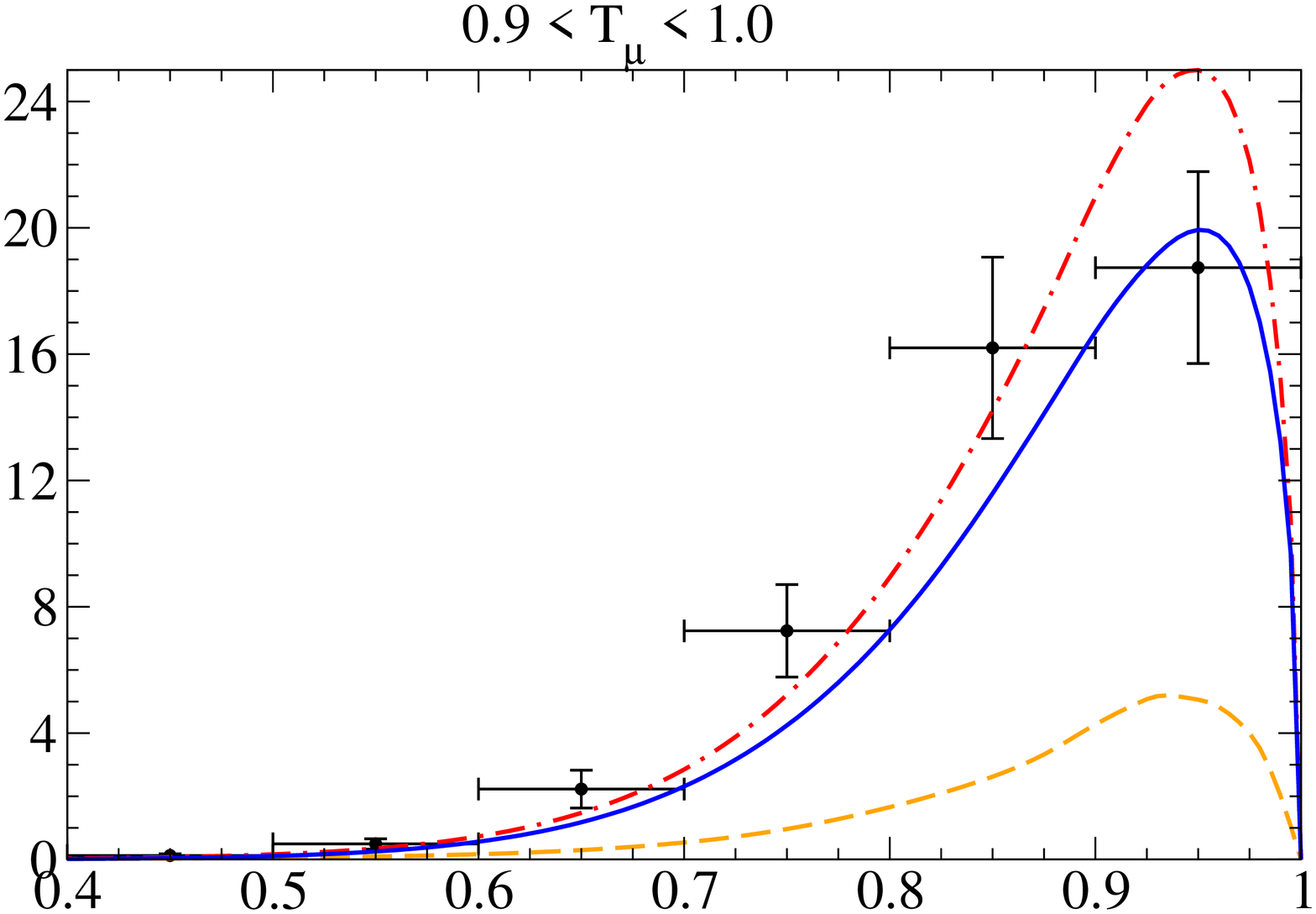}\hspace*{-0.15cm}\includegraphics[scale=0.22, angle=270]{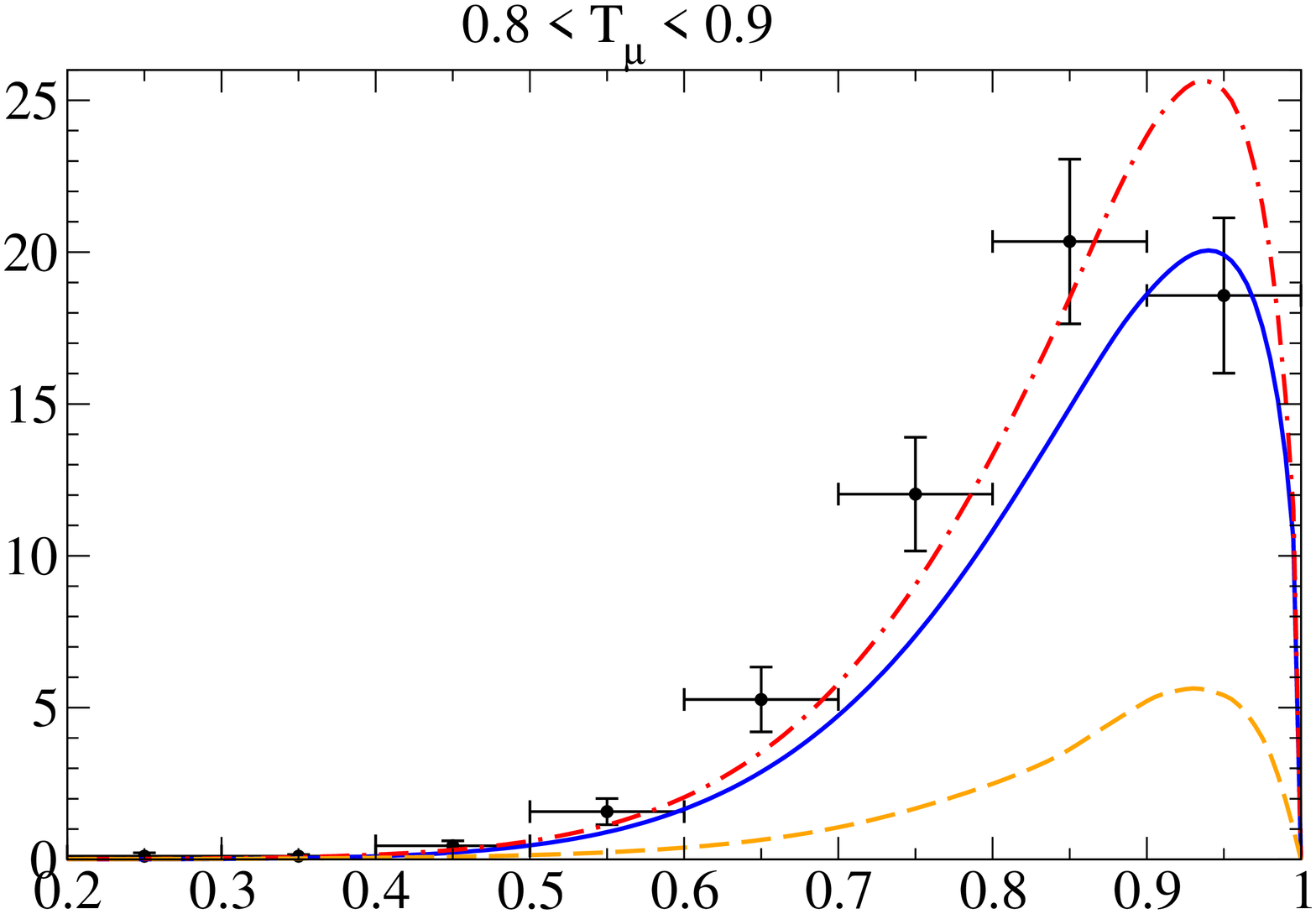}\hspace*{-0.15cm}\\
\includegraphics[scale=0.22, angle=270]{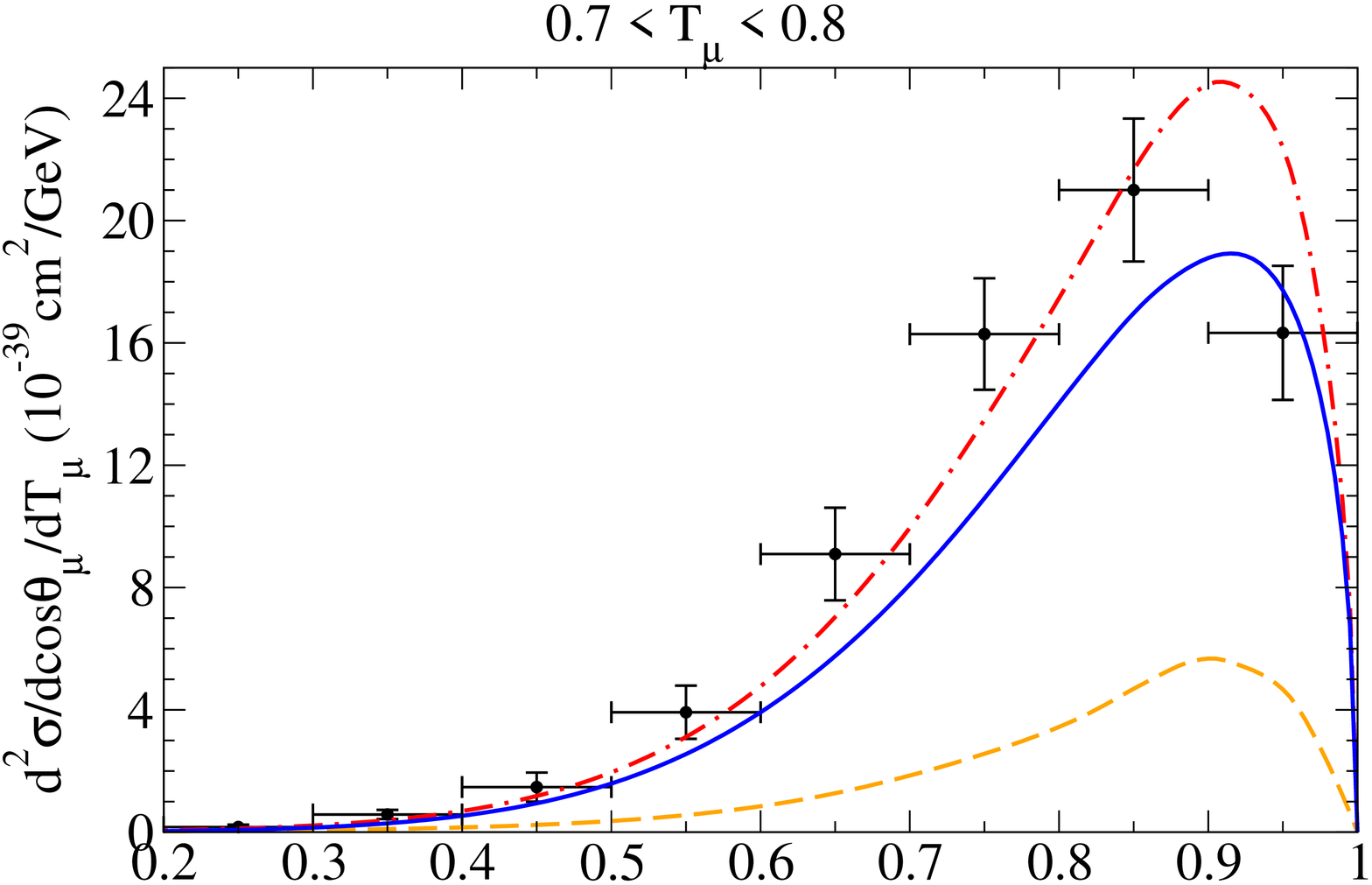}\hspace*{-0.15cm}
\includegraphics[scale=0.22, angle=270]{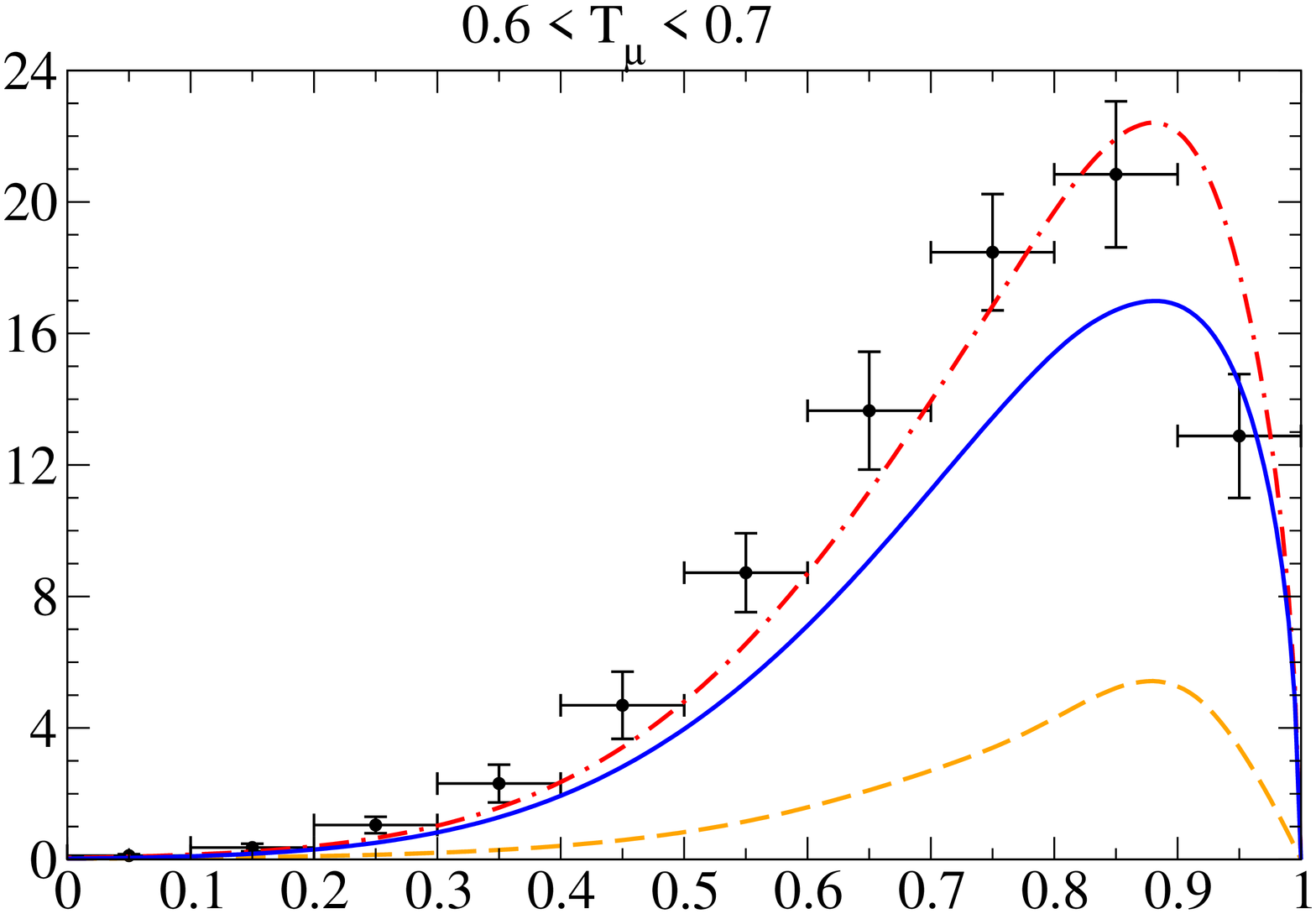}\hspace*{-0.15cm}\includegraphics[scale=0.22, angle=270]{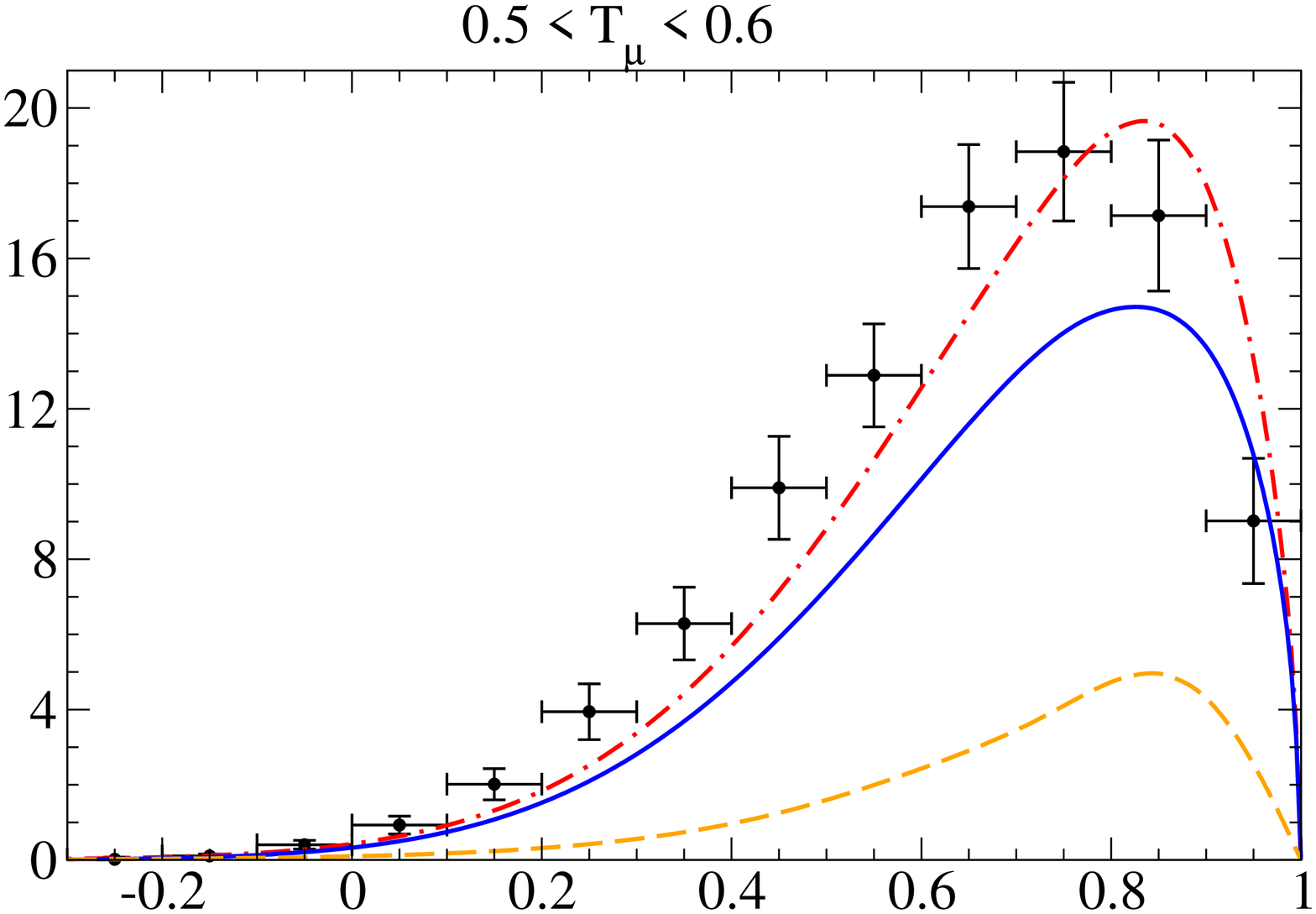}\hspace*{-0.15cm}\\
\includegraphics[scale=0.22, angle=270]{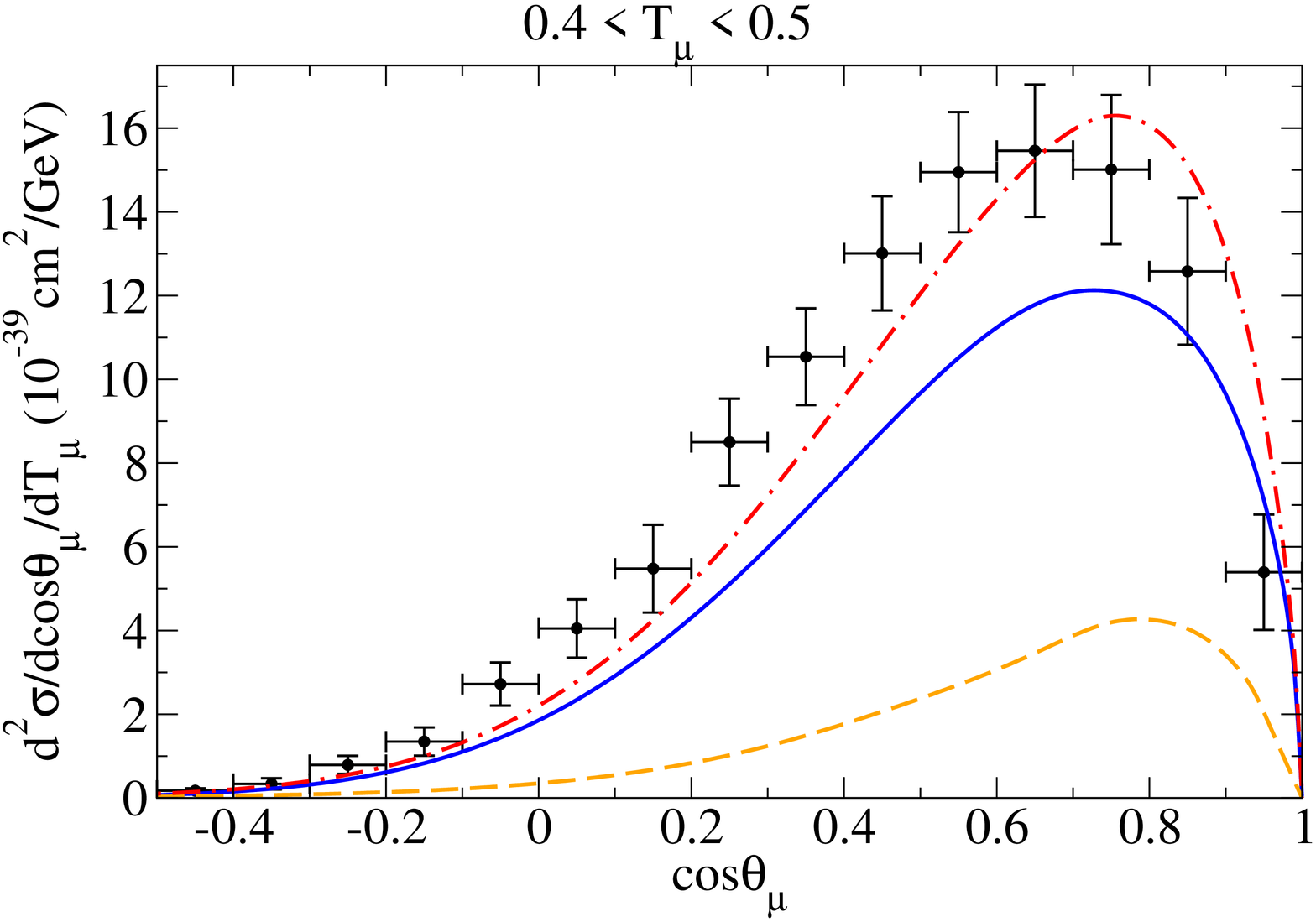}\hspace*{-0.15cm}
\includegraphics[scale=0.22, angle=270]{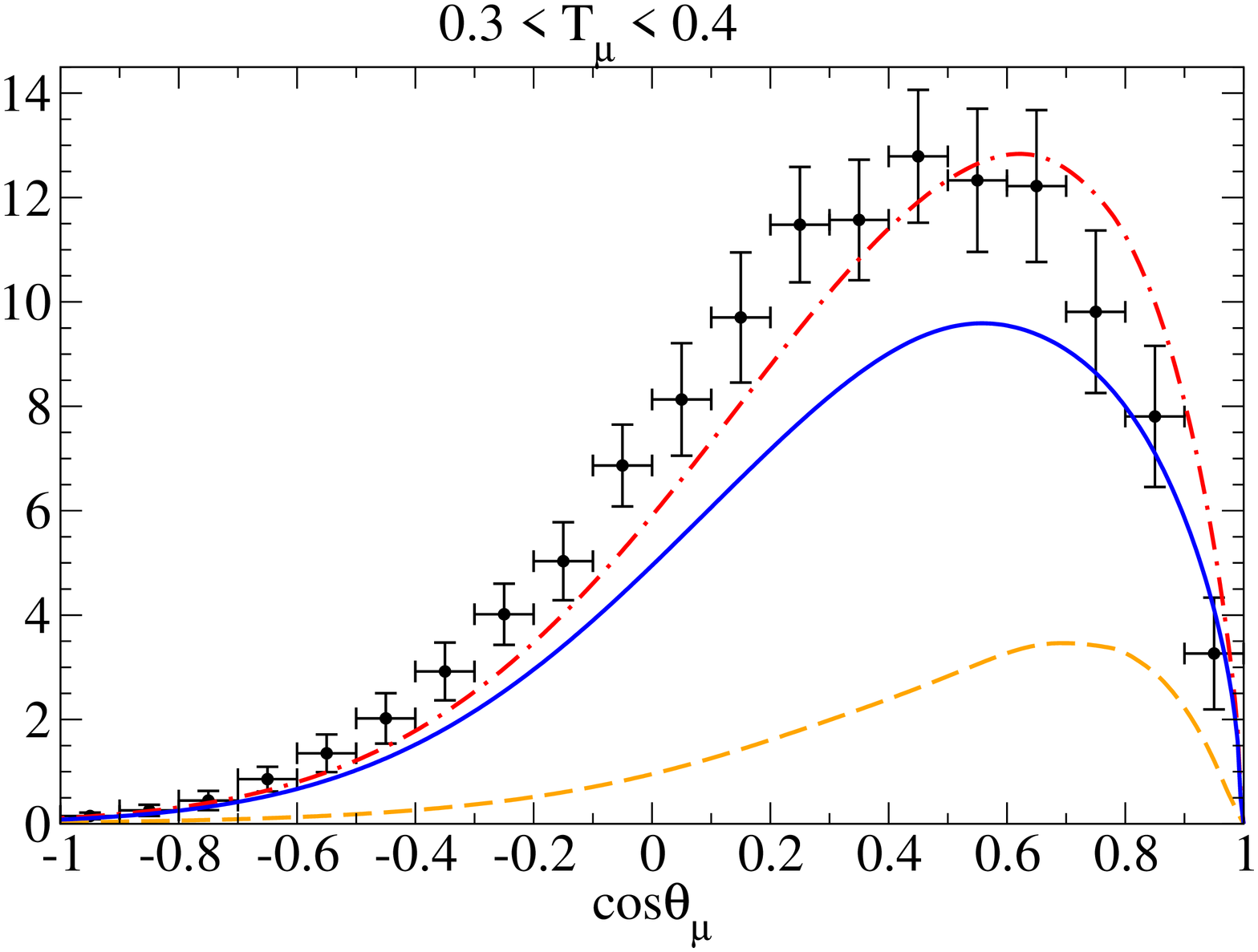}\hspace*{-0.15cm}\includegraphics[scale=0.22, angle=270]{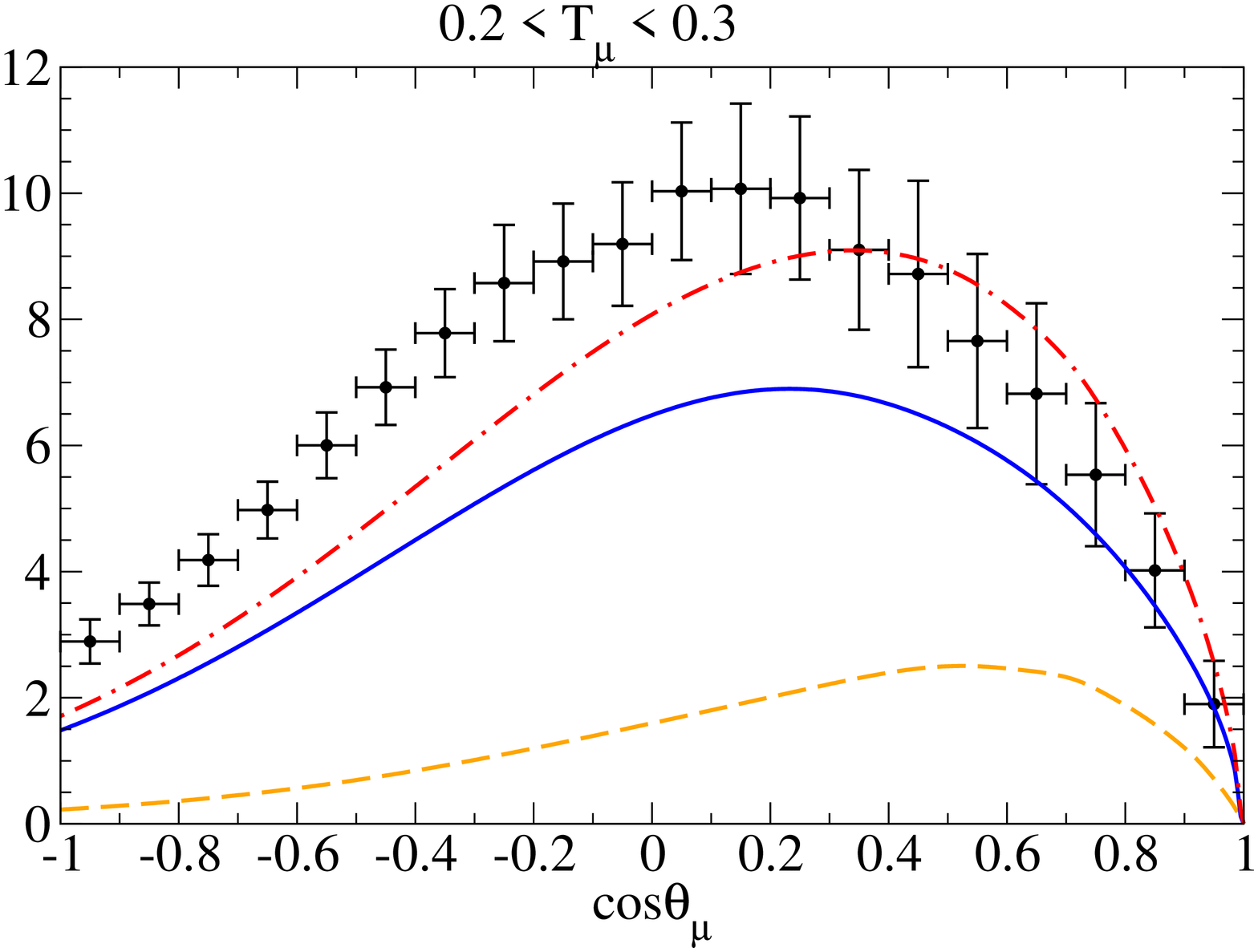}
\begin{center}
\vspace{-1cm}
\end{center}
\end{center}
\caption{(Color online) MiniBoone flux-folded double differential
  cross section per target nucleon for the $\nu_\mu$ CCQE process on
  $^{12}$C displayed versus $\cos\theta_\mu$ for various bins of
  $T_\mu$ obtained within the SuSAv2+MEC approach. QE and 2p-2h MEC
  results are also shown separately.  Data are
  from~\cite{AguilarArevalo:2010zc}.
}\label{CS_thetamu_nu}
\end{figure}

\begin{figure}[H]
\begin{center}\vspace{-1.8cm}
\includegraphics[scale=0.22, angle=270]{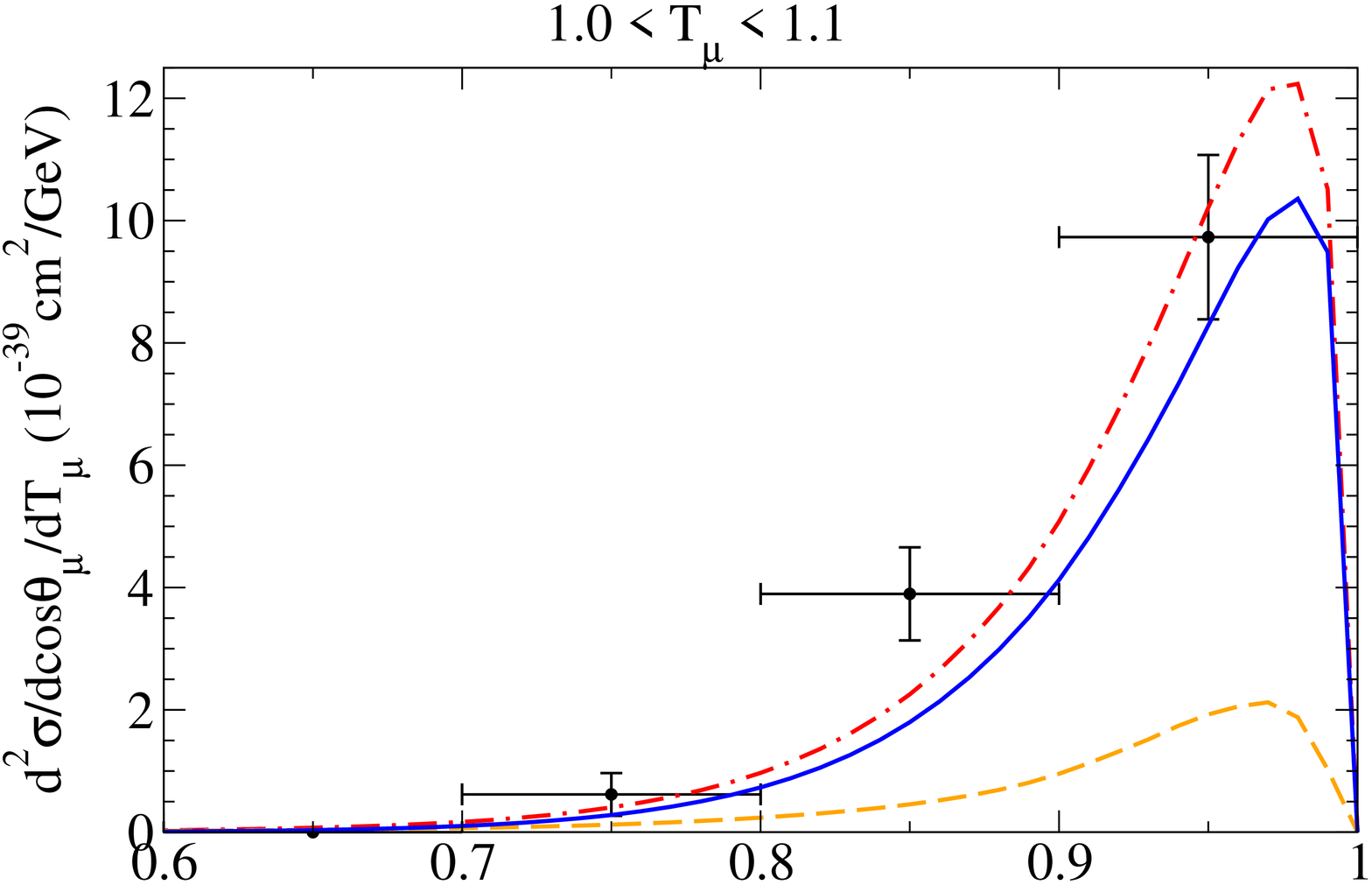}\hspace*{-0.05cm}\includegraphics[scale=0.22, angle=270]{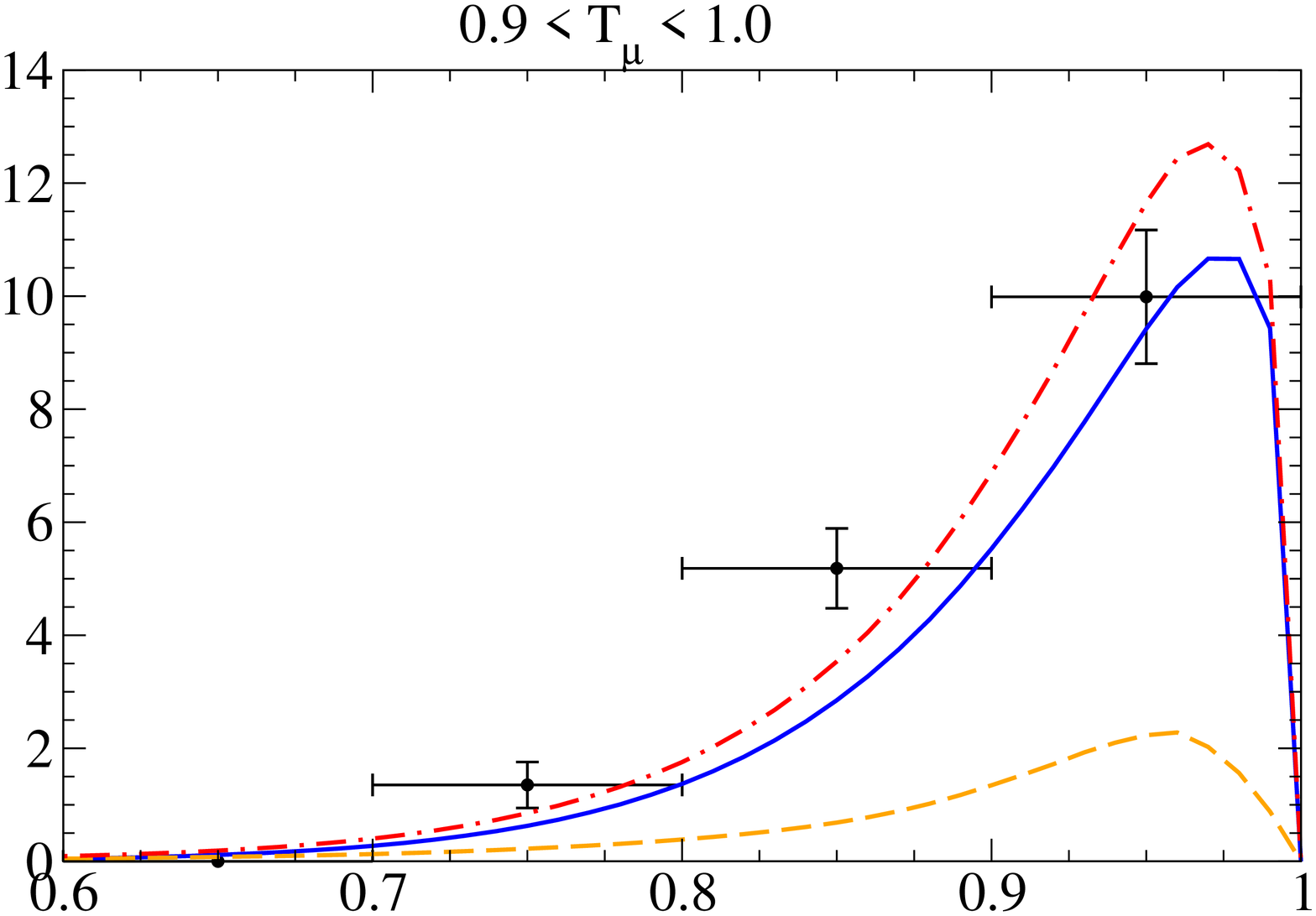}\hspace*{-0.15cm}\includegraphics[scale=0.22, angle=270]{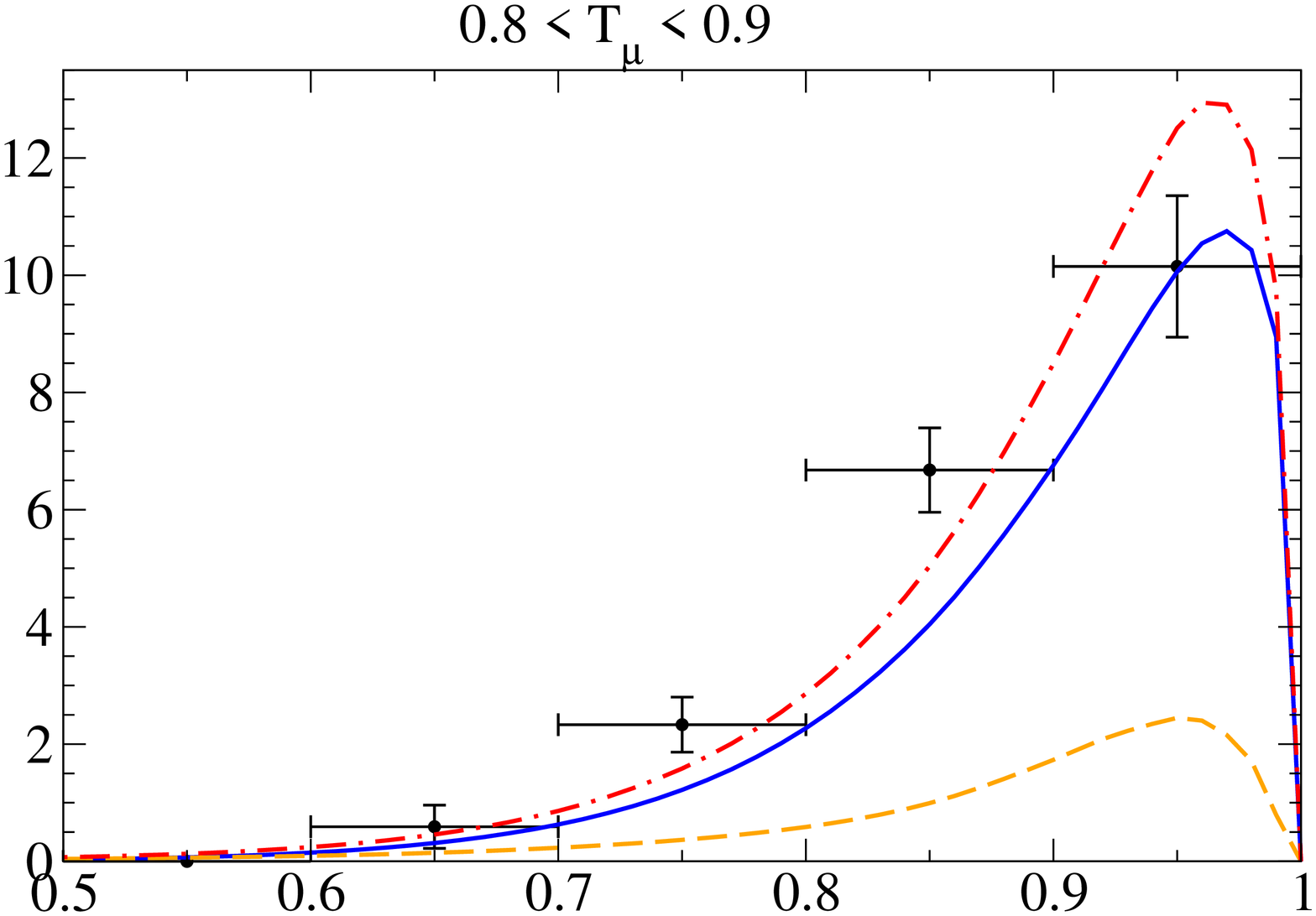}\hspace*{-0.15cm}\\
\includegraphics[scale=0.22, angle=270]{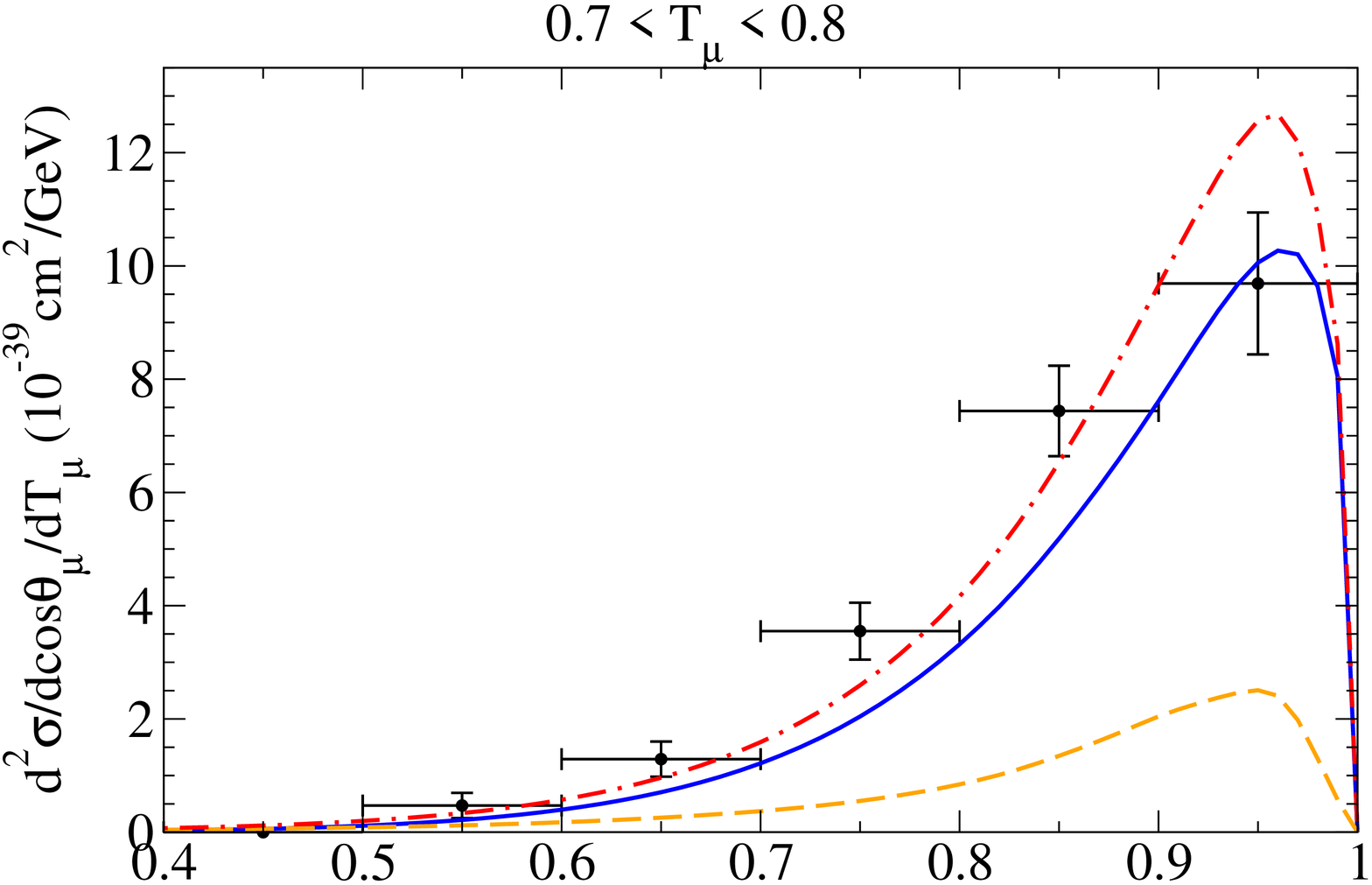}\hspace*{-0.15cm}
\includegraphics[scale=0.22, angle=270]{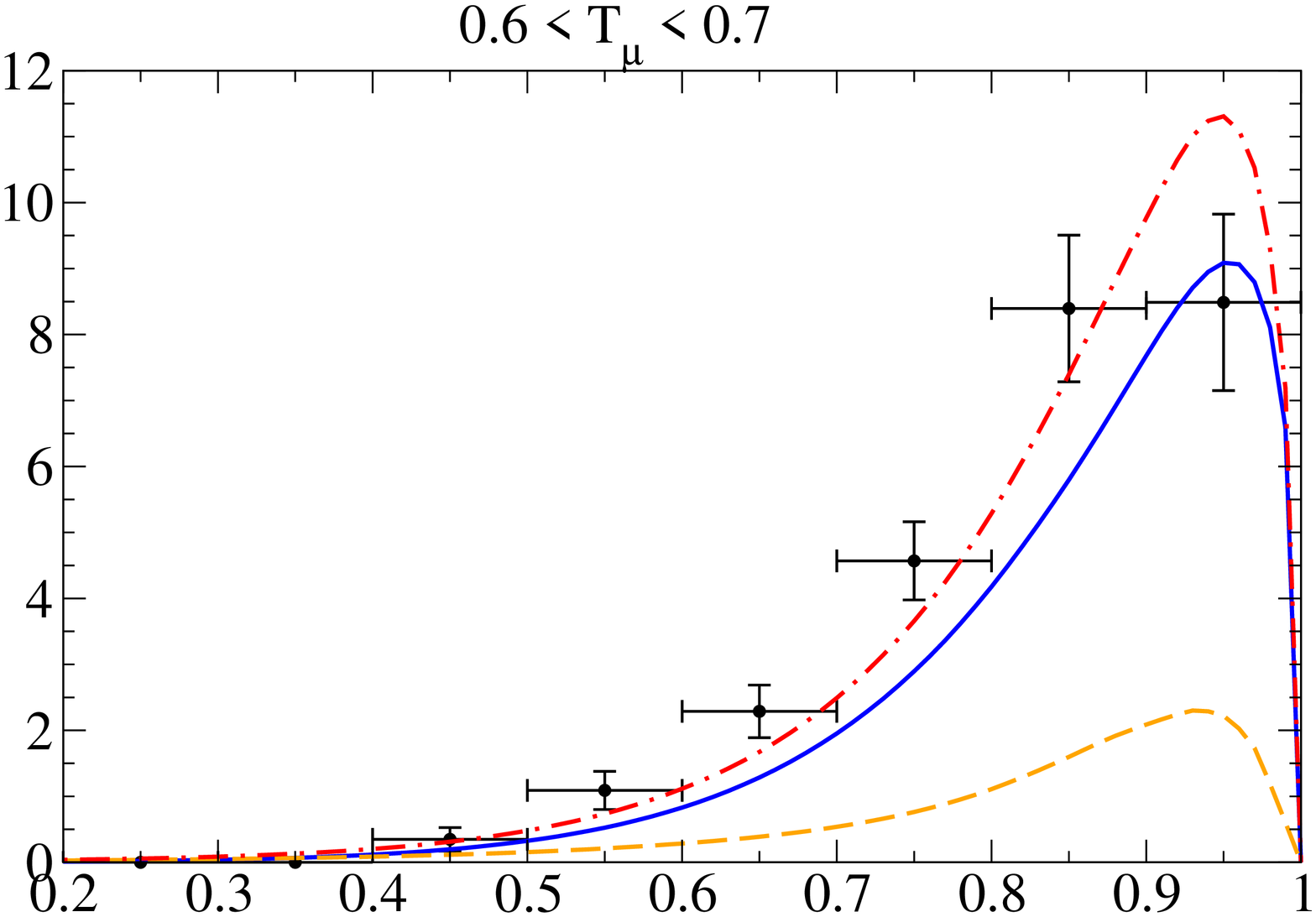}\hspace*{-0.15cm}\includegraphics[scale=0.22, angle=270]{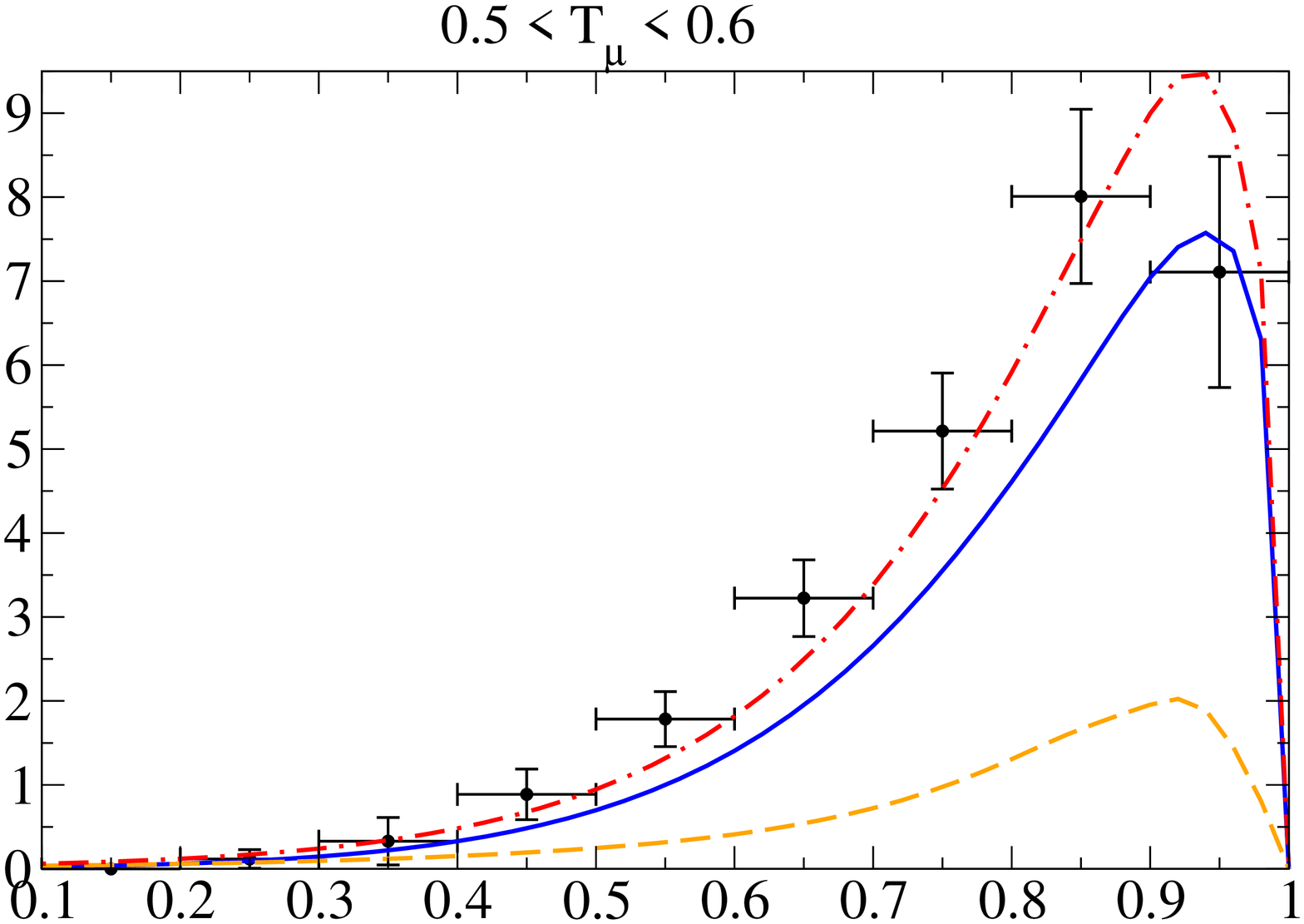}\hspace*{-0.15cm}\\
\includegraphics[scale=0.22, angle=270]{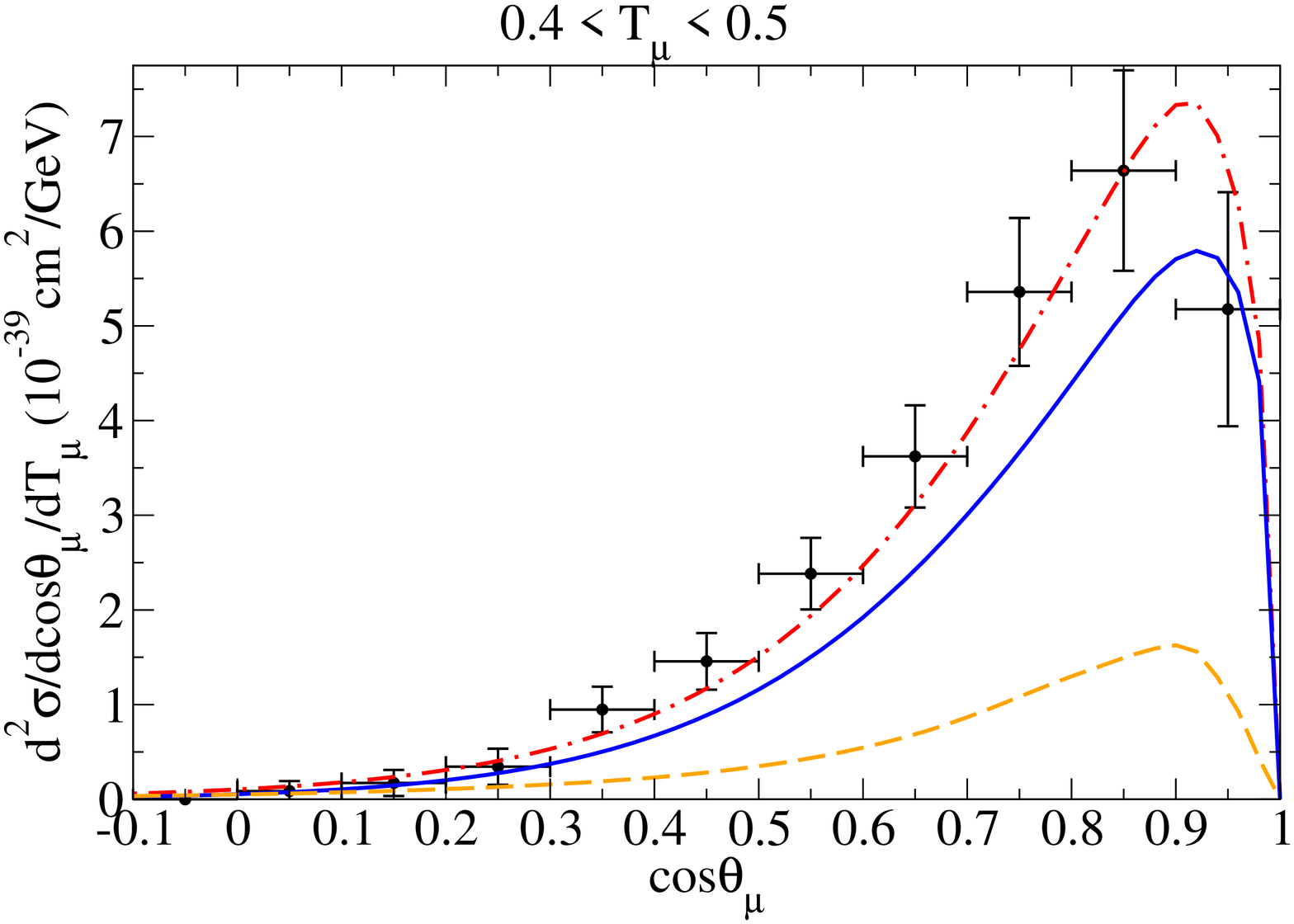}\hspace*{-0.15cm}
\includegraphics[scale=0.22, angle=270]{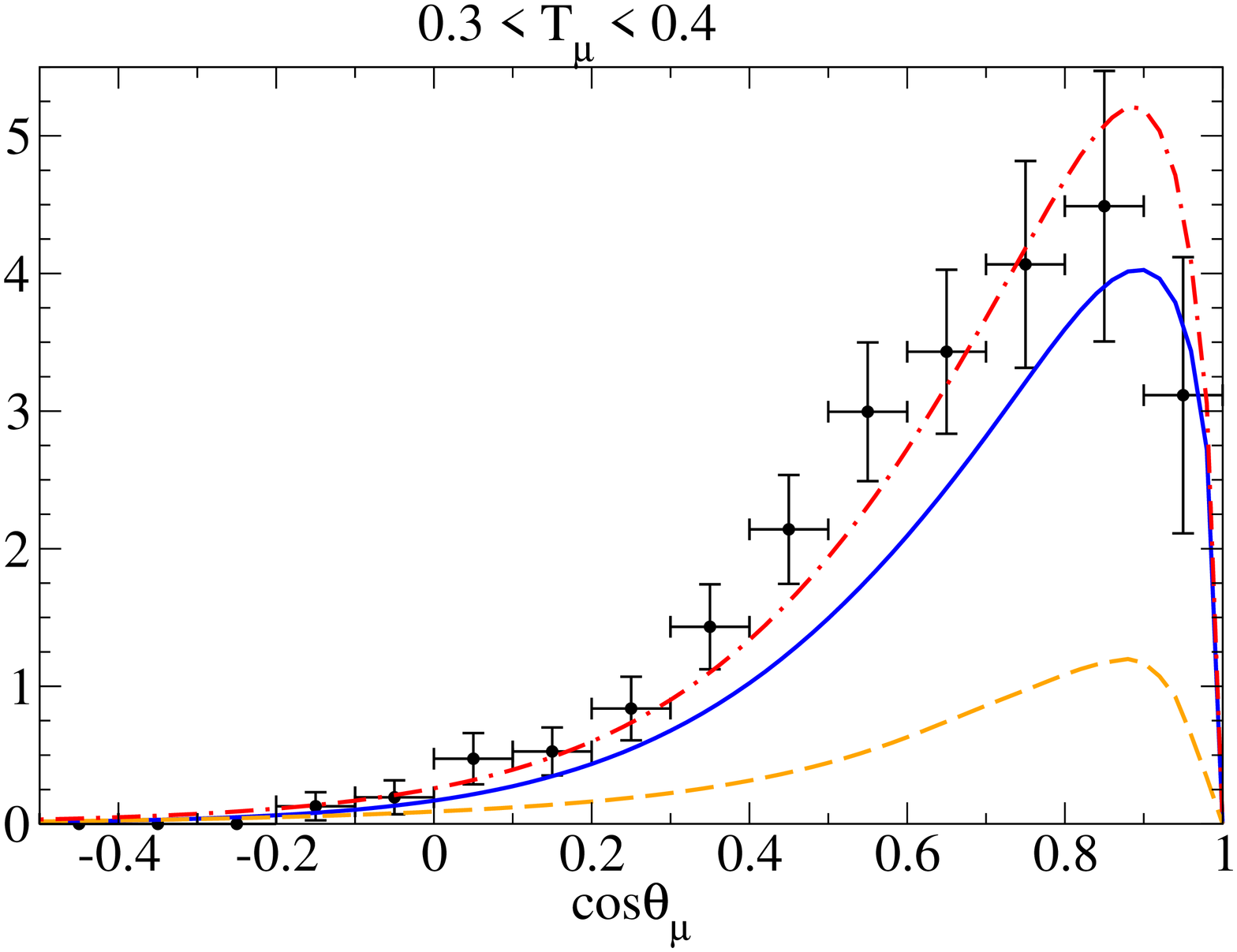}\hspace*{-0.15cm}\includegraphics[scale=0.22, angle=270]{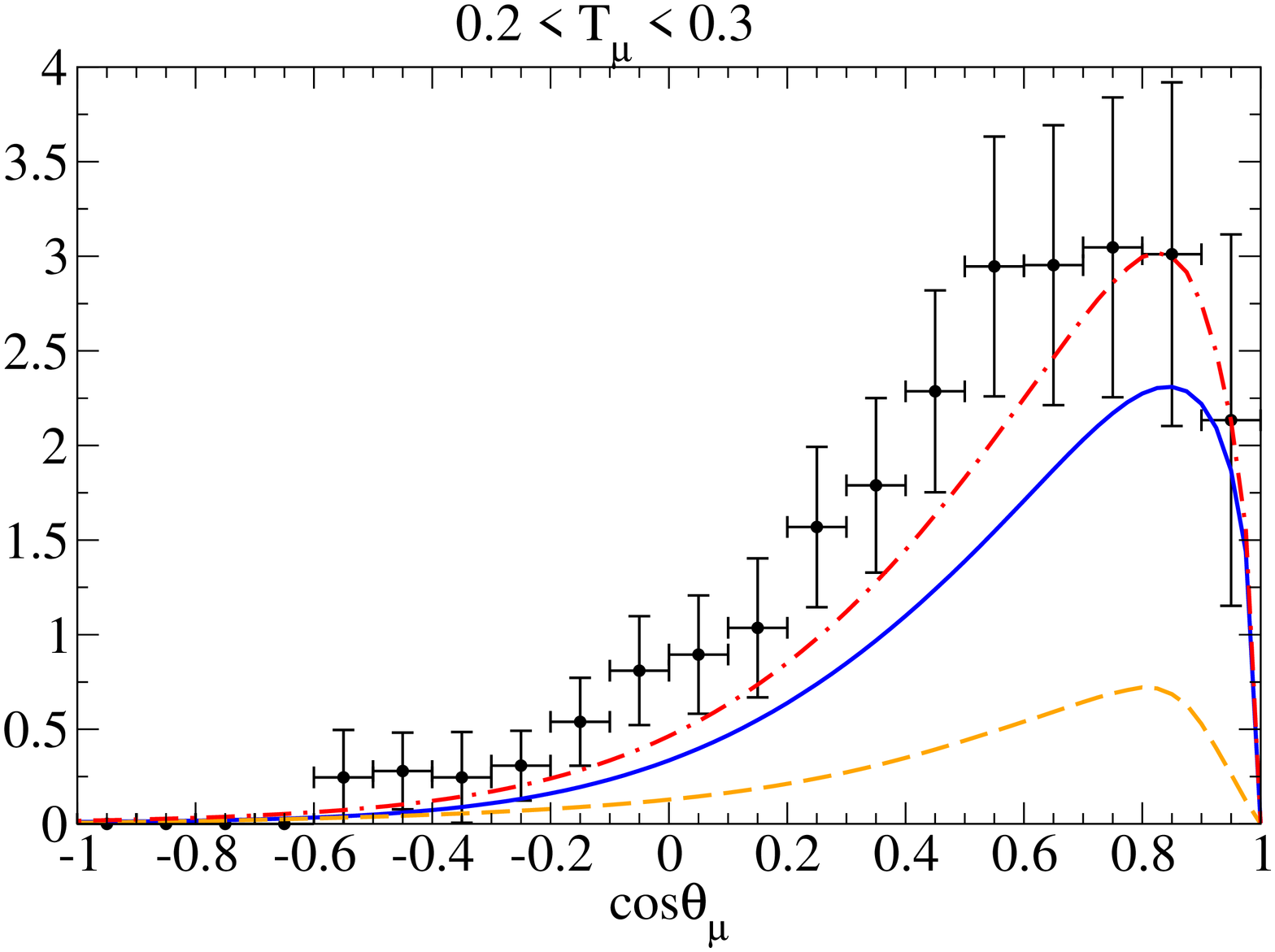}
\begin{center}
\vspace{-1cm}
\end{center}
\end{center}
\caption{(Color online) As for Fig.~\ref{CS_thetamu_nu}, but now for
  the $\bar\nu_\mu$ CCQE process on $^{12}$C. Data are
  from~\cite{AguilarArevalo:2013hm}.
}\label{CS_thetamu_nubar}
\end{figure}


To complete the previous discussion on the double differential cross sections, we present in Figs.  \ref{CS_thetamu_nu} and \ref{CS_thetamu_nubar} the results averaged over the muon kinetic energy bins as functions of the muon scattering angle for
neutrinos and antineutrinos, respectively. These graphs complement the previous ones, and prove the capability of the model to reproduce the data for a large variety of kinematic situations. The 2p-2h MEC contributions increase the pure QE response by $\sim 25-35\%$ (depending on the particular region explored) and are shown to be essential in order to describe the data. As observed,
the total model tends to overpredict the data measured at angles close to zero and $T_\mu$ in the vicinity of $\sim 0.8-1$ GeV. This is consistent with results in previous figures and the inability of the model to describe properly data at very small angles. However, the largest discrepancy between theory and data occurs at the smallest muon kinetic energy bins considered, {\it i.e.,} $0.2<T_\mu<0.4$, in particular, for neutrinos (Fig.\ref{CS_thetamu_nu}) and angles bigger than $90^0$ ($\cos\theta_\mu<0$). As seen, the data are higher by $\sim 25-30\%$ 
than theoretical predictions. This outcome is consistent with the partial results shown in the panels on the bottom in Figs.~\ref{Miniboone_nu} and \ref{Miniboone_nubar}.


\begin{figure}[H]
\begin{center}\vspace{-1.8cm}
\includegraphics[scale=0.262, angle=270]{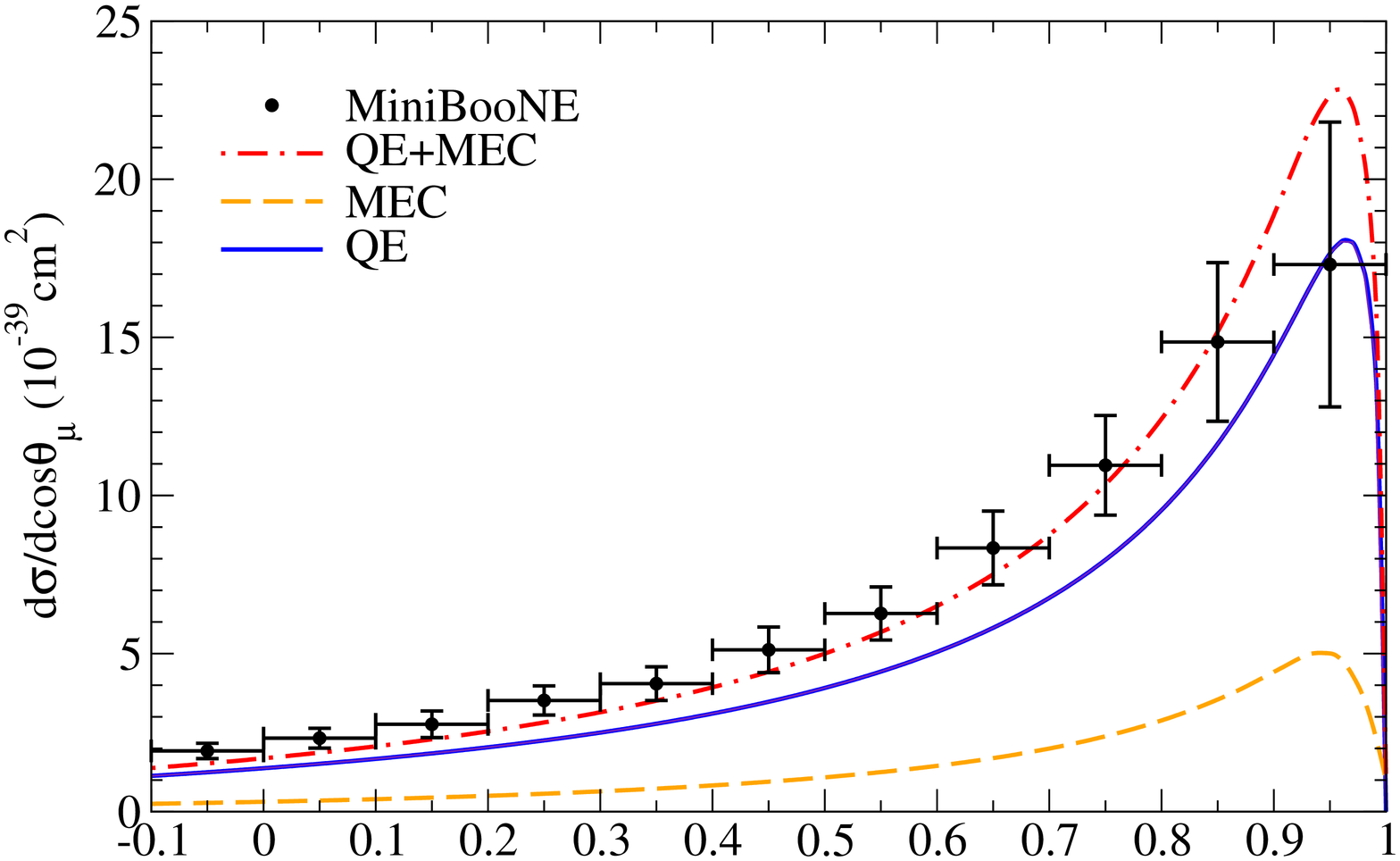}\hspace*{-0.05cm}\includegraphics[scale=0.262, angle=270]{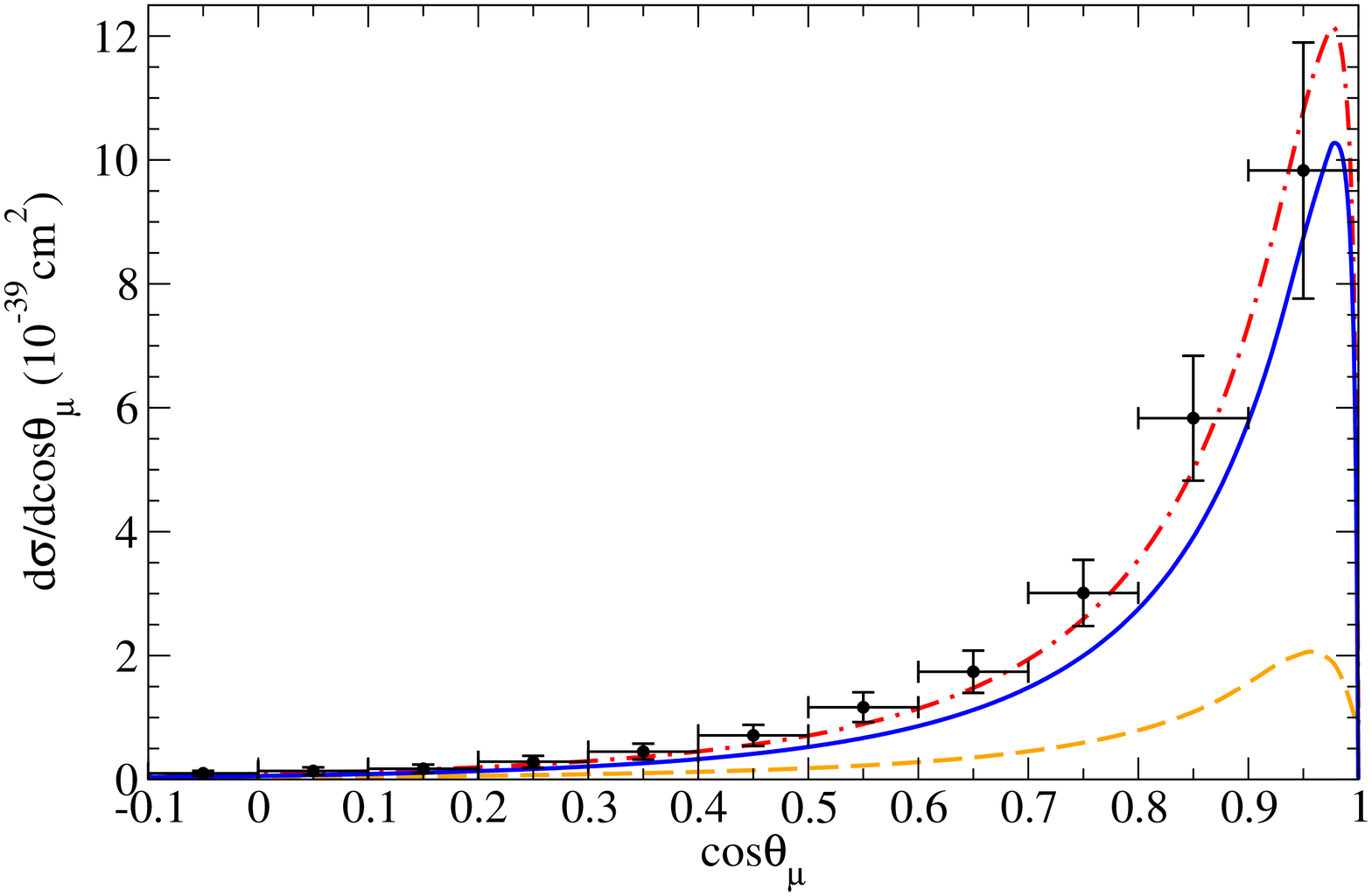}\\
\includegraphics[scale=0.262, angle=270]{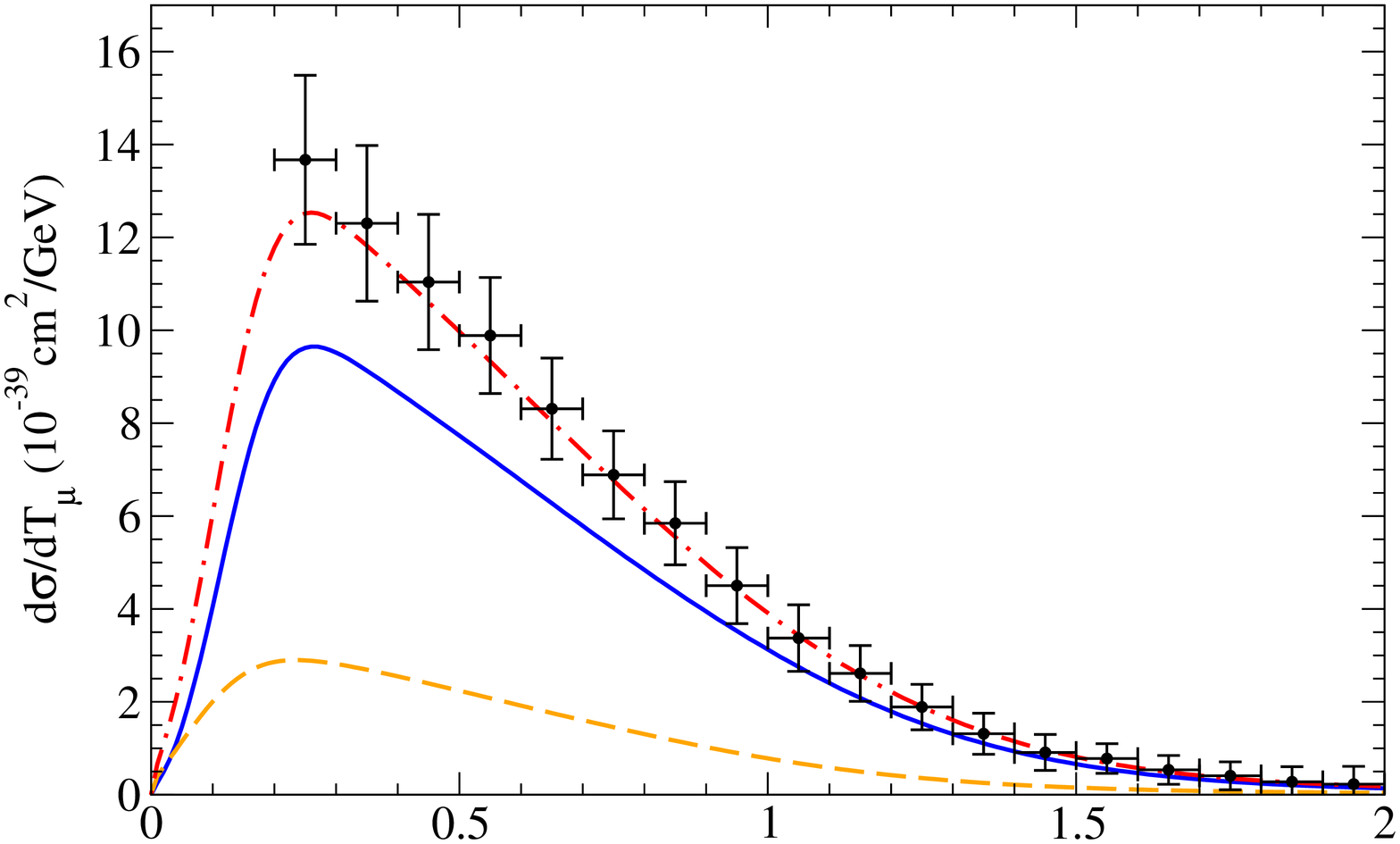}\hspace*{-0.05cm}\includegraphics[scale=0.262, angle=270]{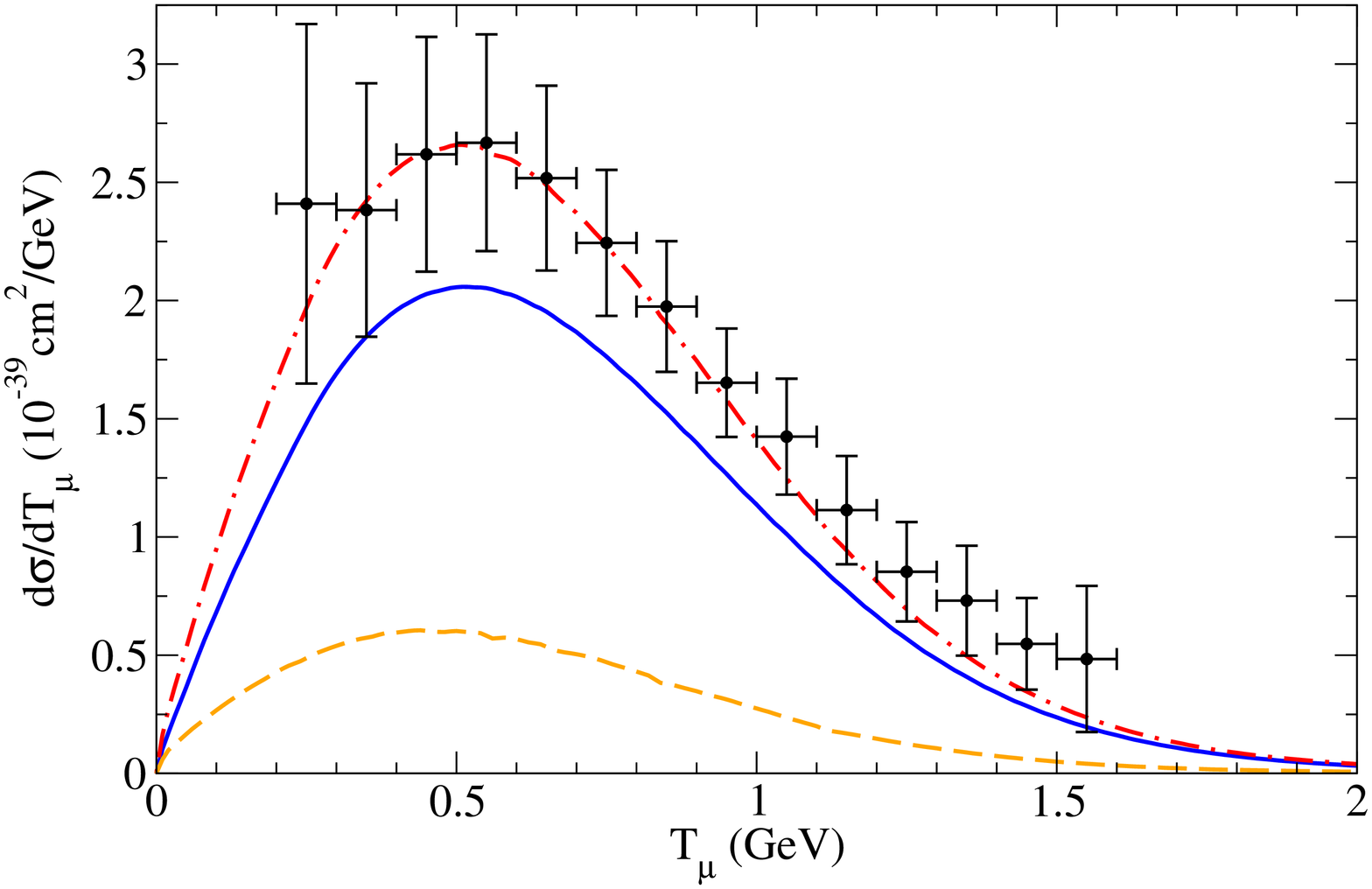}
\begin{center}
\vspace{-1cm}
\end{center}
\end{center}
\caption{(Color online) MiniBooNE flux-averaged CCQE $\nu_\mu$-$^{12}$C ($\bar\nu_\mu$-$^{12}$C) differential cross section per nucleon as a function of the muon scattering angle (top panels) and of
the muon kinetic energy (bottom panels). The left panels correspond to neutrino cross sections and the right ones to antineutrino reactions. Data are from~\cite{AguilarArevalo:2010zc,AguilarArevalo:2013hm}}\label{CS_single}
\end{figure}


\begin{figure}[H]
\begin{center}\vspace{-1.8cm}
\includegraphics[scale=0.262, angle=270]{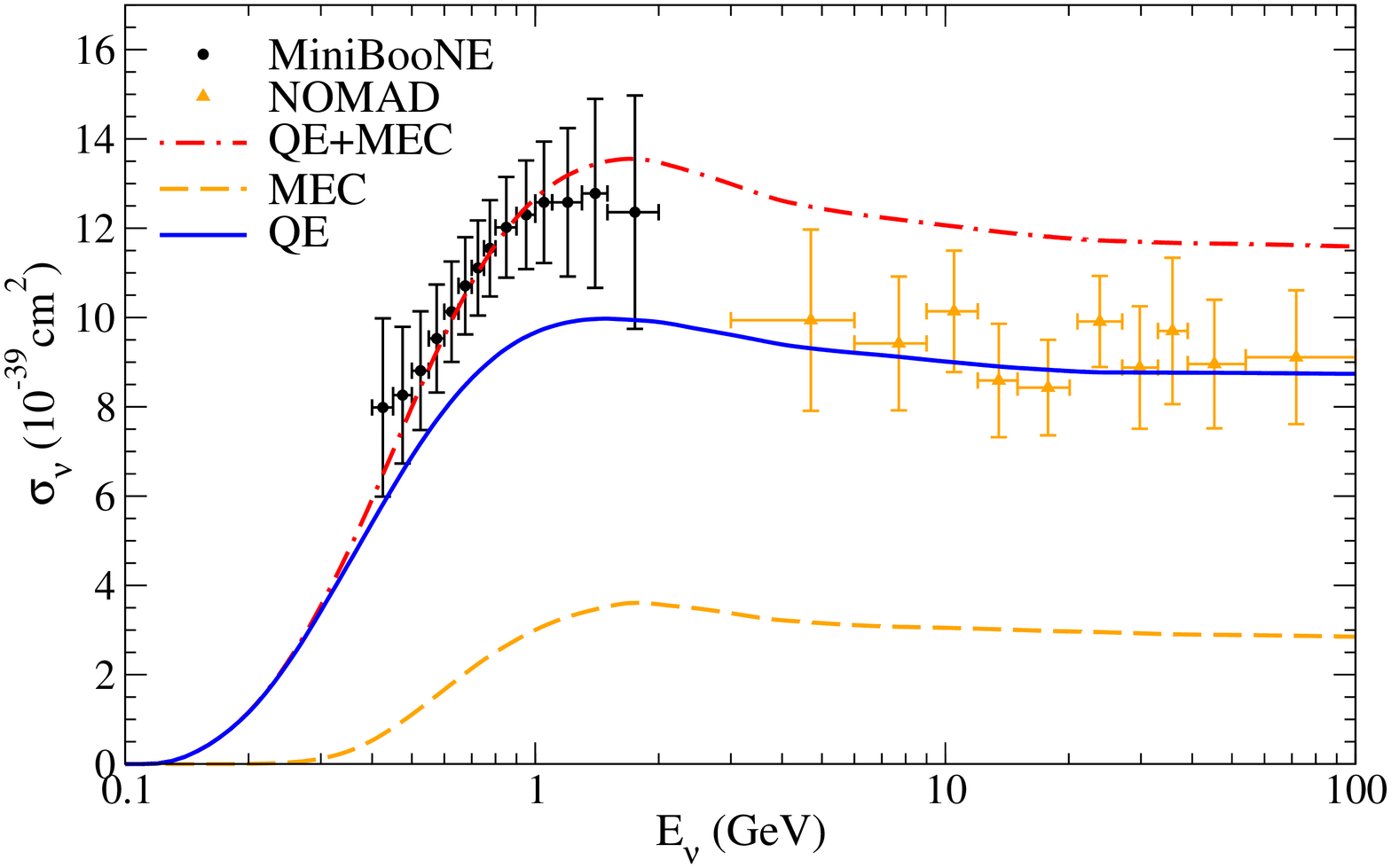}\hspace*{-0.05cm}\includegraphics[scale=0.262, angle=270]{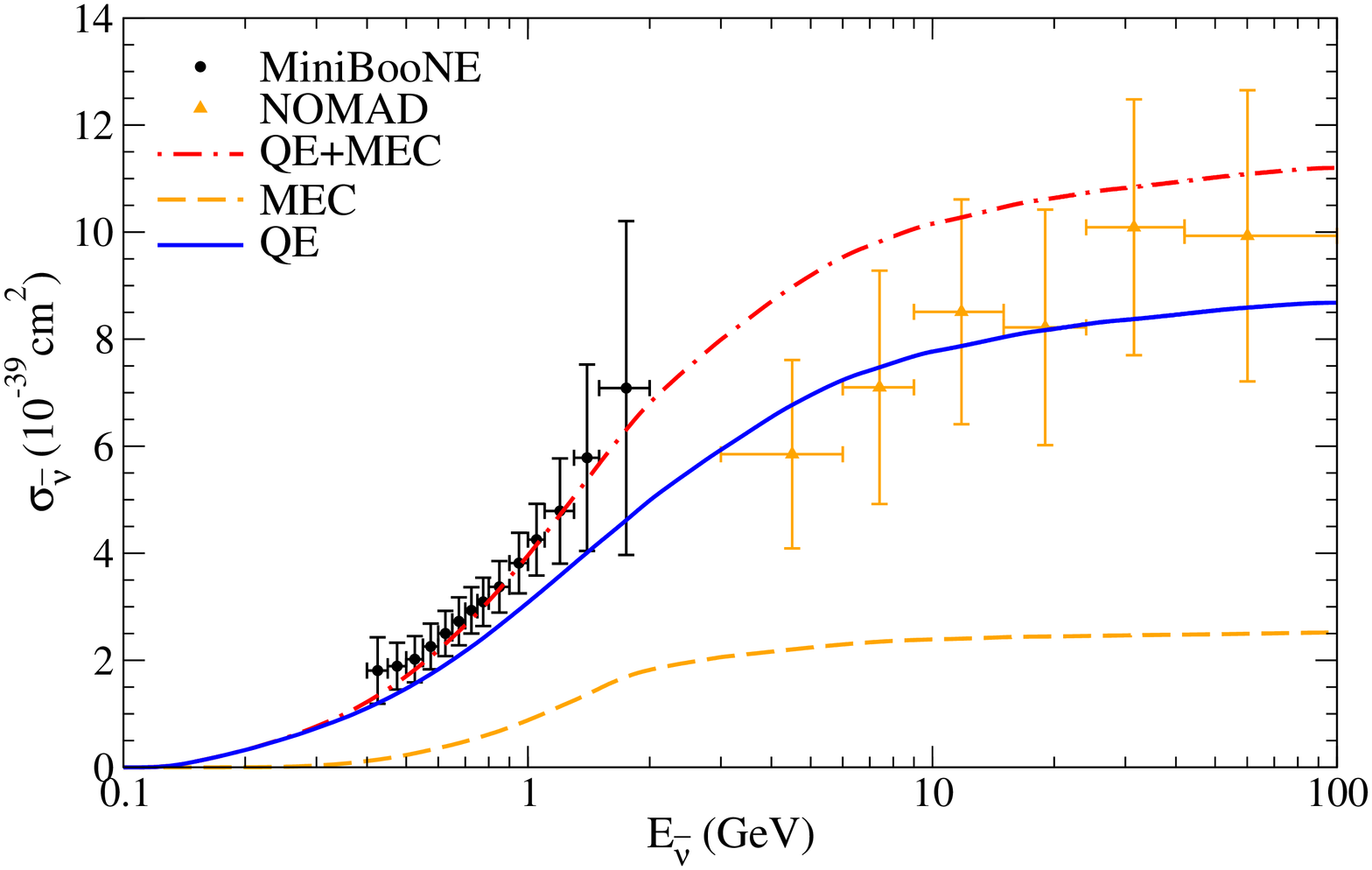}
\begin{center}
\vspace{-1cm}
\end{center}
\end{center}
\caption{(Color online) MiniBooNE CCQE $\nu_\mu$-$^{12}$C ($\bar\nu_\mu$-$^{12}$C) total cross section per nucleon as a function of the neutrino energy. The left panel correspond to neutrino cross sections and the right one to antineutrino reactions. Data are from~\cite{AguilarArevalo:2010zc,AguilarArevalo:2013hm}}\label{total_CS}
\end{figure}


In Fig.~\ref{CS_single} results are presented for the MiniBooNE flux averaged CCQE $\nu_\mu (\overline{\nu}_\mu)-^{12}$C differential cross section per nucleon as a function of the muon scattering angle (top panels) and the muon kinetic energy (bottom panels). The integration over the muon kinetic energy has been performed in the range 0.2 GeV $<T_\mu<2.0$ GeV. Panels on the left (right) correspond to neutrinos (antineutrinos). As shown, and in consistency with previous results, the SuSAv2-MEC model is capable of reproducing the magnitude as well as the shape of the experimental cross section in all of the cases. For completeness, we also show in Fig.\ref{total_CS} the total flux-unfolded integrated cross section per nucleon versus the neutrino (left panel) and antineutrino (right) energies. The energy range has been extended to 100 GeV and data are shown for the MiniBooNE and NOMAD experiments. Whereas 2p-2h MEC contributions are needed in order to reproduce MiniBooNE data (in consistency with the discussion applied to previous figures), the NOMAD experiment seems to be in accordance with the pure QE response. As observed, the role of 2p-2h MEC is very significant at all neutrino (antineutrino) energies, getting an almost constant value for $E_\nu$ ($E_{\overline{\nu}}$) greater than $1-2$ GeV. At these values the pure QE cross section is
increased by $\sim 30-35\%$ due to 2p-2h MEC. It is important to point out that, in spite of the very large neutrino (antineutrino) energies involved in NOMAD experiment, the main contribution to the cross section, about $\sim 90\%$, comes from momentum and energy transfers below $\sim 1$ GeV/c and $\sim 0.5$ GeV, respectively. 


\begin{figure}[H]
\begin{center}\vspace{-1.8cm}
\includegraphics[scale=0.3, angle=270]{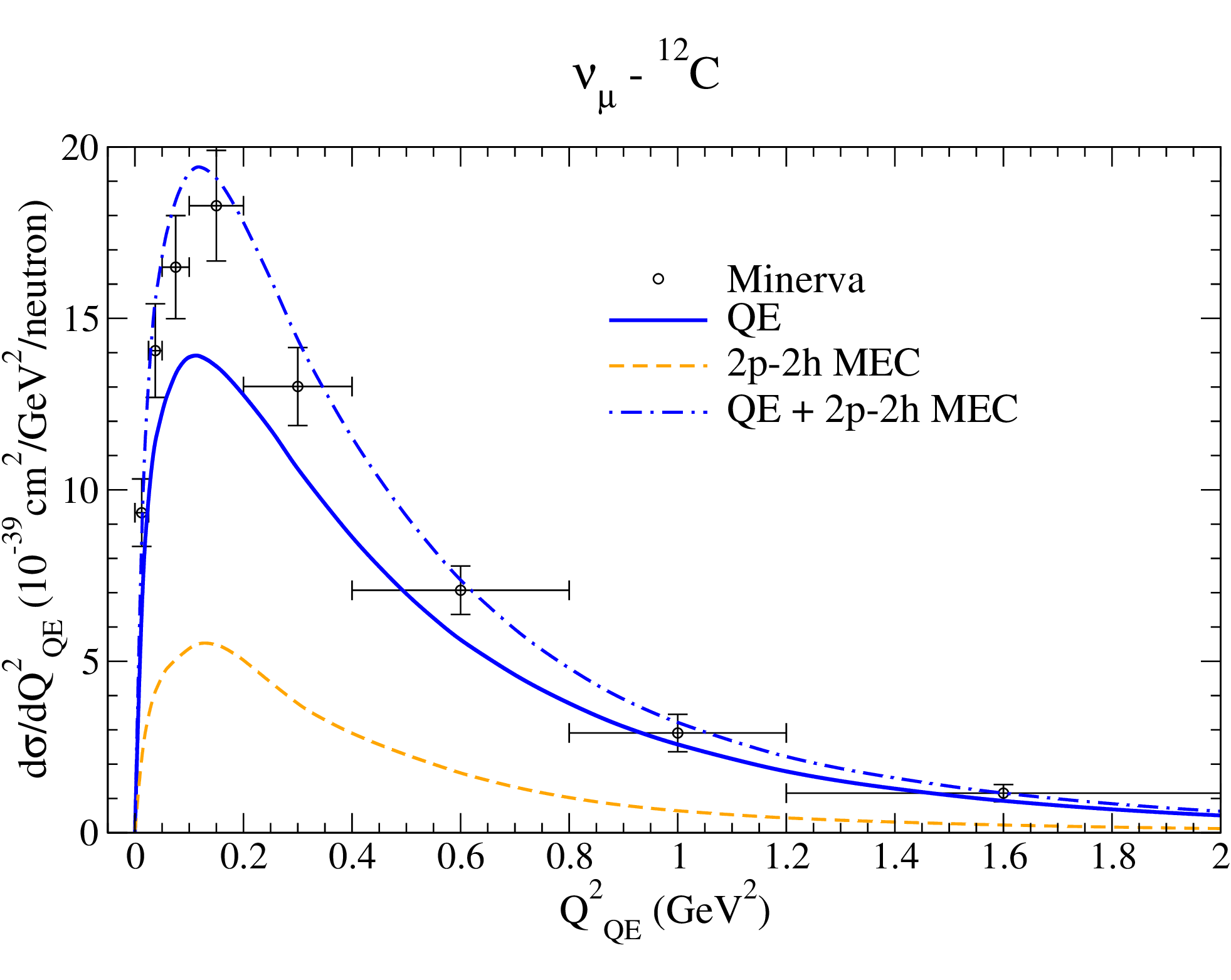}\\
\includegraphics[scale=0.3, angle=270]{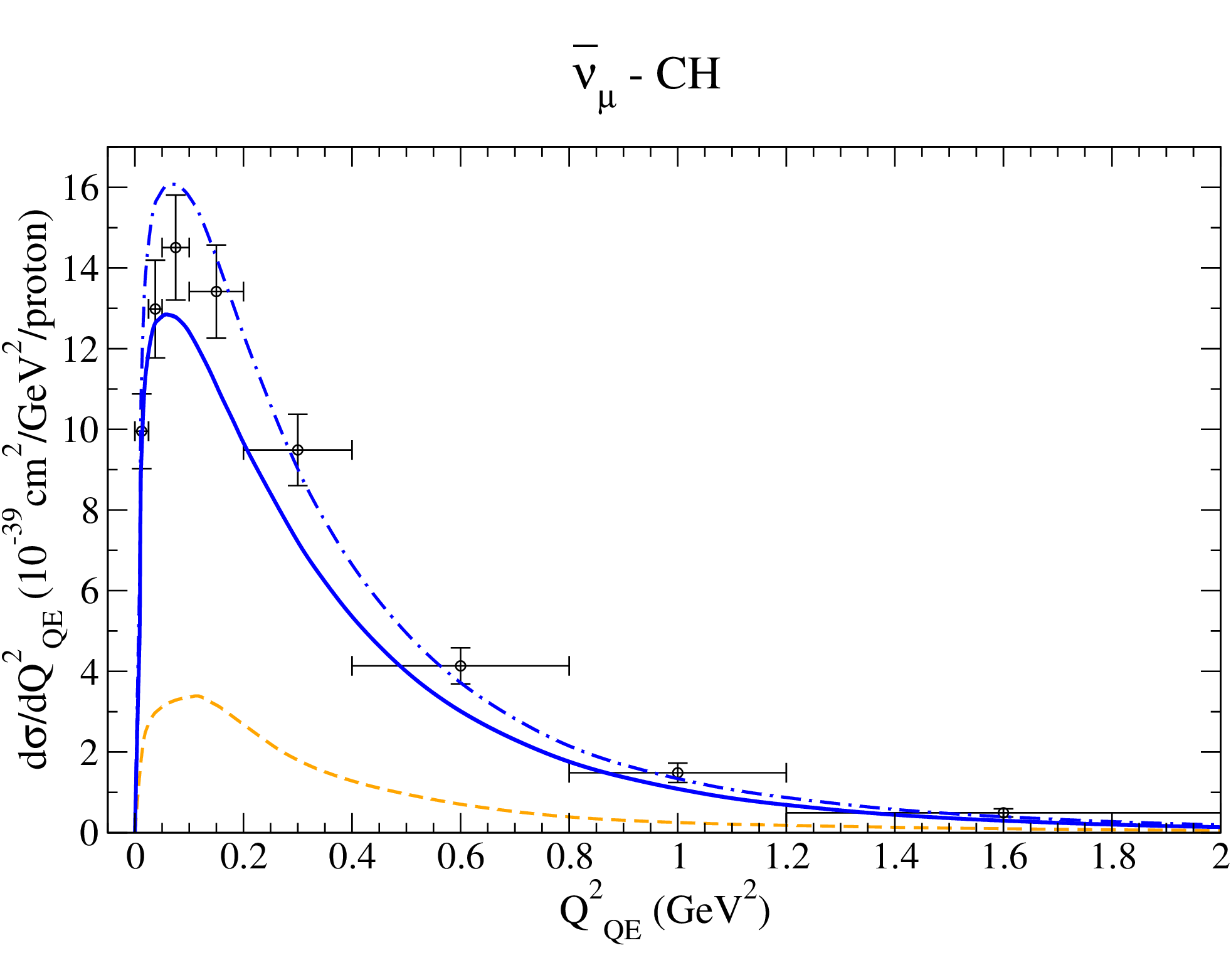}
\begin{center}
\vspace{-1cm}
\end{center}
\end{center}
\caption{(Color online)  Flux-folded $\nu_\mu-^{12}$C CCQE (upper panel) and $\bar\nu_\mu-$CH (lower panel) scattering cross section per
target  nucleon  as  a  function  of $Q^2_{QE}$ and  evaluated  in  the
SuSAv2  and  SuSAv2+MEC  models.    MINER$\nu$A  data  are from~\cite{privatecommunication}.}
\label{Minerva_numu}
\end{figure}



\begin{figure}[H]
\begin{center}\vspace{-1.8cm}
\includegraphics[scale=0.32, angle=270]{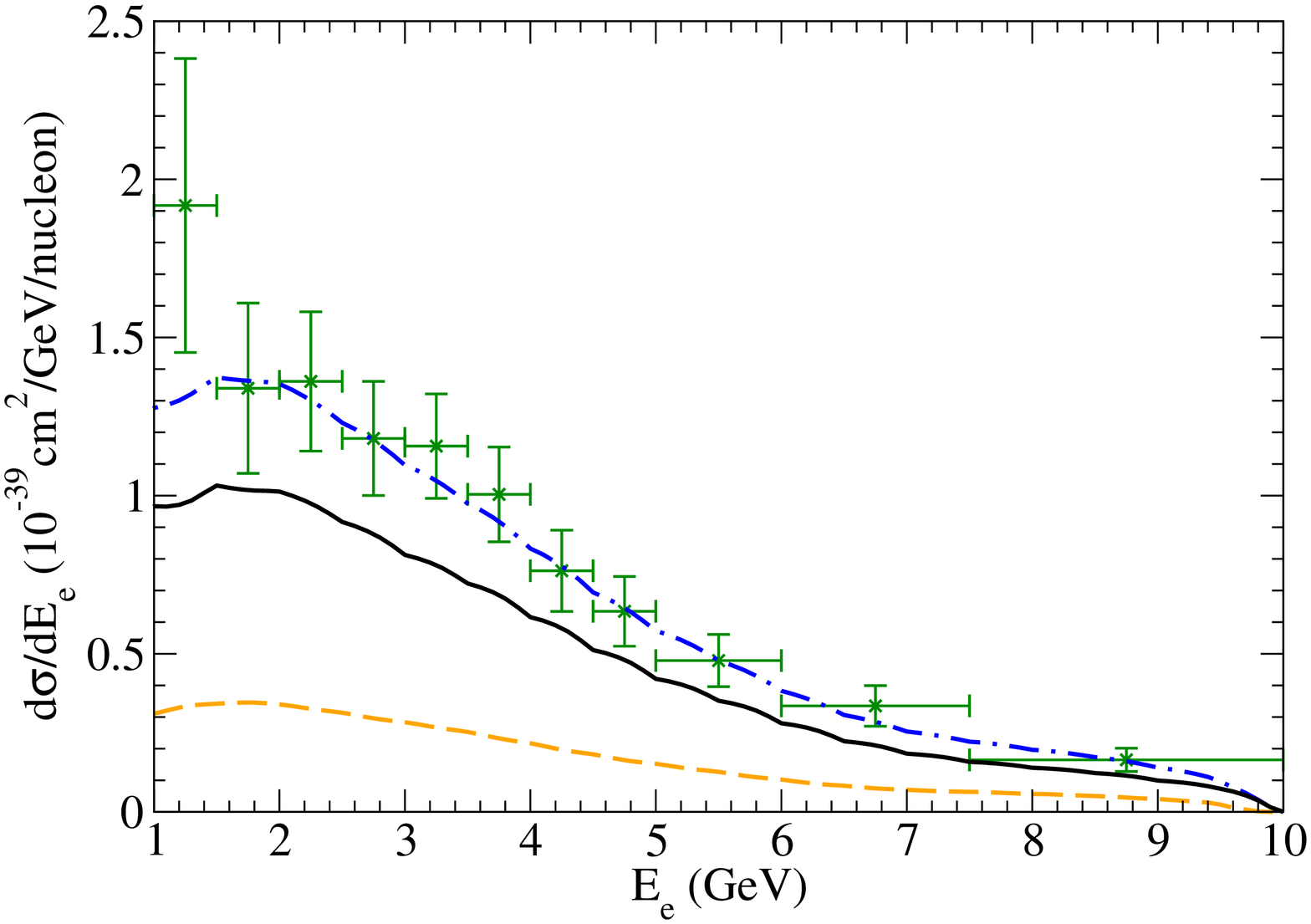}\includegraphics[scale=0.32, angle=270]{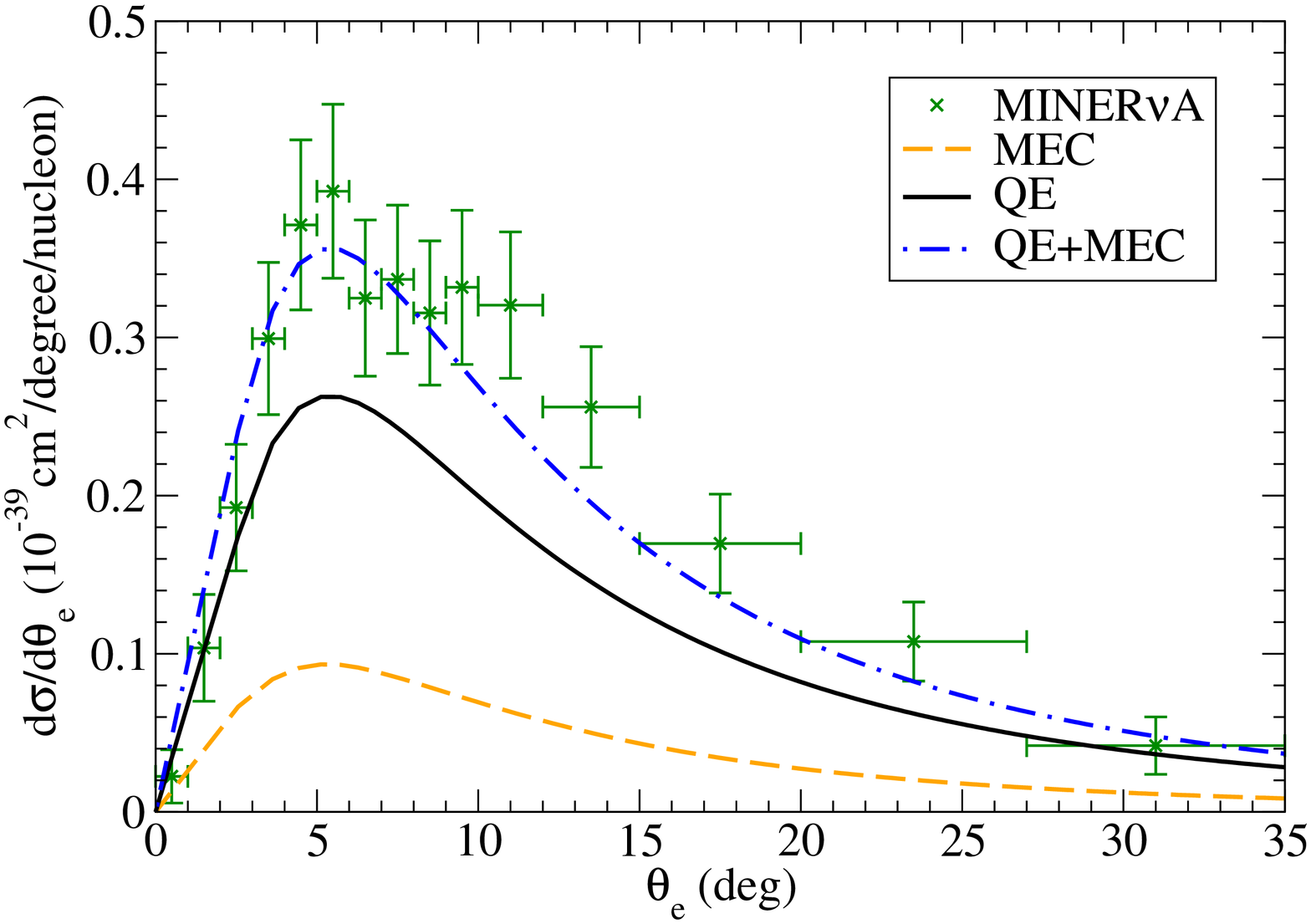}\\
\includegraphics[scale=0.32, angle=270]{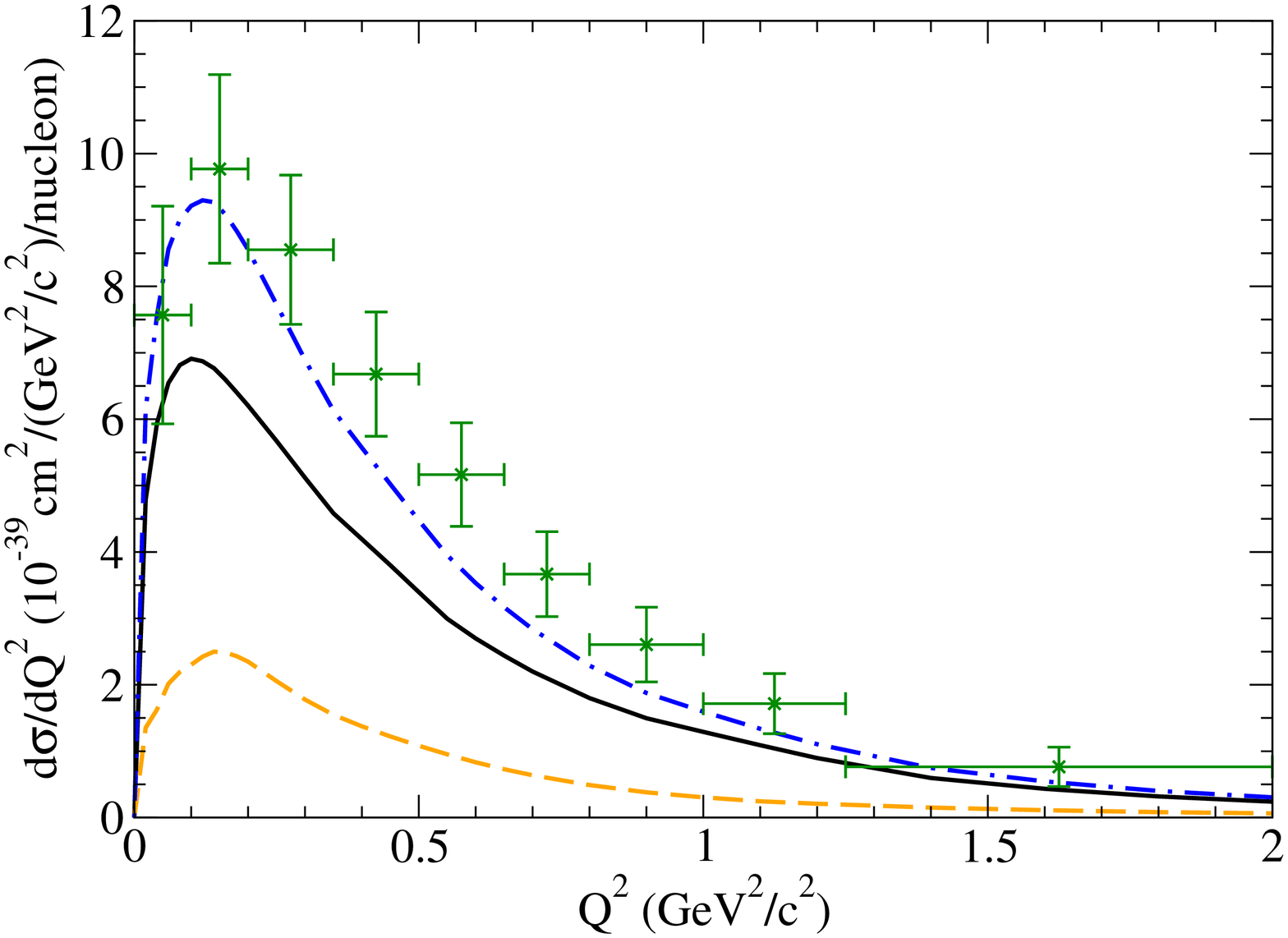}\hspace*{-0.05cm}\includegraphics[scale=0.32, angle=270]{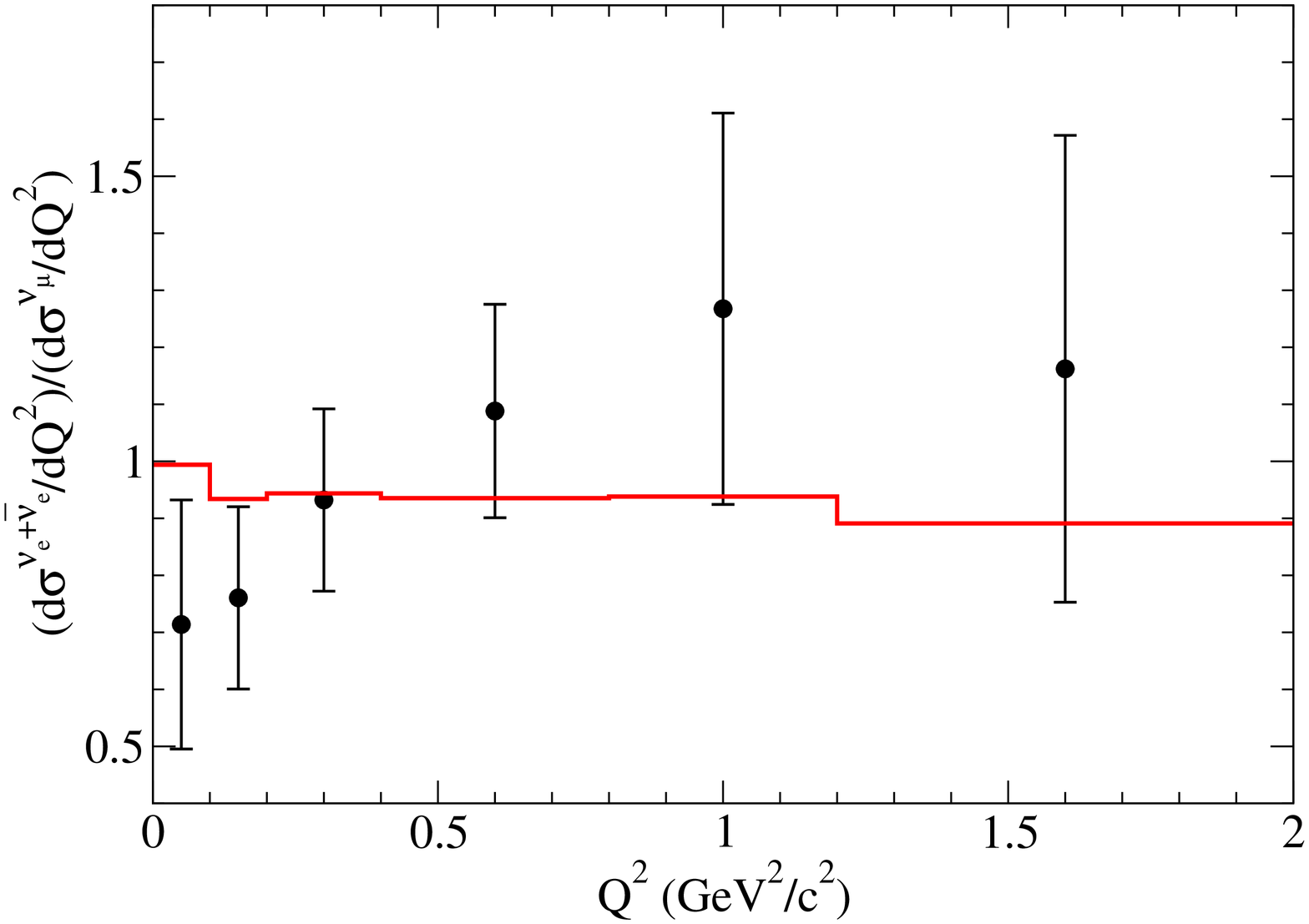}\hspace*{-0.05cm}
\begin{center}
\vspace{-1cm}
\end{center}
\end{center}
\caption{(Color online) MINER$\nu$A flux-integrated differential $\nu_e$ CCQE-like  cross  section  vs.   electron  energy  (top left)  and
electron angle (top right).  The cross section vs. $Q^2_{QE}$ is shown in the bottom panel. Data are from~\cite{Minervanue}.}
\label{Minerva_nue}
\end{figure}


The results in Fig.~\ref{Minerva_numu} correspond to the MINER$\nu$A flux averaged CCQE $\nu_\mu (\overline{\nu}_\mu)$ differential cross section per nucleon as a function of the reconstructed four-momentum $Q^2_{QE}$ (see \cite{privatecommunication} for details). The top panel refers to $\nu_\mu-^{12}$C whereas the bottom panel contains predictions and data for $\overline{\nu}_\mu-$CH. The mean energy of the MINER$\nu$A flux is much higher than the MiniBooNE one, about 3 GeV for both $\nu_\mu$ and $\overline{\nu}_\mu$. As observed, significant contributions of the 2p-2h MEC, of the order of $\sim 35-40\%$ ($\sim 25\%$) at the maxima for $\nu_\mu$ ($\overline{\nu}_\mu$), are needed in order to reproduce the experimental data that correspond to a new analysis performed by the MINER$\nu$A collaboration~\cite{privatecommunication}. These data exceed by $\sim 20\%$ the ones already presented in previous publications~\cite{Minerva1,Minerva2} that, on the other hand, were consistent with calculations based exclusively on the impulse approximation (see \cite{Megias:2014kia}). Thus, the new MINER$\nu$A analysis shows its consistency with the MiniBooNE data. In spite of the very different muon neutrino (antineutrino) energy fluxes in the two experiments, 2p-2h MEC effects remain very significant (on average, $25-35\%$) being their contribution essential in order to fit the data. 

Similar comments apply to the case of electron neutrinos~\cite{Minervanue}. In Fig.\ref{Minerva_nue} we present the MINER$\nu$A flux averaged CCQE $\nu_e$ differential cross section per nucleon as a function of the electron energy (top-left panel), electron angle (top-right) and reconstructed four-momentum (bottom-left). 
Compared to the muon neutrino (antineutrino) fluxes, the $\nu_e$ and $\overline{\nu}_e$ ones
have roughly the same shape in the region of the peak but the tail region is significantly higher in the electronic case.
In all of the situations, results are shown for the pure QE response based on the IA (black line), the 2p-2h MEC contribution (orange dashed line) and the total response (blue dot-dashed). In all the cases the contribution at the maximum coming from the 2p-2h MEC is roughly $30-35\%$ compared with the pure QE response. These results are similar to the ones already presented for muon neutrinos (antineutrinos), and they show the importance of 2p-2h effects in order to explain the behavior of data. As observed, the model is capable of reproducing successfully the data. For completeness, we present in the right-bottom panel the results corresponding to the ratio between the flux averaged CCQE $\nu_e+\overline{\nu}_e$ and $\nu_\mu$ cross sections versus the reconstructed four-momentum. We compare the predictions of the model (red curve) with the data. However, the large error bars presented by the data make this particular analysis rather questionable.


\begin{figure}[H]
\begin{center}\vspace{-1.0cm}
\includegraphics[scale=0.23, angle=270]{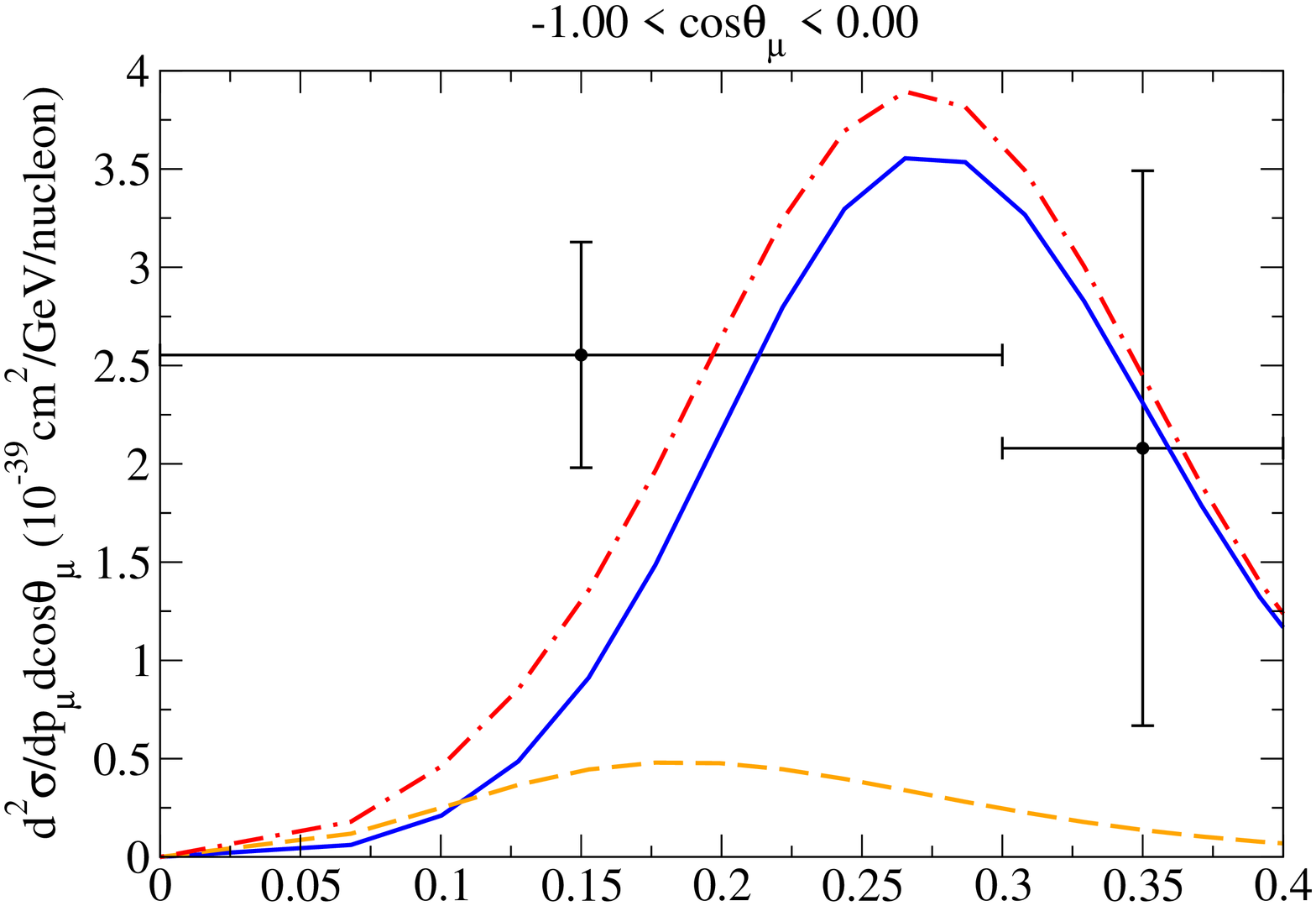}\includegraphics[scale=0.23, angle=270]{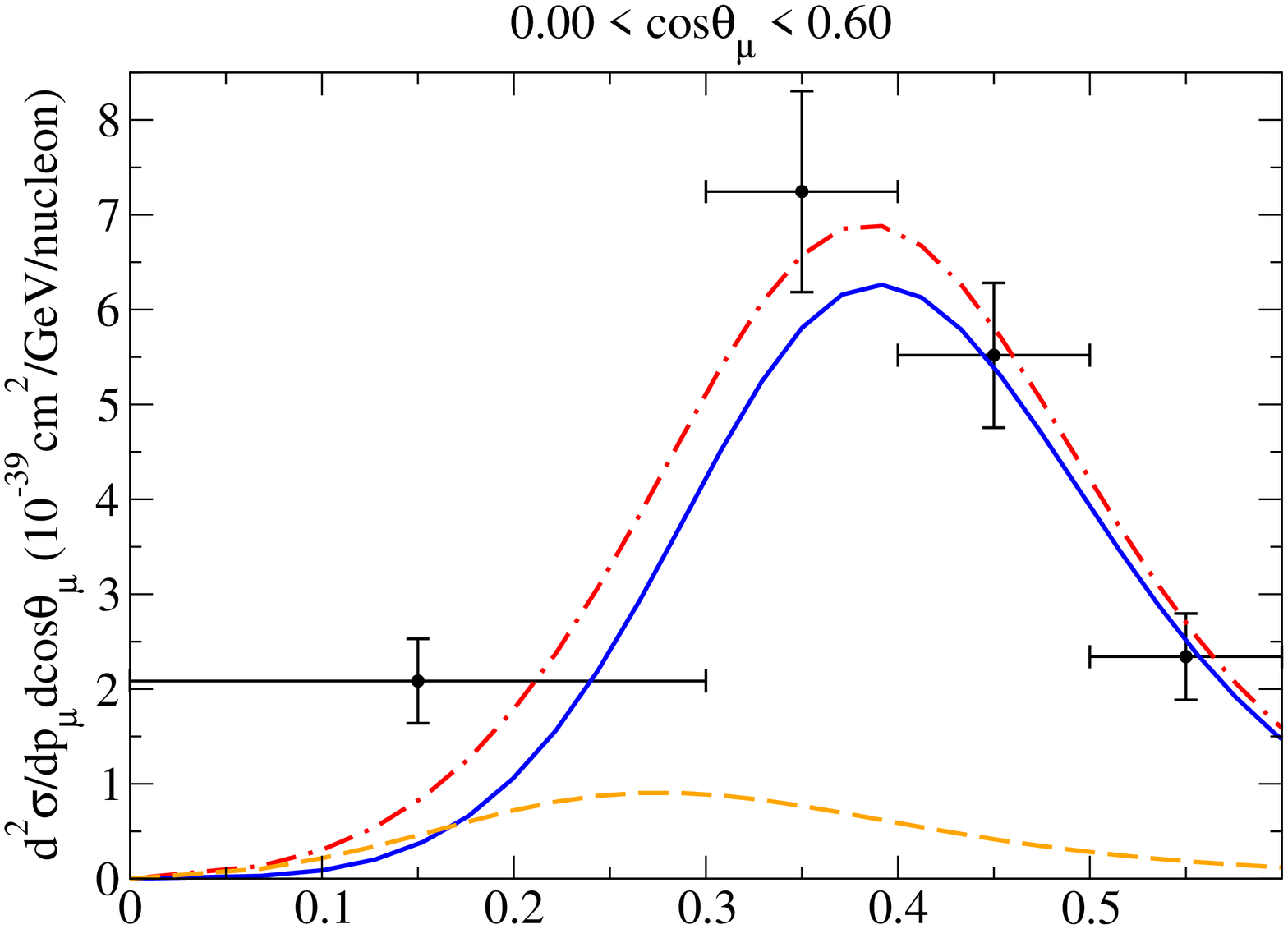}\includegraphics[scale=0.23, angle=270]{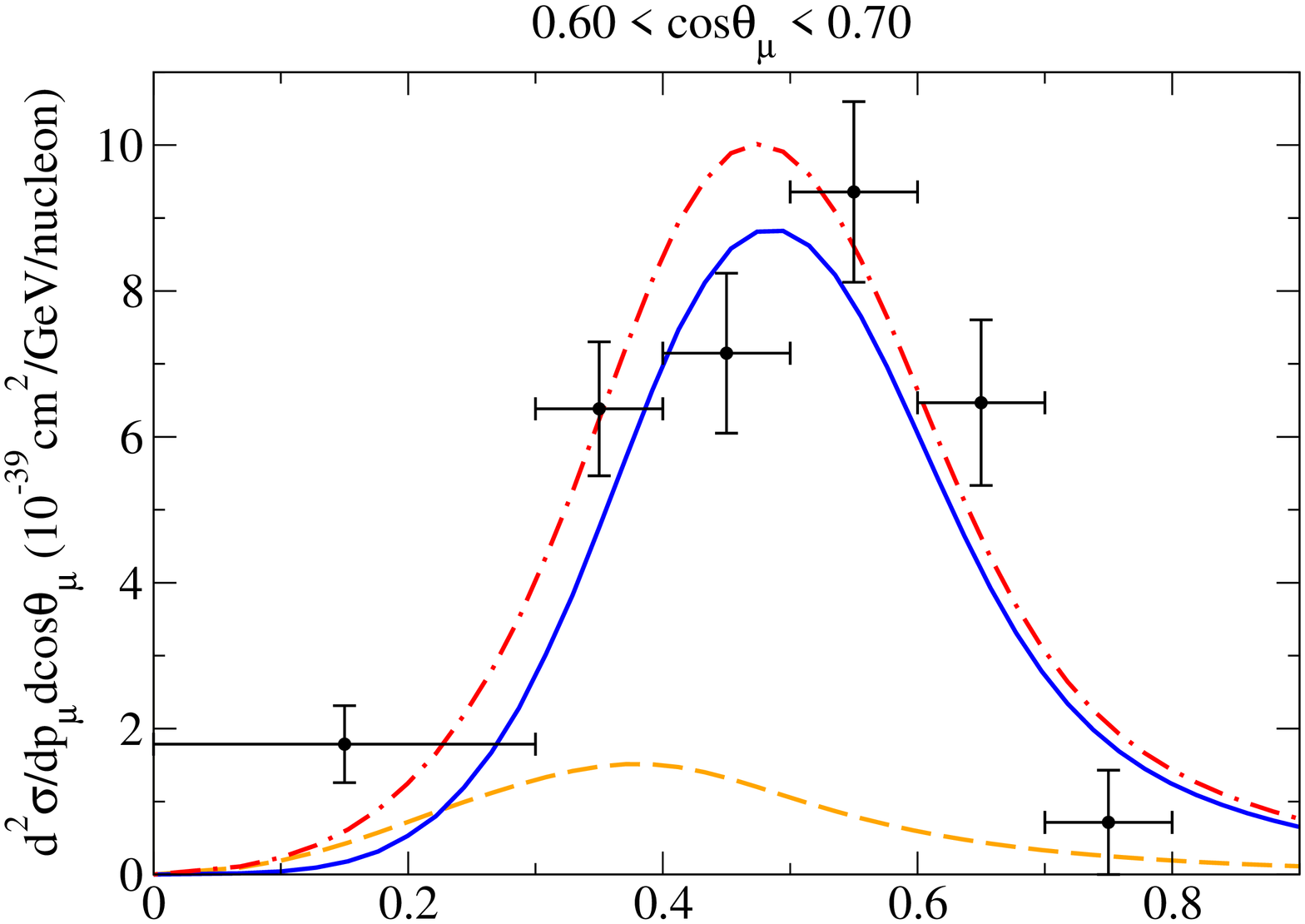}\\\vspace{-0.9cm}
\includegraphics[scale=0.23, angle=270]{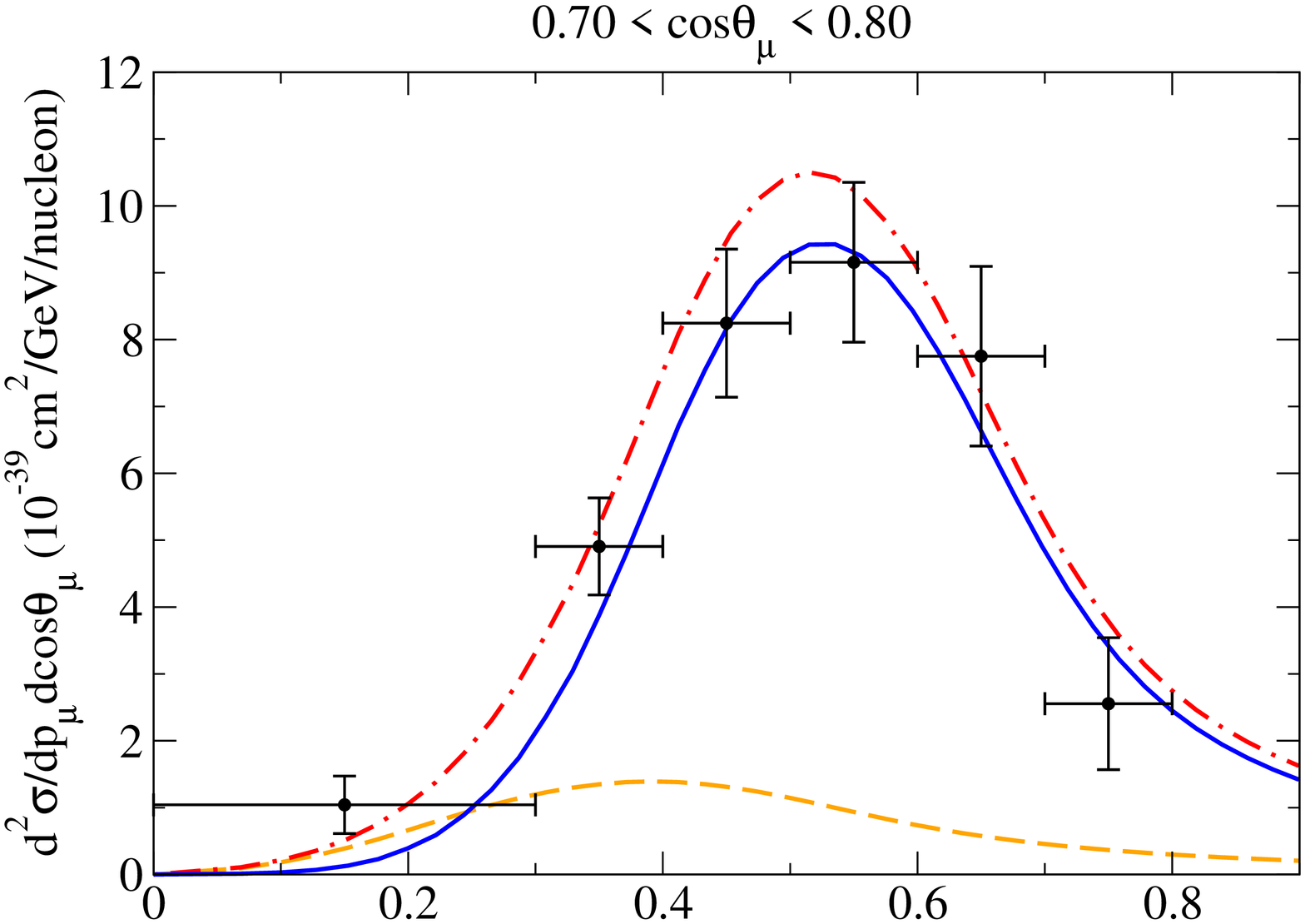}\includegraphics[scale=0.23, angle=270]{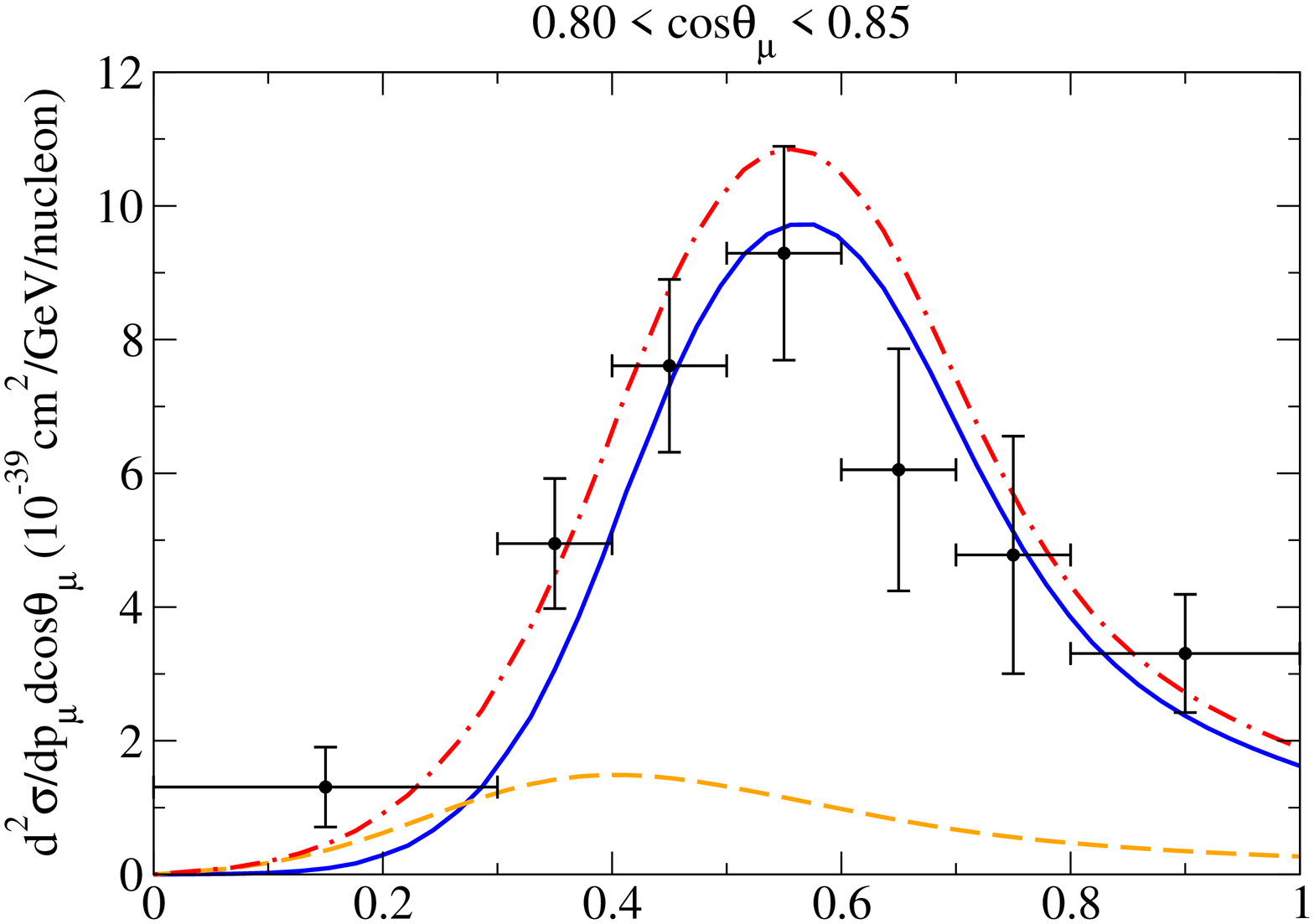}\includegraphics[scale=0.23, angle=270]{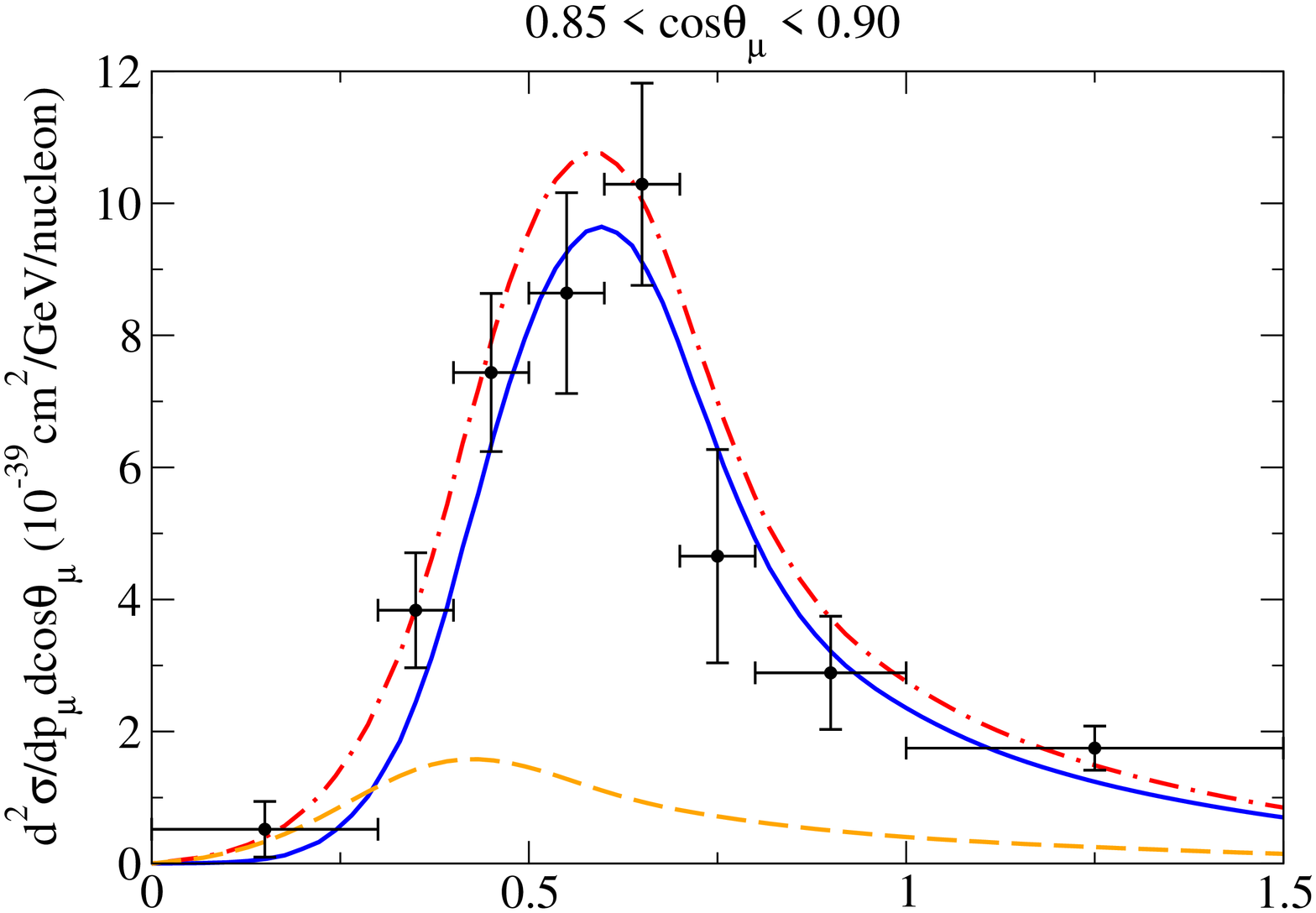}\\\vspace{-0.9cm}
\includegraphics[scale=0.23, angle=270]{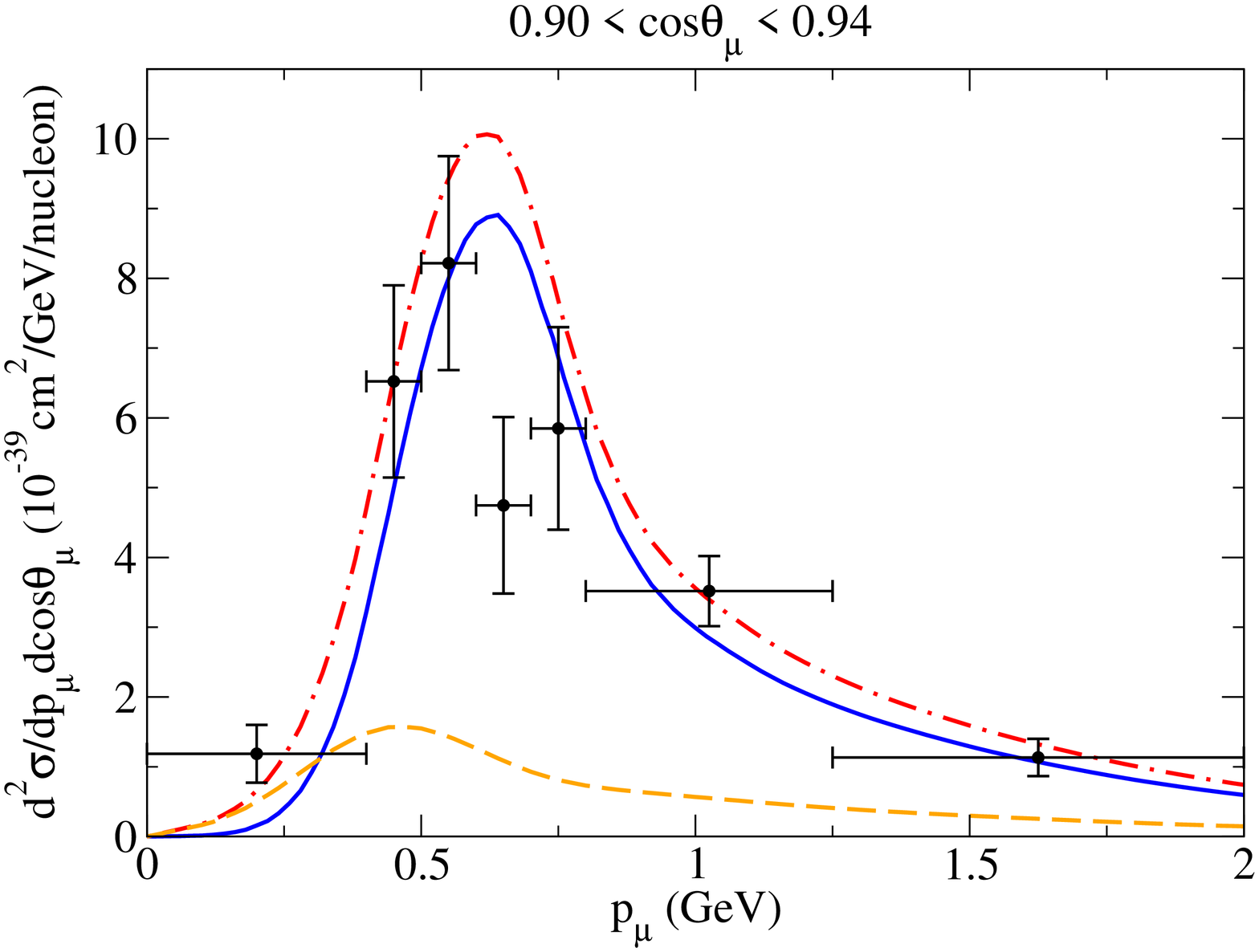}\includegraphics[scale=0.23, angle=270]{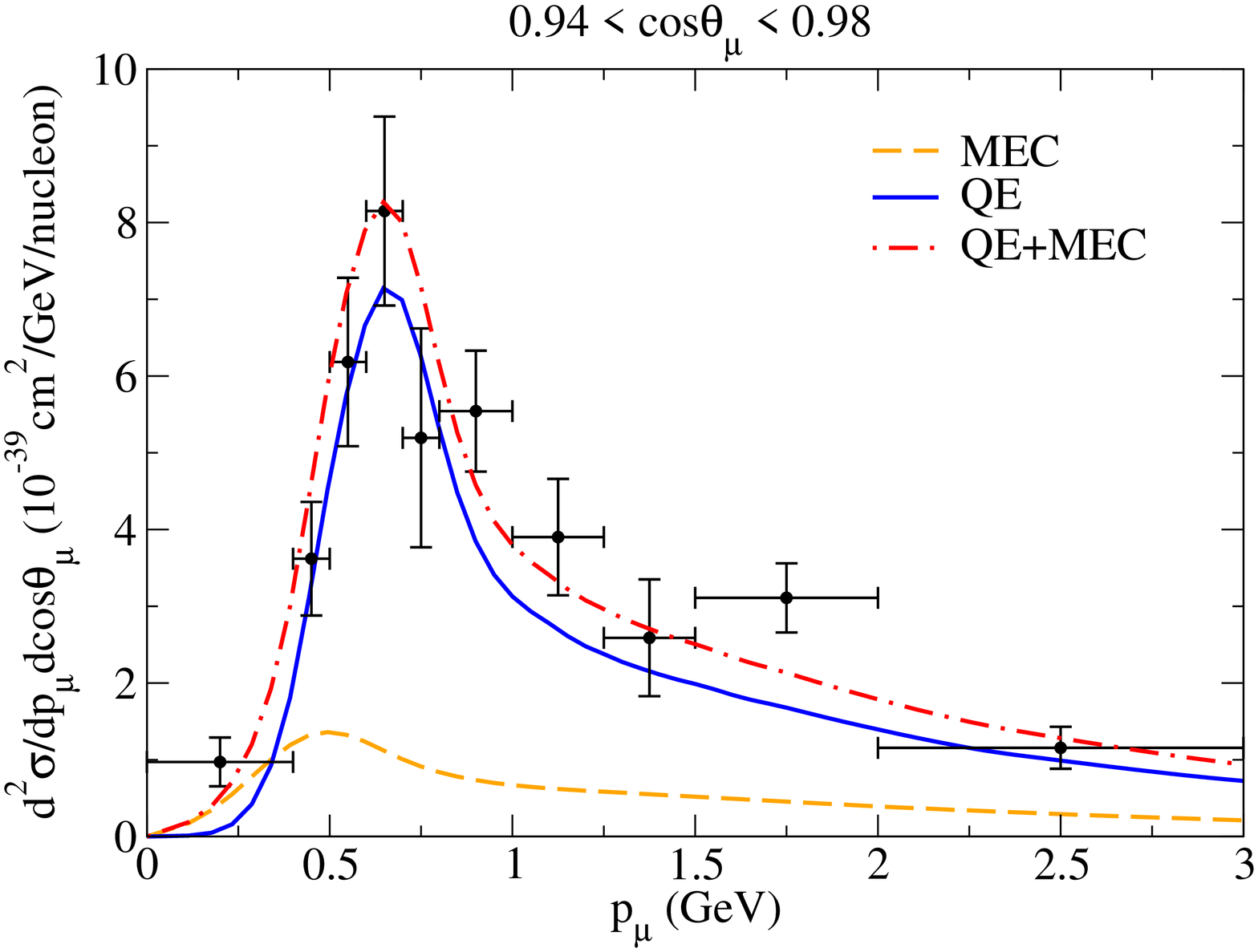}\includegraphics[scale=0.23, angle=270]{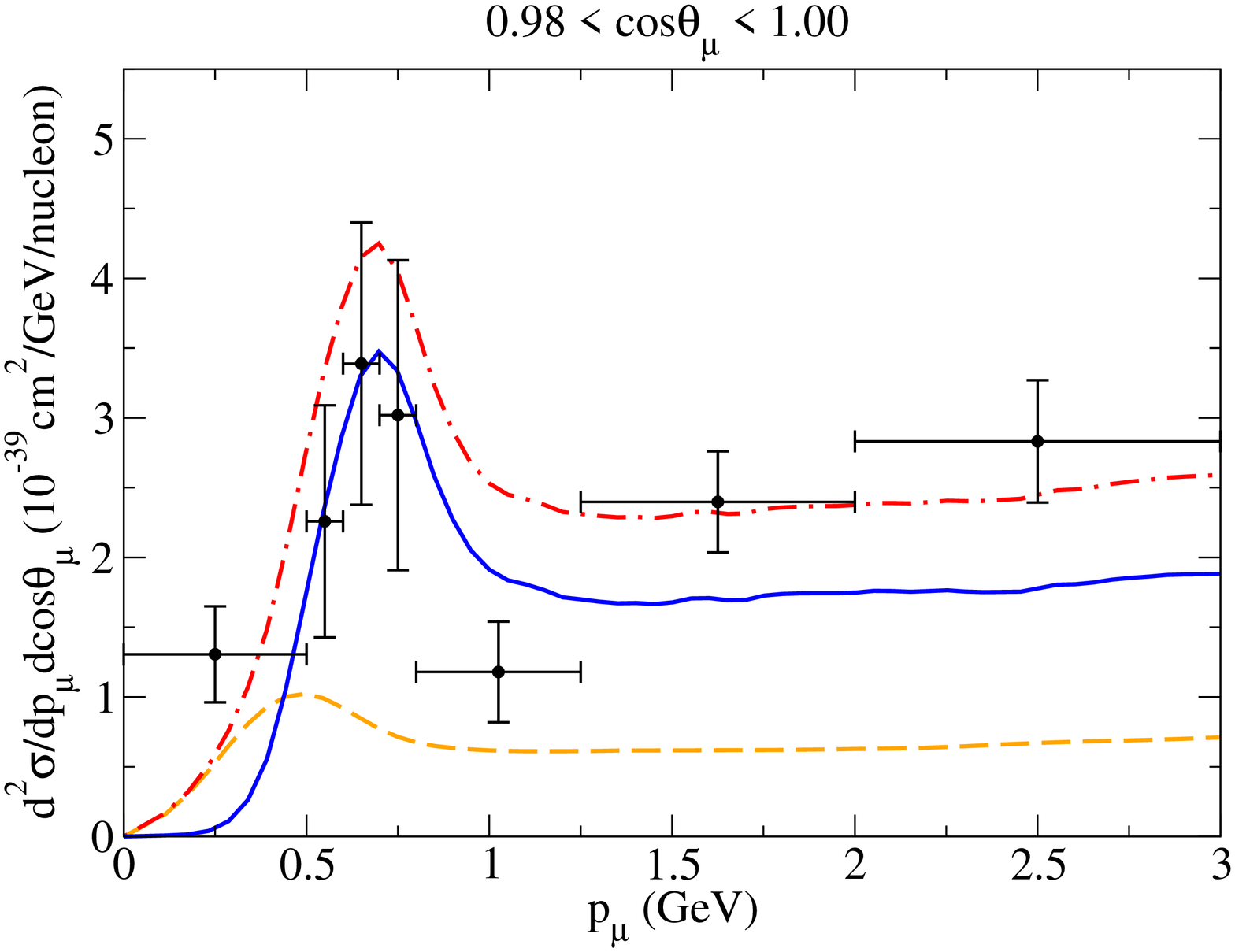}
\begin{center}
\vspace{-1cm}
\end{center}
\end{center}
\caption{(Color online) T2K flux-folded double differential cross section per target nucleon for the $\nu_\mu$ CCQE process on $^{12}$C displayed versus the $\mu^-$ momentum $p_\mu$ for various bins of $\cos\theta_\mu$ obtained within the SuSAv2+MEC approach. QE and 2p-2h MEC results are also shown separately. Data are from~\cite{T2Kcc0pi}.}\label{T2K}
\end{figure}


In Fig.~\ref{T2K} we present the flux-averaged double differential cross sections corresponding to the T2K experiment~\cite{T2Kcc0pi}. The graphs are plotted against the muon momentum, and each panel corresponds to a bin in the scattering angle. As in previous cases, we show the separate contributions of the pure QE, the 2p-2h MEC and the sum of both. Contrary to the MiniBooNE and MINER$\nu$A experiments, the T2K data show a larger dispersion with significant error bands. Concerning the theoretical predictions, in the present case the relative contribution of the 2p-2h MEC compared with the pure QE is significantly smaller than in the previous cases; of the order of $\sim 10\%$ at the maximum of the peak. This can be connected with the T2K neutrino flux that, although with an averaged neutrino flux similar to MiniBooNE, shows a much narrower distribution. Hence 2p-2h MEC contribute less to the differential cross section.

As observed, the theoretical model is capable of reproducing the data although, contrary to the previous experiments, the addition of the 2p-2h MEC does not seem to improve in a clear way the comparison with data. Due to the large error bands and great dispersion shown by T2K data in most of the kinematical situations, both the pure QE as well as the total, QE+2p-2h MEC, predictions are in accordance with the experiment. It is interesting to point out the results for the most forward angles, {\it i.e.,} the panel on the right-bottom corner. Notice that the QE and 2p-2h MEC contributions are stabilized to values different from zero for increasing muon momenta as a consequence of the high energy tail of the T2K neutrino flux. This is at variance with all remaining situations where the cross sections decrease significantly as the muon momentum $p_\mu$ goes up.


\section{Inclusive $\nu$-$^{12}$C cross sections}

The whole analysis presented in the previous section has been restricted to the case of CCQE cross sections, {\it i.e.,} only considering the contributions coming from the pure QE peak and the 2p-2h MEC effects. Here we extend our study by including the inelastic contributions. We restrict our discussion to the effects associated with the $\Delta$ resonance. The analysis of higher inelasticities is still in progress and it will be presented in a forthcoming publication. The addition of inelastic channels is essential in order to explain inclusive charged-current neutrino cross sections. This is the case of recent data taken by the T2K collaboration~\cite{T2Kincl,T2Kinclelectron}, both for muon and electron neutrinos, as well as the SciBooNE experiment~\cite{SciBooNEincl}.


\begin{figure}[H]
\begin{center}\vspace{-1.8cm}
\includegraphics[scale=0.3, angle=270]{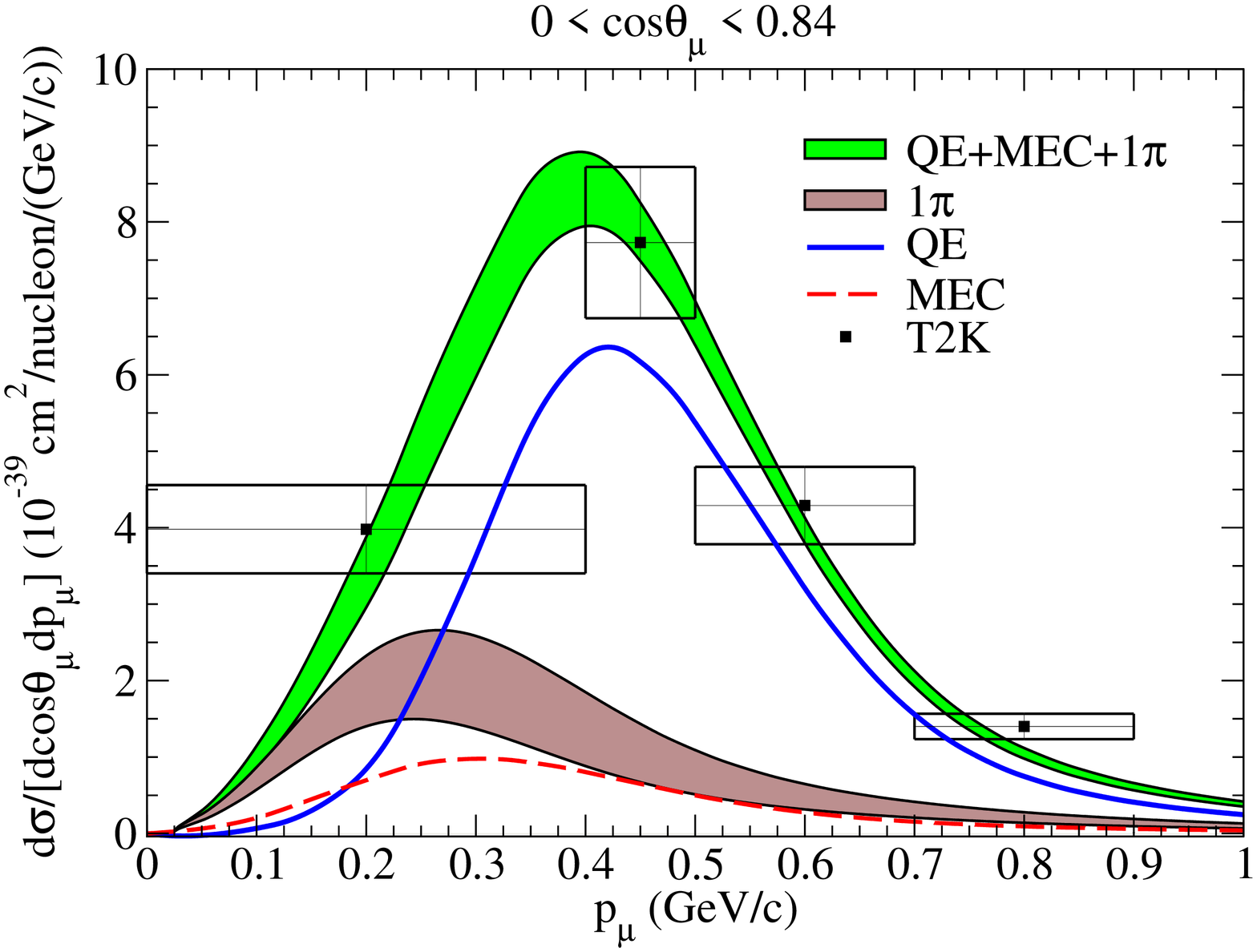}\includegraphics[scale=0.3, angle=270]{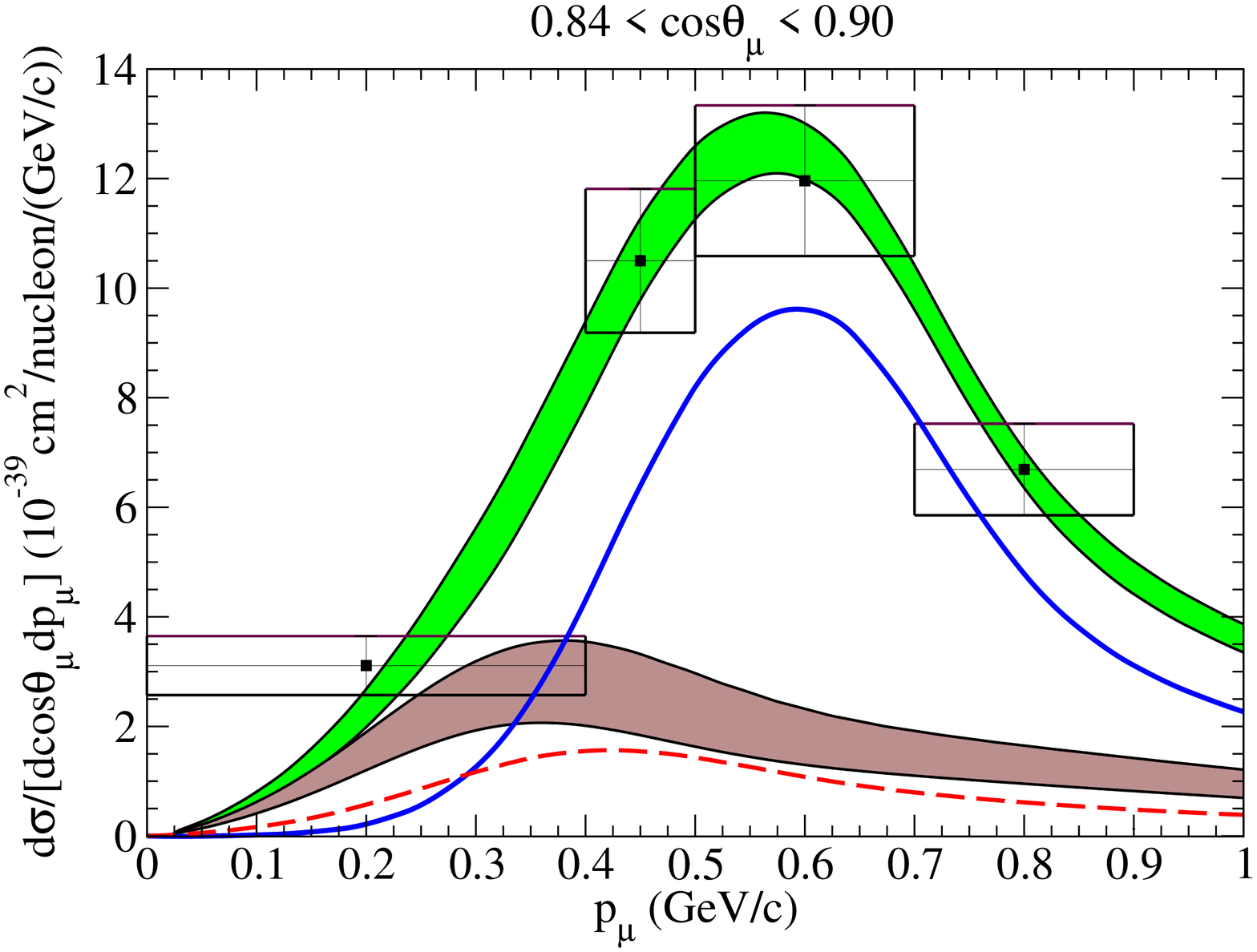}\\
\includegraphics[scale=0.3, angle=270]{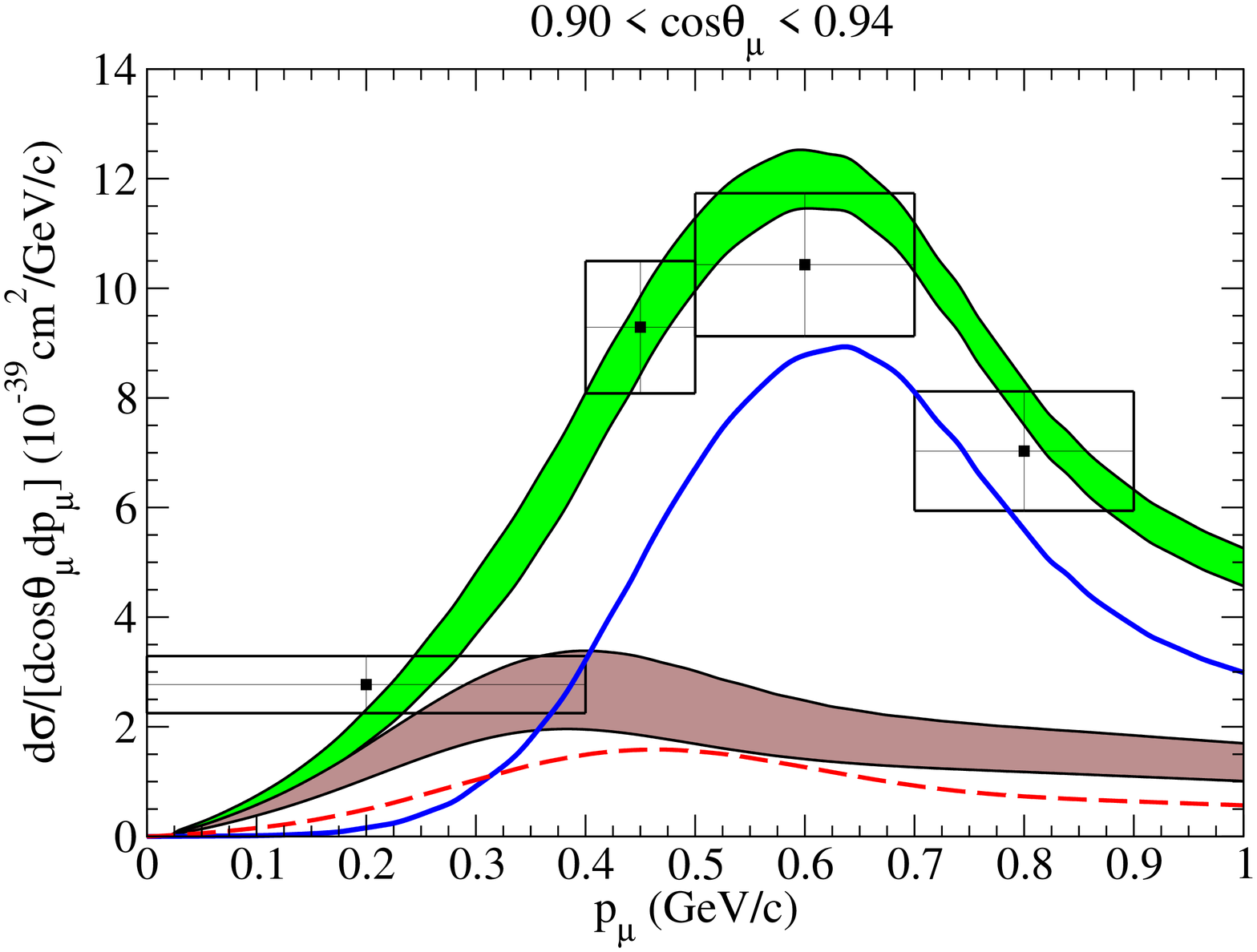}\includegraphics[scale=0.3, angle=270]{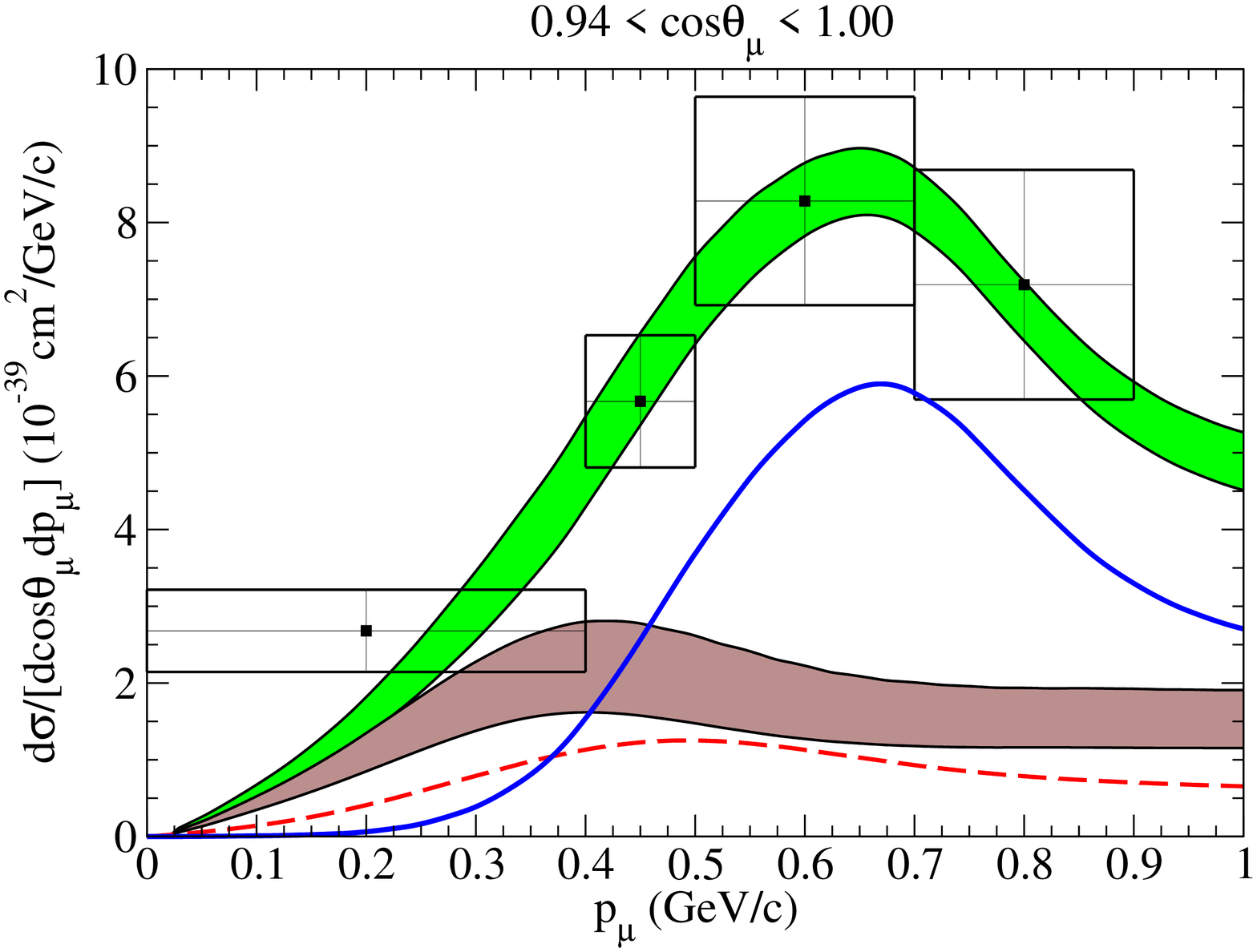}
\begin{center}
\vspace{-1cm}
\end{center}
\end{center}
\caption{(Color online) The CC-inclusive T2K flux-folded $\nu_\mu-^{12}$C double-differential cross section per nucleon evaluated in
the SuSAv2+MEC model is displayed as a function of the muon momentum for different bins in the muon angle.  The separate
contributions of the QE, 1$\pi$
and 2p-2h MEC are displayed. The data are from~\cite{T2Kincl}}\label{inclusive_T2K_mu}
\end{figure}

Fig.\ref{inclusive_T2K_mu} contains the data and theoretical predictions corresponding to the T2K flux-averaged inclusive double differential cross sections for muon neutrinos. Results are shown as function of the muon momentum and averaged over particular muon angular bins (each panel). The separate contribution of the QE (solid blue line), 2p-2h MEC (dashed red) and the $\Delta$ resonance (brown band) are presented. The global response is shown by the green band. The band in the pion contribution takes care of the uncertainty associated with the description of the $\Delta$ scaling function as discussed in detail in \cite{Ivanov:2016Delta}. As observed, 
the model provides a very nice description of data once all contributions are included, {\it i.e.,} QE, 2p-2h MEC and pion. This is consistent with the kinematics implied by the present T2K experiment being the $\Delta$ resonance the main response (almost the only one) within the inelastic region. This was already discussed in detail in \cite{Ivanov:2016Delta} where a similar figure 
was presented, although based on the original SuSA model and with incomplete 2p-2h MEC calculations.
The main difference between the two calculations is the inclusion, in the new results, of the axial 2p-2h contribution. Whereas in \cite{Ivanov:2016Delta} the purely vector MEC were found to be negligible at these kinematics, in Fig. \ref{inclusive_T2K_mu} it is shown that the axial two-body currents give a contribution almost as large as the one associated with the $\Delta$ resonance. The experimental error bars are too large to allow one to discriminate between the two results and both calculations are compatible with the data.


\begin{figure}[H]
\begin{center}\vspace{-1.8cm}
\includegraphics[scale=0.362, angle=270]{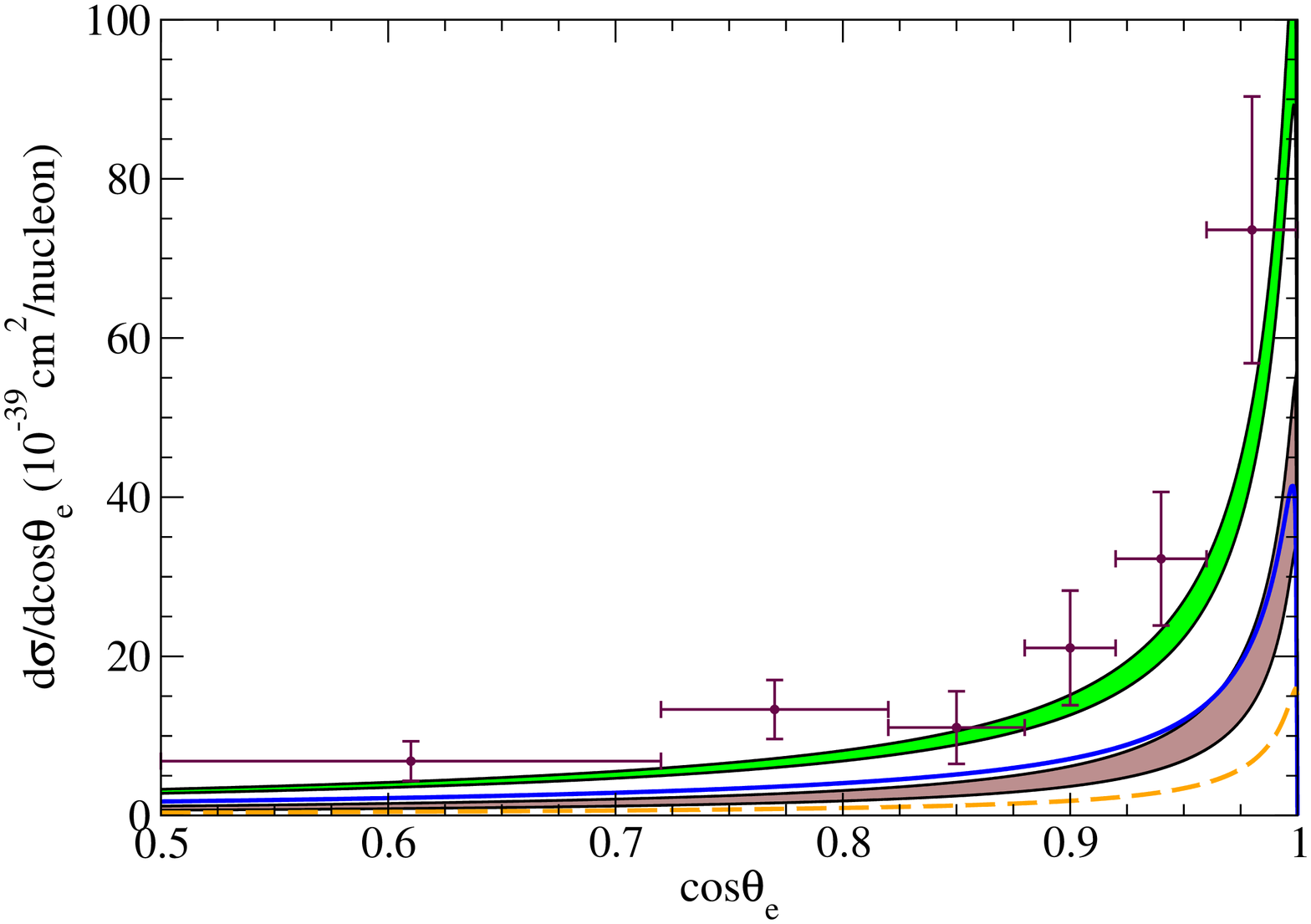}\\
\includegraphics[scale=0.362, angle=270]{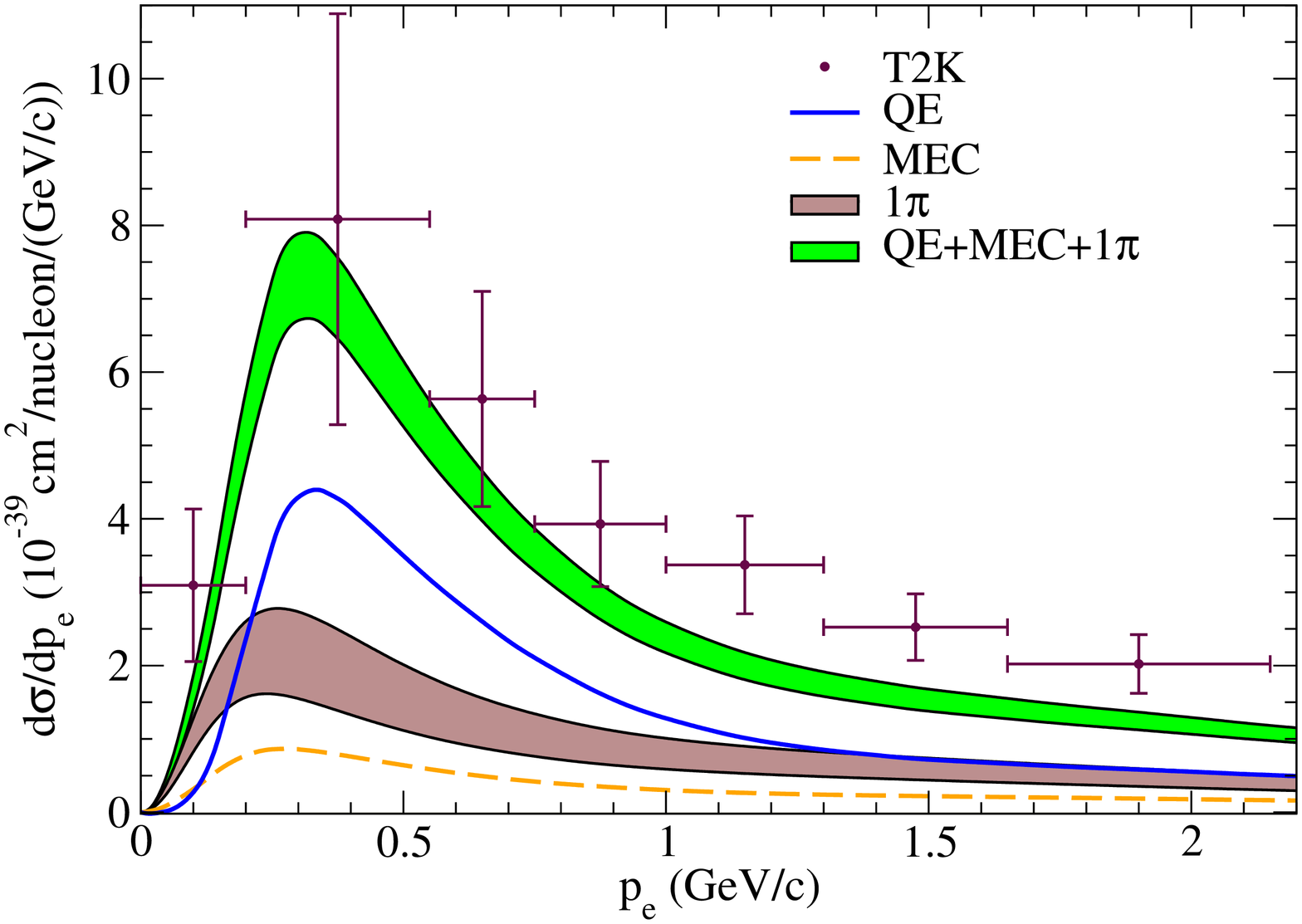}\hspace*{-0.05cm}\\
\includegraphics[scale=0.362, angle=270]{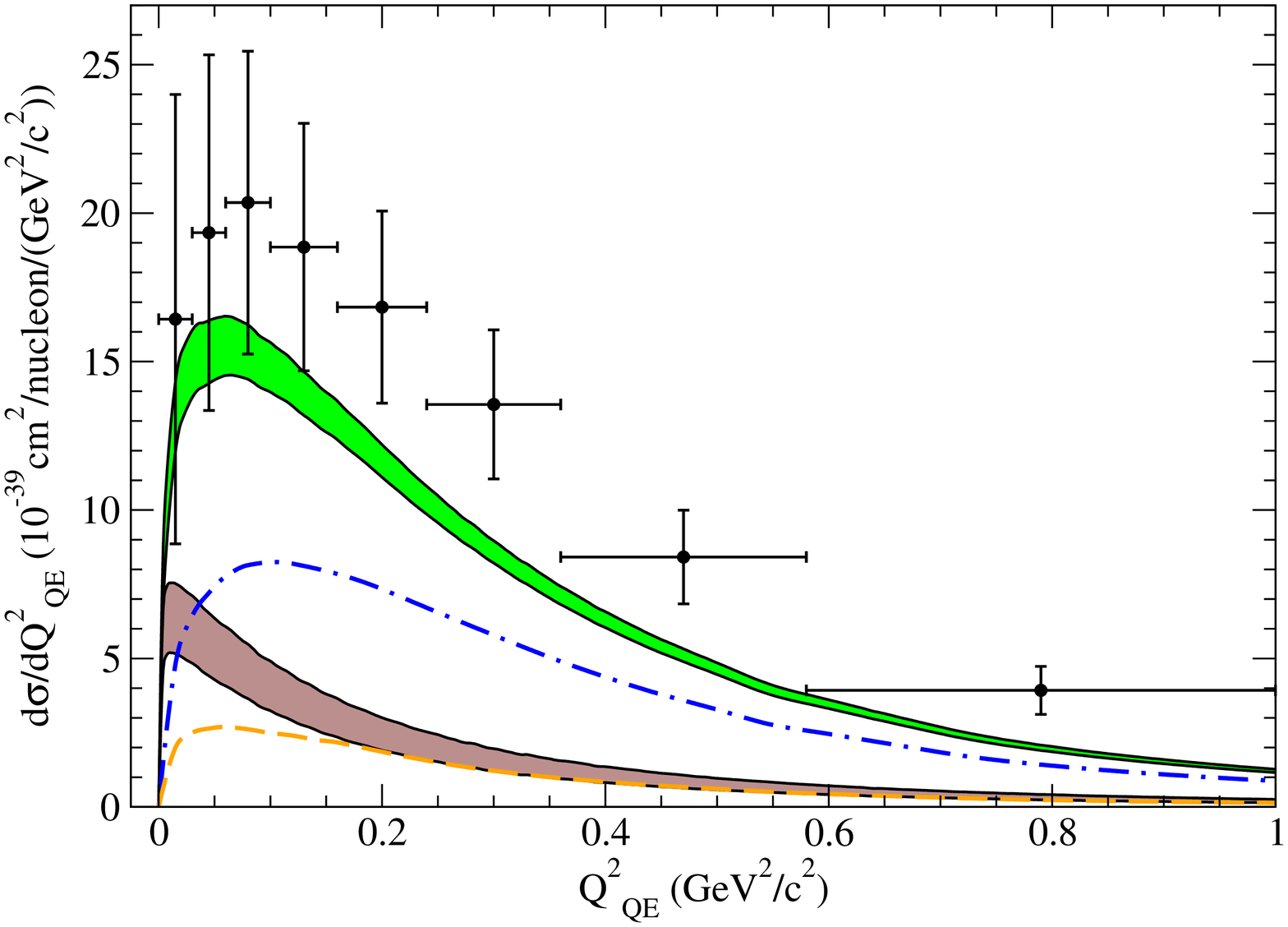}
\begin{center}
\vspace{-1cm}
\end{center}
\end{center}
\caption{(Color online) The CC-inclusive T2K flux-folded $\nu_e-^{12}$C differential cross section per nucleon evaluated in the SuSAv2+MEC model is displayed as a function of the electron momentum (top), $\cos\theta_e$ (middle) and $Q^2_{QE}$ (bottom). The separate contributions of the QE, 1$\pi$
and 2p-2h MEC are displayed. The data are from~\cite{T2Kinclelectron}.}\label{inclusive_T2K_e}
\end{figure}

The inclusive T2K experiment for electron neutrinos is analyzed in Fig.\ref{inclusive_T2K_e} where the flux-averaged single differential cross sections are shown. Results are presented against the electron scattering angle (top panel), the electron momentum (middle) and the reconstructed four-momentum (bottom). In the three cases we show the separate contributions corresponding to the QE response (blue line), the 2p-2h MEC (red dashed), pionic (brown band) and the total response (green region). Although, as noted, the role associated with the $\Delta$ resonance is essential, the data are located above the model predictions. This implies that other higher nucleon resonances, not taken into account in the present description, may also have a significant role in explaining T2K $\nu_e$ data. This is particularly true for increasing values of the electron momentum (see results in the middle panel) and/or the reconstructed four-momentum transfer (bottom panel). Work along this line is presently in progress.



\begin{figure}[H]
\begin{center}\vspace{-1.8cm}
\includegraphics[scale=0.3, angle=270]{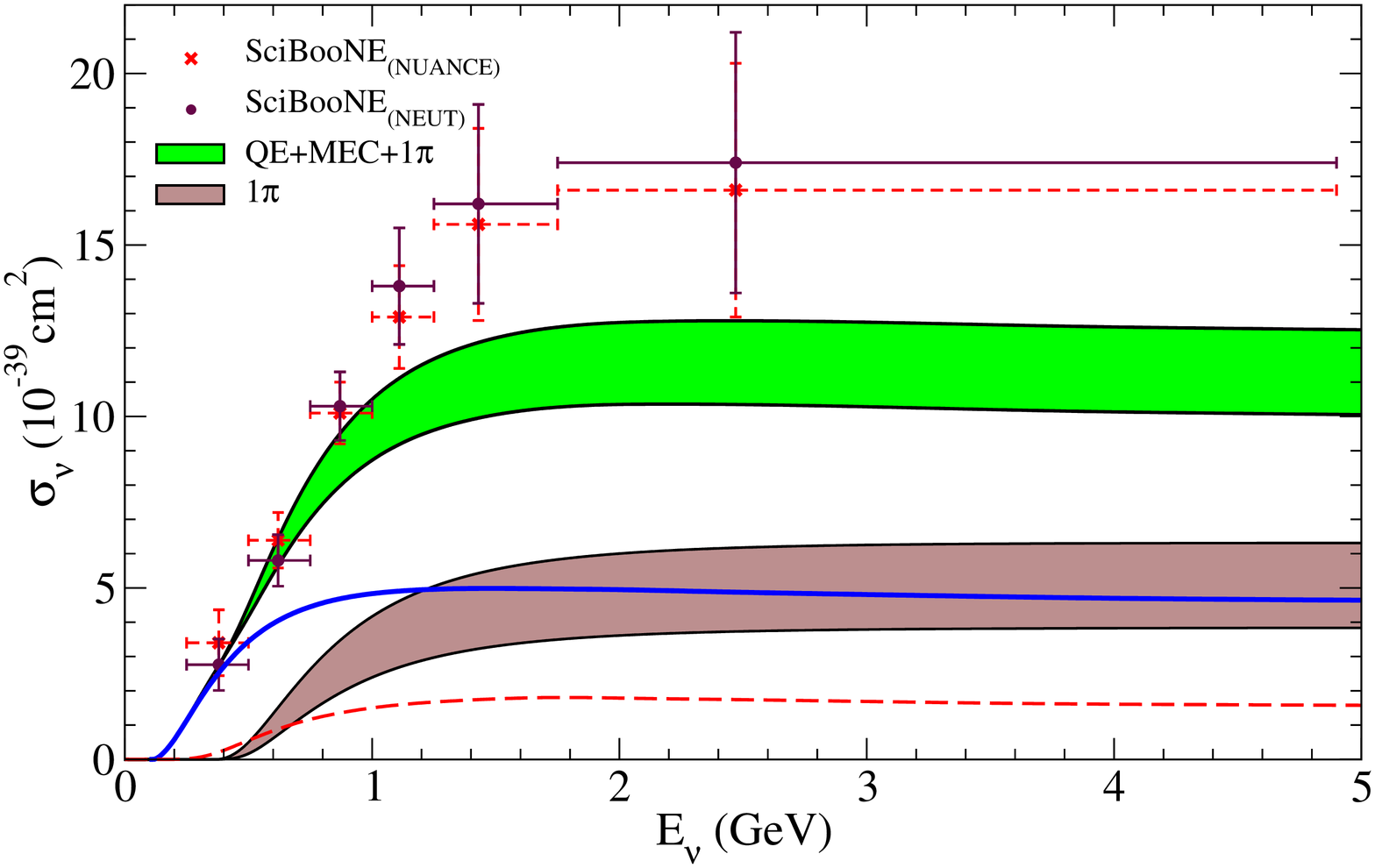}\\
\includegraphics[scale=0.3, angle=270]{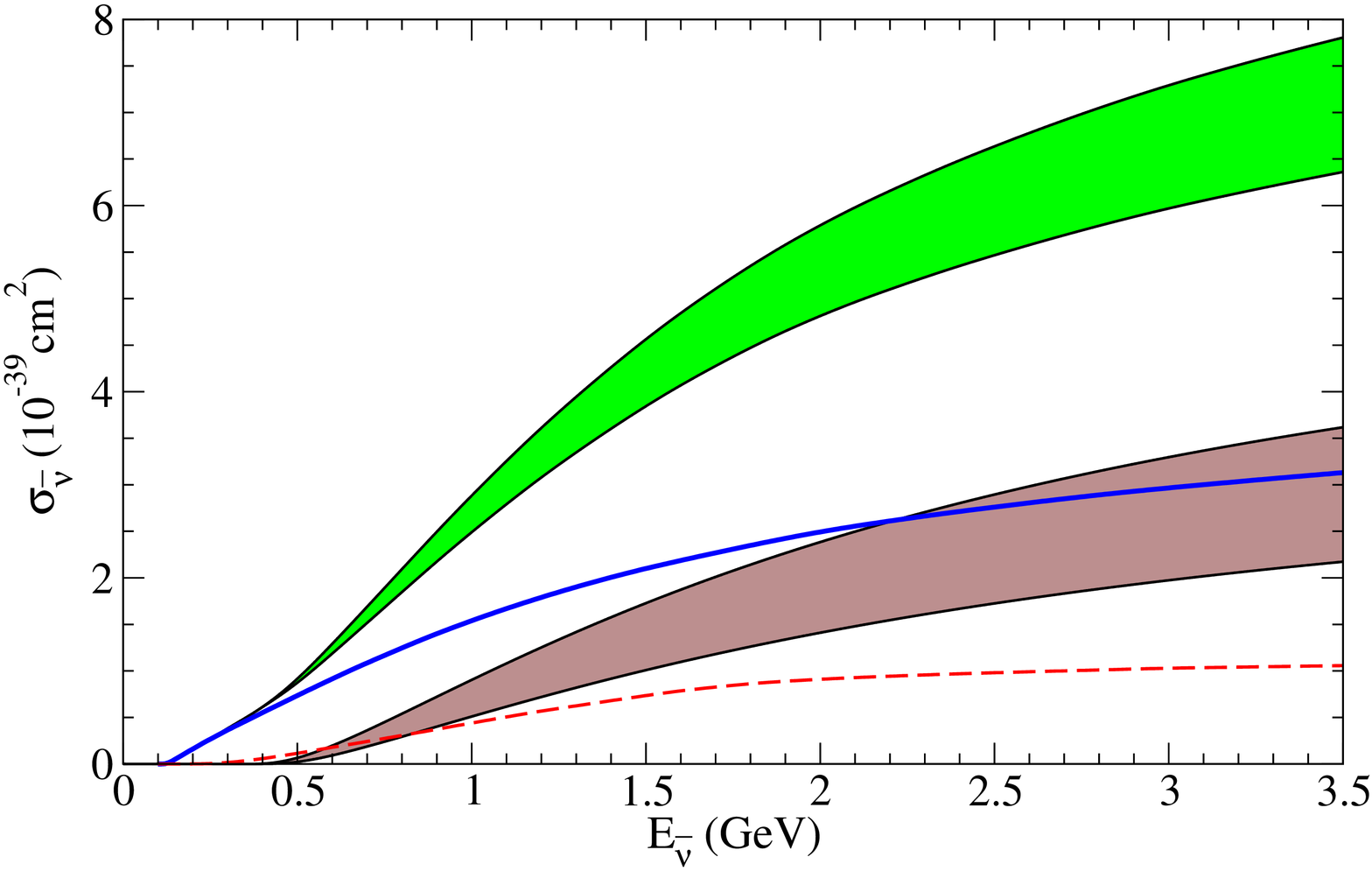}
\begin{center}
\vspace{-1cm}
\end{center}
\end{center}
\caption{(Color online) The CC-inclusive $\nu_\mu$  (left) and $\bar\nu_\mu$ (right) cross section on a polystyrene target (C$_8$H$_8$) per nucleon evaluated in the SuSAv2+MEC
model as a function of the neutrino energy. The SciBooNE
data are from~\cite{SciBooNEincl}.}\label{SciBoone}
\end{figure}

To conclude, we present in Fig.~\ref{SciBoone} the results for the
SciBooNE experiment~\cite{SciBooNEincl}. These correspond to CC
$\nu_\mu$ ($\overline{\nu}_\mu$) scattering on a polystyrene
target. The data are presented as a total unfolded integrated cross
section as a function of the neutrino energy. Because of the unfolding
procedure to reconstruct the neutrino energy, one should be very
cautious in the comparison between data and theoretical predictions
(see discussion in \cite{SciBooNEincl}). The case of neutrinos is presented in the left
panel of Fig.\ref{SciBoone} in comparison with available data, whereas
the predicted cross section for antineutrinos is referred to the right
panel. Results in Fig.\ref{SciBoone} are similar to the ones shown in
Fig. 7 of \cite{Ivanov:2016Delta}, although there the 2p-2h MEC
calculation only included the pure vector contributions.  In fact, one
can observe how the complete 2p-2h MEC calculation, now including also
the axial currents, leads to a much more significant contribution,
bringing the global predictions closer to the data. Contrary to the
analysis in \cite{Ivanov:2016Delta}, here the model reproduces the
neutrino data up to 1 GeV. However, for higher energies the model
still underpredicts the data by a significant amount. This result
clearly indicates that new channels and higher nucleon resonances, in
addition to the resonant pion production, should be added to the
model. Finally, for completeness, we also show the results obtained
for electron antineutrinos with the separate contributions of the
different channels. Notice that the role ascribed to the 2p-2h MEC
effects is of the order of $\sim 15\%$ ($\sim 20\%$) for neutrinos
(antineutrinos), approximately twice compared with the values
discussed in \cite{Ivanov:2016Delta}.

\section{Conclusions}

We have performed a joint calculation of quasielastic and 2p-2h
contribution to neutrino and antineutrino scattering cross sections in
$^{12}$C, using the SuSAv2 model for the quasielastic responses and
the relativistic Fermi gas model for the 2p-2h meson exchange currents
in the weak sector. The model has been validated in the vector sector
by describing the full set of inclusive electron scattering $^{12}$C
data.  We have analyzed the published data from the
experiments MiniBooNE, T2K, MINERvA, NOMAD and SciBooNE, spanning a
wide range of neutrino energies from hundreds of MeV to hundreds of GeV.
For comparison with inclusive data we have used an extension of the 
SuSAv2 model to the $\Delta$ production region to model resonant pion 
production.
We find that the 2p-2h channel is large, contributing about
15--25\% depending on the kinematics, and it is essential to describe
a great amount of experimental data. 

This model is a promising candidate for analyzing the forthcoming neutrino
experiments; work is in progress to extend it to higher inelasticities, to provide the separate
charge channel contributions, pn, pp and nn emission \cite{Simonupn}, and
to describe the cross section of asymmetric nuclei ($Z\ne N$).

\section*{Acknowledgments}
This work was supported by Spanish Direccion General de Investigacion
Cientifica y Tecnica and FEDER funds (grants No. FIS2014-59386-P and
No. FIS2014-53448-C2-1), by the Agencia de Innovacion y Desarrollo de
Andalucia (grants No. FQM225, FQM160), by INFN under project MANYBODY,
and part (TWD) by U.S. Department of Energy under cooperative
agreement DE-FC02-94ER40818. GDM acknowledges support from a Junta de Andalucia fellowship 
(FQM7632, Proyectos de Excelencia 2011). IRS acknowledges support from a Juan de
la Cierva fellowship from Spanish MINECO.

\bibliography{iopart-num}

\end{document}